\documentclass[12pt]{article}
\usepackage{epsfig}
\usepackage{a4}
\usepackage{latexsym}
\usepackage{cite}
\usepackage{color}

\usepackage{pslatex}
\usepackage[latin1]{inputenc}
\usepackage[T1]{fontenc}

\graphicspath{{bigPics/}}
\DeclareGraphicsExtensions{.eps,.ps}

\def\z#1{{\zeta_{#1}}}

\newcommand{\sspr}{s^{\,\prime}}

\def\MSbar{{$\overline{\mbox{MS}}\,$}}

\newcommand{\gsim}{\raisebox{-0.07cm}{$\:\:\stackrel{>}{{\scriptstyle \sim}}\:\: $} }
\newcommand{\lsim}{\raisebox{-0.07cm}{$\:\:\stackrel{<}{{\scriptstyle \sim}}\:\: $} }

\newcommand{\beq}{\begin{equation}}
\newcommand{\eeq}{\end{equation}}
\newcommand{\bea}{\begin{eqnarray}}
\newcommand{\eea}{\end{eqnarray}}
\newcommand{\nn}{\nonumber}

\newcommand{\hspn}{{\hspace*{-3mm}}}
\newcommand{\hspp}{{\hspace*{3mm}}}

\def\muf{{\mu^{}_{\:\!\!f}}}
\def\mufs{{\mu^{\,2}_{\:\!\!f}}}
\def\mur{{\mu^{}_r}}
\def\murs{{\mu^{\,2}_r}}
\def\mus{{\mu^{\,2}}}

\def\ar{{a_{\rm s}}}
\def\ars{{a_{\rm s}^{\,2}}}
\def\art{{a_{\rm s}^{\,3}}}
\def\as{{\alpha_{\rm s}}}
\def\alss{{\alpha_{\rm s}^{\,2}}}
\def\alst{{\alpha_{\rm s}^{\,3}}}

\def\frct#1#2{\mbox{\large{$\frac{#1}{#2}\:\!$}}}

\def\lnx{\ln x}

\def\lnbeta{\ln \beta}
\def\lnbetas{\ln^{\,2} \beta}
\def\lnbetac{\ln^{\,3} \beta}
\def\lnbetaf{\ln^{\,4} \beta}

\def\lntwo{{\ln 2}}
\def\lntwos{{\ln^{\,2} 2}}
\def\lntwoc{{\ln^{\,3} 2}}

\def\ca{C^{}_A}
\def\cas{C^{\,2}_A}

\def\cf{C^{}_F}
\def\cfs{C^{\,2}_F}
\def\nf{n^{}_{\! f}}

\def\tf{T_F}

\def\Lmmu{\,L_{\mu}\,}
\def\Lmmus{\,L_{\mu}^{\,2}\,}
\def\LQm{\,L_Q\,}
\def\LQms{\,L_Q^{\,2}\,}
\def\LQmt{\,L_Q^{\,3}\,}
\def\H(#1){{\rm H}_{#1}}

\def\Qs{{Q^{\, 2}}}

\begin{document}

\setlength{\parskip}{0.2cm}
\setlength{\baselineskip}{0.52cm}

\begin{titlepage}

\noindent
KEK-TH 1378 \hfill May 2012\\
LTH 944\\
DESY 12-050\\
LPN 12-048 \\
SFB/CPP-12-21
\vspace{1.7cm}
\begin{center}
\Large{\bf
On the next-to-next-to-leading order QCD corrections \\ 
to heavy-quark production in deep-inelastic scattering}\\
\vspace{1.5cm}
\large
H. Kawamura$^{\, a}$, N.A. Lo Presti$^{\, b}$, S. Moch$^{\, c}$ and 
A. Vogt$^{\, b}$\\
\vspace{1.5cm}
\normalsize
{\it $^a$KEK Theory Center \\
Tsukuba, Ibaraki 305-0801, Japan}\\[4mm]

{\it $^b$Department of Mathematical Sciences, University of Liverpool \\
Liverpool L69 3BX, United Kingdom}\\[4mm]

{\it $^c$Deutsches Elektronensynchrotron DESY \\
Platanenallee 6, D--15738 Zeuthen, Germany}\\[2.5cm]
\vfill
\large
{\bf Abstract} 
\vspace{-0.2cm}
\end{center}
The contribution of quarks with masses $m \gg \Lambda_{\rm QCD}$ is the only 
part of the structure functions in deep-inelastic scattering (DIS) which is 
not yet known at the next-to-next-to-leading order (NNLO) of perturbative QCD.
We present improved partial NNLO results for the most important structure 
function $F_{\:\!2}(x,\Qs)$ near the partonic threshold, in the high-energy 
(small-$x$) limit and at high scales $\Qs \gg m^2$; and employ these results 
to construct approximations for the gluon and quark coefficient functions
which cover the full kinematic plane. The approximation uncertainties are
carefully investigated, and found to be large only at very small values,
$x \lsim 10^{\:\!-3}$, of the Bjorken variable.

\vspace{1cm}
\end{titlepage}

\section{Introduction}
\label{sec:Intro}
 
The production of heavy quarks in deep-inelastic lepton-hadron scattering (DIS)
is an important process as it sheds light on a number of interesting issues in 
the theory of the strong interaction, Quantum Chromodynamics (QCD). 
First of all, the production mechanism can be described by standard 
renormalization-group improved perturbation theory for sufficiently large 
values of the momentum transfer $\Qs$ between the lepton and the hadronic 
states.
For this purpose it is a great advantage that one can rely on the 
well-developed framework of the operator product expansion (OPE) which allows 
to understand essential features of heavy-quark DIS.  In addition, the presence
of the heavy-quark mass $m$ introduces a second hard scale into the problem, 
$m^2 \gg \Lambda^2_{\rm QCD\,}$, where $\Lambda_{\rm QCD}$ denotes the QCD 
scale. 
As a consequence, at any fixed order in perturbation theory (and neglecting 
bound-state effects), the heavy-quark mass defines the production threshold for 
the squared center-of-mass (CM) energy $S$, e.g., $S \ge 4\:\! m^2$ for 
heavy-quark pair production.
Moreover the heavy-quark mass acts as a regulator of collinear divergences, 
giving rise to large logarithms of the ratio $\Qs/m^2$ at $\Qs \gg m^2$, i.e., 
in the kinematic regime to be considered when matching QCD with $\nf$ light 
quarks and one heavy flavour to a theory with $\nf+\!1$ light quarks.

Experimentally, heavy-quark production in DIS has been studied in fixed-target 
experiments and, in particular, at the electron-proton collider HERA.
In neutral current reactions, measured with high accuracy at HERA, a 
considerable part of the inclusive DIS cross section at small Bjorken-$x$ is 
due to the production of charm quarks, see, e.g., 
Refs.~\cite{Chekanov:2009kj,Aaron:2009ut,Aaron:2011gp}.
In this kinematic regime heavy-quark DIS is dominated by the photon-gluon 
fusion process $\gamma^{\,\ast}g \rightarrow c \bar{c}\,X$, and the respective 
high-precision measurements can provide invaluable information on 
non-perturbative parameters in the cross sections such as the gluon 
distribution of the proton and the strong coupling constant $\as$.
Given sufficiently accurate data, they can even facilitate determinations of 
the heavy-quark masses in a reaction with space-like momentum transfer.
With more data analyses from HERA run~II still to be finalized, the constraints
from heavy-quark DIS on global fits of parton distribution functions (PDFs) and
its resulting consequences for collider phenomenology are, perhaps, the most 
important aspects in the era of the LHC.

It is thus of great importance to provide as accurate theoretical predictions 
for heavy-quark DIS as possible. Within the standard perturbative approach, 
this requires consideration of higher-order radiative corrections.
At present the theory predictions for neutral current heavy-quark production
rely on the complete next-to-leading order (NLO) QCD corrections 
\cite{Laenen:1992zk} which are often used via the parametrizations of 
Ref.~\cite{Riemersma:1994hv}, see also Ref.~\cite{Harris:1995tu} for a check
and minor corrections.
The complete contribution of the next-to-next-to-leading order (NNLO) is not 
known. However, partial information in various kinematic limits has been 
accumulated over the years.
This includes in particular the logarithmically enhanced terms near threshold 
which are accessible by means of the soft-gluon exponentiation
\cite{Laenen:1998kp}. Also the high-energy limit of heavy-quark DIS has been 
known from small-$x$ resummation to all orders at leading-logarithmic accuracy 
for a long time \cite{Catani:1990eg}.
In the asymptotic regime of $\Qs \gg m^2$ fully analytic results have been 
obtained, and the NLO calculations~\cite{Buza:1995ie,Bierenbaum:2007qe} have 
been extended recently towards NNLO by the computation of a number of lowest 
even-integer Mellin moments 
\cite{Bierenbaum:2008yu,Bierenbaum:2009mv,Bierenbaum:2009zt}, see also
Ref.~\cite{Ablinger:2010ty}.
Finally, the dependence on the renormalization and mass-factorization scales 
is known exactly at NNLO from standard renormalization-group arguments 
\cite{Laenen:1998kp,Alekhin:2010sv} and can be used as an independent 
consistency check in all these limits.

In the present article, we provide approximate NNLO QCD corrections for the 
heavy-quark part of the dominant structure function $F_2^{}$ in photon-exchange
DIS.
This is achieved by extending and, for the first time, combining the available 
NNLO information from the different kinematic regimes, i.e., from threshold
($s \simeq 4\:\!m^2$, where $s$ is the partonic CM energy), high energy 
($s \gg 4\:\!m^2$) and large scales $\Qs \gg m^2$, in particular for the 
dominant gluon-initiated contribution.
Specifically, we present results for all four soft-gluon enhanced logarithms 
near threshold (see Ref.~\cite{Presti:2010pd} for a previous brief account), 
and we employ the all order-result in the high-$s$ limit~\cite{Catani:1990eg} 
to derive an analytic expression for the leading $\ln s$ term at NNLO. Finally,
the known Mellin moments of the heavy-quark operator matrix elements (OMEs) 
\cite{Bierenbaum:2008yu,Bierenbaum:2009mv,Bierenbaum:2009zt}
are employed together with the three-loop results for massless quarks 
\cite{Vermaseren:2005qc} to construct approximate $x$-space expressions for 
the heavy-quark coefficient functions at $\Qs \gg m^2$.
By combining these individual constraints we construct NNLO coefficient 
functions for heavy-quark DIS which, while still being approximate, represent 
the most comprehensive results possible at this point.
Below we will provide a detailed documentation of the required calculations as 
well as estimates of the accuracy of these approximate results.
The resulting improvement of the predictions for heavy-quark DIS and the 
low-$\Qs$ small-$x$ limitations of the present NNLO results are then 
illustrated in a brief phenomenological study.

The remainder of this article is organized as follows:
In Section~\ref{sec:SetStage} we introduce our basic notations. 
Section~\ref{sec:theory} is devoted to the calculation of the new NNLO results 
for the gluon coefficient function near threshold 
(Section~\ref{sec:Thresresum}), at small-$x$ (Section~\ref{sec:small-x}) and 
for scales $\Qs \gg m^2$ (Section~\ref{sec:asy}). In the latter two 
cases we also address the light-quark coefficient function.
In Section~\ref{sec:Coefficientfunctions} we combine these results and provide 
an approximate NNLO expression for both coefficient functions for which we 
perform a number of quality checks. 
Section~\ref{sec:Phenomenology} addresses their impact on the charm- and
bottom-production structure functions $F_2^{\,c}$ and $F_2^{\,b}$ at NNLO and 
the limitations of the present results.  
Finally we conclude and briefly discuss possible future improvements in 
Section~\ref{sec:Conclusions}. 
The appendices contain, in Appendix~\ref{sec:appA}, the derivation of the 
analytical result for the gluon coefficient function at NLO in the limit 
$s \,\simeq\, 4 m^2$ and, in Appendix~\ref{sec:appB}, the (up to two OMEs) 
exact, if somewhat lengthy results for the heavy-quark coefficient functions 
at asymptotic values $\Qs \gg m^2$.

\renewcommand{\theequation}{\thesection.\arabic{equation}}
\setcounter{equation}{0}
\section{Setting the stage}
\label{sec:SetStage}

We start by briefly reviewing the main results on heavy-quark production
in deep-inelastic scattering mediated by neutral-current exchange.
For the scattering of a charged lepton $e$ off a proton $P$ this reaction, 
\begin{equation}
\label{eq:hqborn}
  e(l) \,+\, P(p) \:\:\rightarrow\;\:  
  e(l^{\,\prime}) \,+\, q_h^{}(p_1^{}) \,+\, \bar{q}_h^{}(p_2^{}) \,+\, X \; ,
\end{equation}
is dominated by virtual photon exchange if the momentum transfer 
$\Qs=-q^2=-(l-l^{\,\prime})^2$ is much smaller than the $Z$-boson mass, 
$\Qs\ll M_Z^{\,2}$.
The quark pair $q_h^{}\bar{q}_h^{}$ in the final state is heavy,
$m^2 \gg \Lambda^2_{\rm QCD\,}$. The cross section, integrated over the
momenta of the outgoing heavy (anti-)\,quarks, is written in terms of the 
heavy-flavour structure functions $F_k(x,\Qs,m^2)$ with $k=2,L$ as
\begin{eqnarray}
\label{eq:cross-section}
\frac{d^2\sigma}{dx\,d\Qs} \;= \;
\frac{2\:\!\pi\,\alpha^2}{x\, Q^{\,4}} 
 \left[\{1+(1-y)^2\}\, F_2^{}(x,\Qs,m^2) - y^2 F_L^{}(x,\Qs,m^2)\right] \ ,
\end{eqnarray}
where $\alpha$ is the electromagnetic coupling constant,
$m$ the mass of the heavy quark, and the well-known DIS variables $x$ and $y$
are defined as $x=\Qs/(2p\cdot q)$ and $y=(p\cdot q)/(p\cdot l)$, respectively.

Disregarding power corrections, the heavy-quark contribution 
(\ref{eq:cross-section}) to the DIS structure functions can be written in
terms of a convolution of PDFs and coefficient functions as, see 
Ref.~\cite{Riemersma:1994hv},
\begin{equation}
  \label{eq:totalF2c}
  F_k^{}(x,\Qs,m^2) \;=\;
  {\as\, e_h^{\:\!2}\, \Qs \over 4\:\! \pi^{\:\!2}\, m^2} \,\,
  \sum\limits_{i \,=\, q,{\bar{q}},g} \,\,
  \int_{x}^{\,z^{\,\rm max}}
  {dz \over z} \: f_{i}^{}\left({x \over z},\, \mufs \right)\,
  c_{k, i}^{}\left(\eta(z),\,\xi,\,\mufs,\,\murs \right)
  \; ,
\end{equation}
where $z^{\,\rm max} = 1/(1 + 4\:\! m^2/\Qs)$ and $e_h^{}$ is the heavy-quark 
charge. 
The strong coupling constant at the renormalization scale $\mur$ is denoted by
$\as$, and the PDFs for the parton of flavour $i$ are $f_{i}^{}(x,\mufs)$ at 
the factorization scale $\muf$.
The kinematic variables $\eta$ and $\xi$ in Eq.~(\ref{eq:totalF2c}) are given 
by
\begin{equation}
  \label{eq:eta-xi-def}
  \eta \;=\; {s \over 4\:\! m^2}\: - 1 \quad ,
  \qquad
  \xi \;=\; {\Qs \over m^2} \;\; ,
\end{equation}
and the partonic CM energy is $s = \Qs (1/z-1)$.
Instead of $\eta$, which measures the distance to the partonic threshold,
one often uses the heavy-quark velocity $\beta$ or the variable $\rho$,
\begin{equation}
  \label{eq:beta-def}
  \beta \;=\; \sqrt{1-4\:\!m^2/s}
  \quad ,\qquad
  \rho \;=\; 4\:\! m^2/s
  \;\; .
\end{equation}
The coefficient functions of the hard partonic scattering process can be 
expanded in $\as$ as
\begin{eqnarray}
  \label{eq:coeff-exp}
  c_{k, i}^{}(\eta,\xi,\mus)
  \; = \;
  \sum\limits_{j=0}^{\infty}\, (4\:\! \pi\, \as)^j \, 
  c^{\,(j)}_{k,i}(\eta, \xi, \mus)
  \; = \;
  \sum\limits_{j=0}^{\infty}\, (4\:\! \pi\, \as)^j \,
  \sum\limits_{l=0}^{j} c^{\,(j,\ell)}_{k,i}(\eta, \xi)\: 
      \ln^{\,\ell}\frac{\mus}{m^2} \:\; ,\quad
\end{eqnarray}
where we have identified the renormalization and factorization scales,
$\mu\,=\,\muf\,=\,\mur$.
This can easily be undone by expanding $\as(\muf)$ in terms of $\as(\mur)$
using the standard QCD beta-function.

In the above normalization, the coefficient functions at the leading 
order (LO) read~\cite{Witten:1975bh,Gluck:1979aw,Laenen:1992zk}
\begin{eqnarray}
  \label{eq:Born2}
  c_{2,g}^{\,(0)}(\eta,\xi) &=& 
  c_{T,g}^{\,(0)}(\eta,\xi) + c_{L,g}^{\,(0)}(\eta,\xi)
  \; ,
  \\[1mm]
  \label{eq:BornT}
  c_{T,g}^{\,(0)}(\eta,\xi) &=&
  {\pi\:\! \tf \over 2 (1 + \eta + \xi/4)^3}\,
  \Bigl\{ - 2 \, ( (1 + \eta - \xi/4)^2 + \eta + 1)\, \beta
  \nn \\[-2mm] & & \qquad\qquad\qquad\quad\mbox{}
  + \left( 2 (1 + \eta)^2 + \xi^2/8 + 2\eta + 1 \right) L(\beta) \Bigr\}
  \; ,
  \\[-1mm]
  \label{eq:BornL}
  c_{L,g}^{\,(0)}(\eta,\xi) &=&
  {\pi\:\! \tf\, \xi \over 2 (1 + \eta + \xi/4)^3}\,
  \Bigl\{ 2 \, \beta \, (1 + \eta) - L(\beta) \Bigr\}
\end{eqnarray}
with $L(\beta) = \ln((1+\beta)/(1-\beta))$.
Here $c_{T,g}^{\,(0)}$ and $c_{L,g}^{\,(0)}$ denote the contributions from 
transverse and longitudinal photon polarizations, and $\tf= \frac{1}{2}$ for 
the colour group $SU(N_c)$.
The radiative corrections to Eq.~(\ref{eq:coeff-exp}) at NLO have been known 
for a long time from Ref.~\cite{Laenen:1992zk}.
Unlike the massless coefficient functions, the massive NLO coefficient 
functions $c_{k,i}^{\,(1)}$ cannot be expressed in a simple analytic form.
Instead, compact parametrizations for these functions were presented in 
Ref.~\cite{Riemersma:1994hv}, with minor corrections provided later in
Ref.~\cite{Harris:1995tu} (see Ref.~\cite{Alekhin:2003ev} for parametrizations 
in Mellin-$N$ space including complex values of $N$ as required for a
numerical Mellin inversion).

The NNLO coefficient functions $c_{k,i}^{\,(2,0)}$ in Eq.~(\ref{eq:coeff-exp}) 
are not fully known, although, as outlined above, substantial partial 
information is available.
On the other hand, all scale-dependent terms at this order, i.e., $c_{k,i}
^{\,(2,1)}$ and $c_{k,i}^{\,(2,2)\!}$, have been constructed by means of 
renormalization-group arguments~\mbox{\cite{Laenen:1998kp,vanNeerven:2000uj}}
and are completely known in numerical form, using the parametrized NLO results
and the well-known expressions for the NLO splitting functions, 
see Refs.~\cite{Laenen:1998kp,Alekhin:2010sv}.
In the following we will thus focus on improved predictions for the
scale-independent parts of the NNLO coefficient functions.
To be precise, we will confine ourselves to the gluon coefficient function 
$c_{2,g}^{\,(2,0)}\!$, which is by far the most important contribution to the 
heavy-quark structure function $F_2$ after the convolution with the gluon 
distribution in Eq.~(\ref{eq:totalF2c}), and to the pure-singlet light-quark 
coefficient function $c_{2,q}^{(2,0)}$ for the heavy-quark contribution 
proportional to $e_h^{\,2}$.

For further reference we finally introduce the short-hand notations
\begin{equation}
\label{eq:LQmLmmdef}
  \LQm \:=\: \ln\, \frac{\Qs}{m^2}
             \quad , \quad
  \Lmmu \:=\: \ln\, \frac{m^2}{\mus}
             \quad \mbox{ and } \quad
  \ar \:=\; {\as \over 4\:\!\pi}
\:\; ,
\end{equation}
and we also note that throughout this article we will employ the on-shell 
scheme for the heavy-quark mass $m$, i.e., the pole mass.
Heavy-quark DIS with a running mass in the \MSbar\ scheme has been considered 
in Refs.~\cite{Bierenbaum:2009mv,Alekhin:2010sv}.

\renewcommand{\theequation}{\thesection.\arabic{equation}}
\setcounter{equation}{0}
\section{The NNLO corrections in three kinematic regions}
\label{sec:theory}
In this section we provide improved predictions for the above two NNLO 
coefficient functions $c_{2,g}^{(2,0)}$ and $c_{2,q}^{(2,0)}\!$. We address,
in this order, the threshold region $s \simeq 4\:\! m^2$, the high-energy 
regime $s \gg 4\:\! m^2$, and the high-scale region $\Qs \gg m^2$.
In all these limits the  higher-order corrections exhibit well-known features 
and regularities which we briefly discuss and exploit.
The results will be assembled into approximate coefficient functions for the
whole kinematic plane in Section~\ref{sec:Coefficientfunctions}.
  
\subsection{Threshold limit and soft-gluon resummation}
\label{sec:Thresresum}

Partonic cross sections generally receive large logarithmic corrections near
threshold, which appear as the result of an incomplete cancellation of the 
soft (and collinear) contributions between the real and virtual corrections 
due to the reduced phase-space near threshold.
In the present case of massive quarks, the corrections appear as a function of 
the heavy-quark velocity $\beta$ at each order of the perturbation series in 
the form $\alpha_{\,\rm s}^{\:\!n}\, \ln^{\,m}\beta$ with $m\leq 2n$.
The production threshold corresponds to $\beta\rightarrow 0$, recall
Eq.~(\ref{eq:beta-def}), and the highest logarithms can be resummed to all
orders in perturbation theory, see, e.g., Refs.~\cite{Sterman:1986aj,%
Catani:1989ne,Contopanagos:1996nh,Kidonakis:1997gm,Bonciani:1998vc}.
In addition to the logarithmic terms, heavy-quark pair production also receives
corrections of the form $\alpha_{\rm s}^{\:\!n}/\beta^{\,-m}\ln^{\,\ell}\beta$ 
with $m\leq n$ at all orders from the Coulomb-exchange of gluons between the 
heavy quarks. Also these terms can be resummed~\cite{Hoang:2000yr}.

The {\it raison d'$\,\hat{e}$tre} of phenomenology based on threshold 
resummation derives from the well-known fact, see, e.g., 
Refs.~\cite{Gluck:1993dpa,Vogt:1996wr,Laenen:1998kp}, that, at not too large 
values of $\Qs$, the convolution of the coefficient function for $F_2$ with the
gluon PDF in Eq.~(\ref{eq:totalF2c}) is often dominated by rather low partonic 
CM energies.
Therefore the NNLO predictions of the threshold resummation can provide
useful information on this numerically important contribution to $F_2^{}$.
Previous research has already determined exactly the two highest 
\cite{Laenen:1998kp} and approximately the third \cite{Alekhin:2008hc} 
threshold logarithms at this and all higher orders.
In what follows, we derive exact expressions for all four logarithmically 
enhanced terms together with the complete Coulomb corrections for the NNLO 
gluon coefficient function $c^{^{}(2)}_{2,g}$ in Eq.~(\ref{eq:coeff-exp}), a 
result that we have already reported in a numerical form in 
Ref.~\cite{Presti:2010pd}.
The (light-)$\,$quark coefficient function $c_{2,q}^{}$ for heavy-quark 
DIS is non-vanishing only from NLO and does not exhibit any enhancement near 
threshold.

According to the general formula for the threshold resummation 
\cite{Contopanagos:1996nh,Kidonakis:1997gm,Bonciani:1998vc},
the gluon coefficient function in Mellin space 
(recall Eqs.~(\ref{eq:eta-xi-def}) and (\ref{eq:beta-def}) for the definitions 
of $\xi$ and $\rho$) is of the form
\begin{eqnarray}
\label{eq:resummation}
  c_{2,g}^{}(\as,N) &=& \int_0^1 \! d\rho \, \rho^{N-1} c_{2,g}^{}(\xi,\beta)
\nn\\[1mm]
  &=&
  c^{\,(0)}_{2,g}(N) \, \cdot \, g_0^{}(\as, N) \, \cdot \,
  \exp\left[ G_N(\as,m^2) \right] \, \cdot \,
  \left( 1 + {\cal O} (\:\!N^{\,-1} \ln^{\,k} N \:\!) \right)
\; .
\end{eqnarray}
The coefficient $c^{\,(0)}_{2,g}(N)$ is the Mellin transform of the LO 
coefficient function of Eq.~(\ref{eq:Born2}). In the large-$N$ limit
(which is dominated by the transverse component $c_{T,g}^{}$) it is given by
\begin{equation}
\label{eq:N-c2g0-thresh}
  c^{\,(0)}_{2,g}(N) \, = \,
  \frac{\pi\:\! \tf}{(1+\xi/4)}\;\frac{\sqrt{\pi}}{2\,N^{\,3/2}}\:
  \left( 1 + {\cal O} (1/N) \right)
\; .
\end{equation}
$g_0^{}(\as, N)$ denotes the matching function to be discussed below, 
cf.~Eq.~(\ref{eq:g0-matching}).
The exponential factor resums the threshold logarithms to all orders in $\as$.
The exponent can be written in the standard form 
\cite{Bonciani:1998vc,Moch:2008qy,Presti:2010pd},
\begin{equation}
  \label{eq:GN-exponent}
  G_N \; = \:
  \int_0^1 dz\;\frac{z^{\,N-1} -1}{1-z} \,
  \bigg\{
    \int_{\mus}^{\,4\:\!m^2(1-z)^2}
    \frac{dq^2}{q^2}\: A_g\left(\as(q^2)\right)
    + D_{\,\gamma^*g\rightarrow q_h\bar{q}_h}
      \left(\as\left(4\:\!m^2[1-z]^2\:\!\right)\right)\!
  \bigg\}
  \, .
\end{equation}
Here the first term includes the gluonic cusp anomalous dimension $A_g$, known 
to order $\alst$~\cite{Moch:2004pa,Vogt:2004mw}, which governs the 
process-independent soft-collinear gluon emission off the initial gluon.
The second term, $D_{\,\gamma^*g\rightarrow q_h\bar{q}_h}$, collects 
process-dependent effects of soft-gluon emission from the initial- and 
final-state particles and is built as
\begin{eqnarray}
  \label{eq:DggamQQ}
  D_{\,\gamma^*g\rightarrow q_h\bar{q}_h}(\as)
  \; = \;
  \frct{1}{2}\, D_{g}(\as) + D_{q_h\bar{q}_h}(\as) 
\end{eqnarray}
from $D_{g}$ and $D_{q_h\bar{q}_h}$ which are, respectively, the soft 
anomalous dimension for Higgs production in gluon-gluon fusion
(known to order $\alst$~\cite{Moch:2005ky,Idilbi:2005ni})
and for the heavy-quark production in the color-octet channel 
(known to order $\alss$~\cite{Beneke:2009rj,Czakon:2009zw,Beneke:2009ye},
see also Refs.~\cite{Becher:2009kw,Ferroglia:2009ep}).
For reference we assemble $D_{\,\gamma^*g\rightarrow q_h\bar{q}_h}$,
employing the usual notation (\ref{eq:LQmLmmdef}) for the strong coupling,
\begin{eqnarray}
\label{eq:D2loop}
 D_{\,\gamma^*g\rightarrow q_h\bar{q}_h}(\as) &\!=\!&
 - 4\,\ca\, \ar
 + 4\, \ca \left[
  \ca  \left(- \frct{547}{27} + \frct{28}{3} \z2 + 5\,\z3\right)
  +  \nf \left(\frct{94}{27} - \frct{4}{3} \z2 \right)
 \right] \ars
 +{\cal O}(\art) \; .
 \nn \\ \, \\[-1.2cm] \nn
\end{eqnarray}
Here and below $\zeta_n$ stands for values of the Riemann zeta-function. 

To next-to-next-to-leading logarithmic (NNLL) accuracy,
as needed for the NNLO corrections we are considering,
the exponent $G_N$ in Eq.~(\ref{eq:GN-exponent}) takes the form
\begin{equation}
\label{eq:exponent2}
  G_N \;=\;
  \ln N \, g_1^{}(\lambda)
  \,+\,g_2^{}(\lambda)
  \,+\,\ar\,g_3^{}(\lambda)
  \,+\,\dots\;
\end{equation}
with $\lambda = \beta_0\,\ar\ln N$ and, again, $\ar = \as/(4 \pi)$.
The functions $g_i~(i=1,2,3,\ldots)$ are defined such that $g_i(0)=0$, with
the first term $\ln N \, g_1^{}$ resumming the leading logarithms and so on.
Explicit expressions for $g_i$ can be obtained from Ref.~\cite{Moch:2008qy},
even with the $\mur$ and $\muf$ dependence separated
(for the computation see, e.g.,~Refs.~\cite{Vogt:2000ci,Moch:2005ba}),
with the obvious replacements
$A_q \to \frac{1}{2}\, A_g$, $D_q \to \frac{1}{2}\, D_g$
and $D_{Q\bar{Q}} \to - D_{q_h\bar{q}_h}$ in Eq.~(A.5) -- (A.9)
of~Ref.~\cite{Moch:2008qy} (note especially the unconventional sign for
the soft anomalous dimension of color-octet heavy-quark production, there
denoted by $D_{Q\bar{Q}}\:\!$).
Below we will expand in terms of $\ln \,\tilde{N} = \ln \,N + \ln\, 
\gamma_{\,\rm e}$, which absorbes the otherwise ubiquitous Euler-Mascheroni 
constant into the logarithms.

The process-specific new information needed for heavy-quark DIS resides 
entirely in the matching function $g_0^{}(\as, N)$ in 
Eq.~(\ref{eq:resummation}) which we need to discuss next. 
This function can be expressed as a product of a hard coefficient 
$g_0^{h}(\as)$ and a Coulomb coefficient $g_0^{c}(\as, N)$, where the latter 
specifically induces a dependence on the Mellin variable $N$,
\begin{eqnarray}
\label{eq:g0-matching}
  g_0^{}(\as, N) &=&
  g_0^{\,h}(\as)\; g_0^{\,c}(\as, N)
  \nn\\
  &=&
  \big(1\:+\: \ar\:g_0^{\,h\,(1)}\: + \: \ars\:g_0^{\,h\,(2)}\:+\;\dots\,\big) 
  \;
  \big(1\:+\: \ar\:g_0^{\,c\,(1)}\: + \: \ars\:g_0^{\,c\,(2)}\:+\;\dots\,\big)
  \nn\\[0.5mm]
  &=&
  1\:+\: \ar \left( g_0^{\,h\,(1)} + g_0^{\,c\,(1)} \right)
  \:+\:\: \ars \left( g_0^{\,h\,(2)} + g_0^{\,c\,(2)} 
  + g_0^{\,h\,(1)}\cdot g_0^{c\,(1)} \right)\: +\;\ldots
  \;\: .
\end{eqnarray}
Such a factorized form for the Coulomb corrections and the threshold 
logarithms has been discussed for top-quark pair production at hadron colliders
before~\cite{Hagiwara:2008df,Kiyo:2008bv}, see also 
Refs.~\cite{Beneke:2009rj,Beneke:2011mq}
for studies in the framework of soft-collinear effective theory.
For heavy-quark DIS a similar factorization applies to the gluon coefficient 
function, since the structure of radiative corrections to the DIS subprocess
$\,\gamma + g \rightarrow q_h+\bar{q}_h+X$ is similar to top-quark 
hadro-production in $\,g+g\rightarrow q_h+\bar{q}_h+X$, i.e., essentially the 
same up to replacements of the corresponding color factors.

The matching coefficients are fixed by comparing the (Mellin-inverted) 
expansion of Eq.~(\ref{eq:resummation}) to known fixed-order results, in our
case those of Ref.~\cite{Laenen:1992zk}.
The calculation required in this case is discussed in Appendix~\ref{sec:appA} 
and leads to
\begin{eqnarray}
  g_0^{\,c\,(1)}(N) \!&\!=\!& 2\:\!\pi^2\,\left( \cf  - \frac{\ca }{2} \right)
  \sqrt{\frac{4N}{\pi}}
  \, , \\[0.5mm]
  g_0^{\,h\,(1)}\phantom{(N)} &\!=\!&
  \left(4\:\!\ln^22 - 4\:\!\ln2 - 40+ 2\:\!\pi^2\right)\,\ca + c_0^{}(\xi)
  + \left[ \:\!8 \,\ca \, - \bar{c}_0^{}(\xi) \right] \Lmmu \; ,
\label{eq:nlo-matching}
\end{eqnarray}
where
\begin{eqnarray}
\label{eq:nloconst}
c_{0}^{}(\xi) &\!\!=\!\!&
  \ca \* \bigg\{50 - \pi^2 + 12\,\frac{\ln(\sqrt{\xi}(y - 1)/2)}{y}
  + 4\,\ln^2(\sqrt{\xi}(y - 1)/2) + \ln^2(1 + \xi/2)
\nonumber\\
&& \mbox{\hspn}
  + 6\,\ln(2 + \xi/2)- 4\,\ln^2(2 + \xi/2) +
  2\,\rm{Li}_2( - \frac{2}{2 + \xi}) + \frac{48}{2 + \xi}
  - 4\,\frac{\ln(2 + \xi/2)}{2 + \xi}
\nonumber\\
&& \mbox{\hspn}
  + 64\,\frac{\ln(2 + \xi/2)}{(2 + \xi)^2}
  - 128\,\frac{\ln(2 + \xi/2)}{(2 + \xi)^2(4 + \xi)}
  - \,\frac{160}{(2 + \xi)(4 + \xi)}
  - 64\,\frac{\ln(2 + \xi/2)}{(2 + \xi)(4 + \xi)}
\nonumber\\
&& \mbox{\hspn}
  + \frac{128}{(2 + \xi)(4 + \xi)^2}
  - 12\,\frac{4 + \,\zeta_2}{4 + \xi}
  - 8\,\frac{\ln^2(\sqrt{\xi}(y - 1)/2)}{4 + \xi} 
  + \frac{64}{(4 + \xi)^2} \bigg\}
\\
&& \mbox{\hspn} \mbox{\hspn}
  + \cf \* \bigg\{- 18 - \frac{2}{3}\,\pi^2
  - 24\,\frac{\ln(\sqrt{\xi}(y - 1)/2)}{y}
  - 8\,\ln^2(\sqrt{\xi}(y - 1)/2) + 2\,\ln^2(1 + \xi/2)
\nonumber\\
&& \mbox{\hspn}
  - 6\,\ln(2 + \xi/2) + 4\,\rm{Li}_2( - \frac{2}{2 + \xi})
  - \frac{48}{2 + \xi} + 8\frac{\ln(2 + \xi/2)}{2 + \xi}
  + \frac{360}{(2 + \xi)(4 + \xi)}
\nonumber\\
&& \mbox{\hspn}
  + 128\,\frac{\ln(2 + \xi/2)}{(2 + \xi)(4 + \xi)} 
  - \frac{544}{(2 + \xi)(4 + \xi)^2}
  + 48\,\frac{\ln^2(\sqrt{\xi}(y - 1)/2)}{4 + \xi}
  - 8\,\frac{\ln^2(1 + \xi/2)}{4 + \xi} 
\nonumber\\
&& \mbox{\hspn}
  + \frac{44 + 40\,\zeta_2}{4 + \xi}
  - 120\,\frac{\ln(2 + \xi/2)}{(2 + \xi)^2}
  + 256\,\frac{\ln(2 + \xi/2)}{(2 + \xi)^2(4 + \xi)}
  - 16\frac{\rm{Li}_2( - \frac{2}{2 + \xi})}{(4 + \xi)}
  - \frac{272}{(4 + \xi)^2}
  \bigg\}
\nonumber \; ,
\\[2mm]
\label{eq:c0bar}
  \bar{c}_0^{}(\xi) &=& 4\, \ca
  \left[2+\ln\left(1+\frac{\xi}{4}\right)\right]-\frac{4}{3}\,\tf
\end{eqnarray}
with $\,y = \sqrt{1 + 4/\xi}\,$ and $\ca=3$, $\cf=4/3$ in QCD.
The additive constants taken out of $c_0^{}(\xi)$ and $\bar{c}_0^{}(\xi)$ in
Eq.~(\ref{eq:nlo-matching}) are conventional; the reason for the above choice 
will become clear below.

The Coulomb coefficient $g_0^{\,c}(\as, N)$ in Eq.~(\ref{eq:g0-matching}) can
be obtained to order $\alss$ by making use of the result of the 
non-relativistic (NR) cross section in $e^+e^-\rightarrow q_h^{}\bar{q}_h^{}$ 
calculated up to this order in Refs.~\cite{Czarnecki:2001gi,Pineda:2006ri}.
The only difference in the present case is the color structure, i.e., the
Coulomb corrections for the color-octet state require the colour factor 
replacement $\cf \to (\cf  - \ca /2)$ in the corresponding $e^+e^-$ result 
\cite{Czarnecki:2001gi,Pineda:2006ri}.
Thus we identify the NR part of the NNLO coefficient function~as
\begin{eqnarray}
\label{eq:nnlo-coulomb} c_{2,g}^{(2)}(\xi,\beta) &\stackrel{\rm NR}=&
 \frac{c_{2,g}^{\,(0)}(\xi,\beta)}{(4\pi)^4} \,\left( \cf  - \frac{\ca }{2} 
 \right)
\left[
  \left(
    \frac{62}{9}\,\ca  - \frac{20}{9}\, \nf  - 4\beta_0\ln(2\beta)
    - 2\beta_0 \Lmmu
  \right) \frac{\pi^{\,2}}{\beta}\right.
\nonumber\\ &&
\left.
  + \,\frac{4}{3}\left(\cf -\frac{\ca }{2}\right)\frac{\pi^{\,4}}{\beta^2}
  - 32\,\cf \,\pi^{\,2}\,\ln\beta
\right] + {\cal O}(\beta)
\, ,
\end{eqnarray}
where $\beta_0 = 11/3 \ca - 2/3 \nf$ is the leading coefficient of the QCD 
beta function and we have removed the term corresponding to the product of 
the NLO hard and NLO Coulomb corrections to avoid double counting.  
Note that the single logarithmic term originates in the non-leading part of the
NR QCD potential~\cite{Beneke:1999qg} and is not included in the soft-gluon 
exponent $G_N$.
Performing the Mellin transform, we obtain $g_0^{\,c\,(2)}$ as
\begin{eqnarray}
\label{eq:nnlo-coulomb-N}
g_0^{\,c\,(2)} &\!=\!\!&
\pi^{\,2}\left(  \cf  - \frac{\ca }{2}  \right)\,
\left[\left(
    \frac{62}{9}\,\ca - \frac{20}{9}\, \nf  - 4\,\beta_0\ln2
    - 2\:\!\beta_0 \Lmmu
  \right)\sqrt{\frac{4N}{\pi}}
\right.
\nonumber\\ &&
\left. \mbox{}
  + 2\:\!\beta_0\sqrt{\frac{4N}{\pi}}\ln\tilde{N}
  + \frac{4}{3}\left( \cf  - \frac{\ca }{2} \right)\pi^{\,2}\cdot (2\,N)
  - 32\,\cf \left(- {1 \over 2}\ln\tilde{N} + 1 - \ln2 \right) \right]
\, . \quad
\end{eqnarray}

The only missing information for the complete NNLL + NNLO resummation is the 
coefficient $g_0^{\,h\,(2)}$ in Eq.~(\ref{eq:g0-matching}), which requires 
the presently unknown NNLO constants in $\beta$, i.e., the two-loop analogues
of $c_0^{}(\xi)$ and $\bar{c}_0^{}(\xi)$ in Eqs.~(\ref{eq:nloconst}) and
(\ref{eq:c0bar}), which can be determined only from a full three-loop 
calculation of the heavy-quark contribution to DIS in the soft-gluon limit.
Setting this constant to zero in $\beta$-space, we find
\begin{eqnarray}
\label{eq:nnlo-matching}
  g_0^{\,h\,(2)} &\!\!=\!&
\left(\,\frct{135416}{27}\,\ln2 - \,\frct{35888}{9}\,\ln^22
  + \,\frct{11672}{9}\,\zeta_3 - \,\frct{88856}{27} + \,\frct{1340}{9}\,\pi^2
  + \,\frct{16}{3}\,\pi^4
\right.
\nonumber\\ &&
\left.
  - \,\frct{932}{3}\,\pi^2\,\ln2 - \,1384\,\zeta_3\,\ln2
  + \,\frct{488}{3}\,\pi^2\,\ln^22 + \,\frct{21584}{9}\,\ln^32
  - \,640\,\ln^42\right)  \cas
\nonumber\\ &&
+ \,\left(\frct{4592}{27} - \,\frct{224}{9}\,\zeta_3- \,\frct{80}{9} \,\pi^2
  - \,\frct{6512}{27}\,\ln2- \,\frct{416}{9}\,\ln^32  + \,8\,\pi^2\,\ln2
  + \frct{1376}{9}\,\ln^22\right) \ca \, \nf
\nonumber\\ &&
+ \, ( - 32\,\ln^22 + 56\,\ln2 + 2\,\pi^2 - 40)\,\ca \,c_0(\xi)
\nonumber\\[1mm] &&
- \left[\left(- \,\frct{152}{3}\,\pi^2\,\ln2
    + \,288\,\ln^32 + \,\frct{6128}{9}\,\ln2 + \,34\,\pi^2
    - \,\frct{1888}{3}\,\ln^22 + \,224\,\zeta_3 - \frct{4160}{9}\right) \cas
\right.
\nonumber\\[1mm] &&
+ \left( \frct{320}{9} + \frct{64}{3}\,\ln^22 - \frct{416}{9}\,\ln2
  - \frct{4}{3}\,\pi^2\right)\,\ca \, \nf
+ \,(8\,\ln2 - 8 )\,\ca \,c_0(\xi)
\nonumber\\ &&
+ \, (2\,\pi^2 - \,32\,\ln^22 + \,56\,\ln2 - \,40)\,\ca \,\bar{c}_0(\xi)
\bigg]\,\Lmmu
+ \left[\left( 4\,\pi^2 - \,32\,\ln^22 - \,\frct{44}{3} + \,\frct{44}{3}\,\ln2
    \right) \cas \right.
\nonumber\\ &&
\left.
  + \left(\frct{8}{3} - \frct{8}{3}\,\ln2\right)\,\ca \, \nf
  + (8\,\ln2 - \,8)\,\ca \,\bar{c}_0(\xi)\right] \Lmmus
\end{eqnarray}
which cancels all NNLO constant terms originating in the inverse Mellin 
transform of the \mbox{$N$-space} resummation formula (\ref{eq:resummation}).
 
Supplemented by an estimate of the missing $\beta$-space constant, e.g., 
from a Pad{\'e} estimate as in Eq.~(\ref{eq:c2g2-const}) below, the results in 
Eqs.~(\ref{eq:exponent2}) -- (\ref{eq:nnlo-matching}) can be used to provide an
approximate NNLL + NNLO exponentiation of the gluon coefficient function in 
heavy-quark DIS, which is analogous to the case of hadronic top-quark pair 
production in Refs.~\cite{Bonciani:1998vc,Moch:2008qy}.

In the present article, though, we are rather concerned with improved 
predictions for $c_{2,g}^{}$ to NNLO, essentially using 
Eq.~(\ref{eq:resummation}) to generate results to fixed order in perturbation 
theory. To that end  we now write down the $\beta$-space threshold behaviour
for $c_{2,g}^{}$ up to NNLO. 
The LO and NLO results \cite{Witten:1975bh,Gluck:1979aw,Laenen:1992zk} are
\begin{eqnarray}
  \label{eq:c2g0-thresh}
  c_{2,g}^{\,(0)}(\xi,\beta) &\!=\!&
  \pi \tf\,\frac{\beta}{1+\xi/4}
  + {\cal O}(\beta^3)
\end{eqnarray}
and 
\begin{eqnarray}
\label{eq:c2g1-thresh}
c_{2,g}^{\,(1)}(\xi,\beta) &\!=\!& \frac{c_{2,g}^{\,(0)}(\xi,\beta)}{(4\pi)^2}
\,
\bigg\{
   16\,\* \ca \* \lnbetas
   + [ 48\,\* \ca \* \ln2 - 40\,\* \ca ] \* \lnbeta
   + (2 \* \cf  - \ca )\, \* \frac{\pi^2}{\beta}
   + 8\, \* \ca \* \lnbeta \* \Lmmu 
\bigg\}
\nonumber\\[1mm] && \mbox{}
 +\, c_{2,g}^{\,(1){\rm const}}(\xi,\beta) 
 \,+\, {\cal O}(\beta^2)
\end{eqnarray}
with
\begin{equation}
  \label{eq:c2g1-const}
  c_{2,g}^{\,(1){\rm const}}(\xi,\beta) \;=\; 
  \frac{c_{2,g}^{\,(0)}(\xi,\beta)}{(4\pi)^2}
  \left\{
  c_0^{}(\xi) \,+\, 36\,\ca\,\ln^22 \,-\, 60\,\ca\,\ln2
   + \Lmmu \* \Big[
   8 \* \ca \* \ln2 - \bar{c}_0^{}(\xi)
   \Big]
  \right\}
\end{equation}     
in terms of $c_0^{}(\xi)$ and $\bar{c}_0^{}(\xi)$ in Eqs.~(\ref{eq:nloconst}) 
and (\ref{eq:c0bar}). The convention for these constants derives from 
Refs.~\cite{Laenen:1992zk,Riemersma:1994hv}, where the logarithmic enhancement
was expressed in terms of $\ln^{\,\ell} (8\:\!\beta^2)$, $\ell = 1,\,2$, and  
$\ln (4\:\!\beta^2)$ for the scale-independent and scale-dependent parts, 
respectively.

The corresponding NNLO result $c_{2,g}^{\,(2)}$ is obtained from our results
above using the Mellin transform in  Appendix~A of Ref.~\cite{Moch:2008qy},
\begin{eqnarray}
\label{eq:c2g2-thresh} 
c_{2,g}^{\,(2)}(\xi,\beta) &\!=\!&
\frac{c_{2,g}^{\,(0)}(\xi,\beta)}{(4\pi)^4}\,
\Bigg\{
   128\, \* \cas \* \lnbetaf
\\ && \hspn
 + \bigg[
   \bigg( 768\, \* \ln2 - \frac{6464}{9}\, \bigg) \* \cas 
   + \frac{128}{9}\, \* \ca \* \nf + 128\,\* \cas\* \Lmmu
   \bigg]\* \lnbetac
\nonumber\\ && \hspn
 + \bigg[
   \bigg( 1728\, \* \lntwos - 3232\, \* \ln2 - \frac{208}{3}\, \* \pi^2 
   + \frac{15520}{9}\, \bigg)\*\cas
\nonumber\\ &&
   + \bigg( 64\, \* \ln2 - \frac{640}{9}\, \bigg) \* \ca \* \nf
   + 16\, \* \ca \* c_0(\xi) + 32\, \* \ca \* \bigg(\cf - \frac{\ca}{2}\bigg) 
     \* \frac{\pi^2}{\beta}
\nonumber\\ &&
   - \bigg\{ \bigg( -512\, \* \ln2 + \frac{1136}{3}\, \bigg) \* \cas 
   - \frac{32}{3}\, \* \ca \* \nf
   + 16\, \* \ca \* \bar{c}_0(\xi) \bigg\} \* \Lmmu
   + 32\, \*\cas\* \Lmmus \bigg] \* \lnbetas
\nonumber\\ && \hspn
 + \bigg[
   \bigg( 1728\, \* \lntwoc - 4848\, \* \lntwos + \frac{15520}{3}\, \* \ln2 
   - 208\, \* \pi^2 \* \ln2
   + 936\, \* \z3 + \frac{608}{3}\, \* \pi^2 
\nonumber\\ &&
   - \frac{88856}{27}\, \bigg) \* \cas
   + \bigg( 96\, \* \lntwos - \frac{640}{3}\, \* \ln2 - \frac{16}{3}\, \* \pi^2
   + \frac{4592}{27}\, \bigg) \* \ca \* \nf
   - 32\, \* \cf \* \bigg(\cf - \frac{\ca}{2} \bigg) \* \pi^2
\nonumber\\ &&
   + (48\, \* \ln2 - 40) \* \ca \* c_0(\xi)
   + \bigg\{ \bigg(-\frac{92}{3} + 32\, \* \ln2 \bigg) \* \ca
   + \frac{8}{3}\, \* \nf \bigg\} \* \bigg(\cf - \frac{\ca}{2}\bigg) 
     \* \frac{\pi^2}{\beta}
\nonumber\\ &&
   - \bigg\{\bigg( -672\, \* \lntwos + 976\, \* \ln2 + \frac{104}{3}\, \* \pi^2
   - \frac{4160}{9}\,\bigg) \* \cas
   + \bigg( - 32\, \* \ln2 + \frac{320}{9}\, \bigg) \* \ca \* \nf
\nonumber\\ &&
   + (48\, \* \ln2 - 40)\, \* \ca \* \bar{c}_0(\xi) - 8\, \* \ca \* c_0(\xi)
   - 16\, \*\ca \* \bigg(\cf - \frac{\ca}{2} \bigg) \* \frac{\pi^2}{\beta}
   \bigg\} \* \Lmmu
\nonumber\\ &&
   + \bigg\{\bigg(64\, \* \ln2 - \frac{44}{3}\,\bigg) \* \cas
   + \frac{8}{3}\, \* \ca \* \nf   - 8\, \* \ca \* \bar{c}_0(\xi) \bigg\} 
     \* \Lmmus \bigg] \* \lnbeta
\nonumber\\ && \hspn
   + \bigg[
   \bigg(8\, \* \lntwos - \frac{68}{3}\, \* \ln2 + \frac{8}{3}\, \* \pi^2 
   - \frac{658}{9}\, \bigg) \*\ca
   + \bigg(\frac{8}{3} \* \ln2 - \frac{20}{9} \bigg) \* \nf  + 2 \* c_0(\xi)
\nonumber\\ &&
   + \bigg(\frac{26}{3}\, \* \ca + \frac{4}{3}\, \* \nf 
   - 2\, \* \bar{c}_0(\xi) \bigg) \* \Lmmu
   \bigg]\* \bigg(\cf  - \frac{\ca}{2} \bigg) \* \frac{\pi^2}{\beta}
   \,+\, \frac{4}{3} \* \bigg(\cf - \frac{\ca}{2}\bigg)^2\, 
   \* \frac{\pi^4}{\beta^2}
\, \Bigg\}
+ {\cal O}(\beta)
\; .
\nonumber
\end{eqnarray}
Several checks can be performed of this result which provides, for the first
time, all logarithmically and Coulomb-enhanced contributions.
First of all, the scale dependence at NNLO is fully known 
\cite{Laenen:1998kp,Alekhin:2010sv} from the renormalization group (see also 
Section~\ref{sec:Coefficientfunctions}),
and we have verified that all terms proportional to powers of $\Lmmu$ in the 
threshold expansion in Eq.~(\ref{eq:c2g2-thresh}) agree.
Another strong check is the agreement of Eq.~(\ref{eq:c2g2-thresh}) with the 
results presented in Eq.~(A.1) of Ref.~\cite{Beneke:2009ye} for the hadronic 
heavy-quark production in the gluon-gluon-fusion channel, 
$gg \to q_h\bar{q}_h$, after replacing one gluon by a photon 
(together with the appropriate substitutions of the colour factors).
 
The NNLO constants in $\beta$ multiplying the Born term are currently unknown.
This includes the scale-dependent term $c_{2,g}^{\,(2,1)}(\xi,\eta)$ which
is only known in a numerical form beyond the threshold logarithms.
Based on Eq.~(\ref{eq:c2g1-const}), one can write down a [0$/$1] Pad{\'e} 
estimate, 
\begin{equation}
  \label{eq:c2g2-const}
  c_{2,g}^{\,(2){\rm const}}(\xi,\beta) \;\approx\; 
  \frac{c_{2,g}^{(0)}(\xi,\beta)}{(4\pi)^4}
  \left(
  c_0^{}(\xi) \,+\, 36\,\ca\,\ln^22 \,-\, 60\,\ca\,\ln2
   + \Lmmu \* \bigg[
   8\, \* \ca \* \ln2 - \bar{c}_0^{}(\xi)
   \bigg]
  \right)^2
  \; ,
\end{equation}     
of which we will employ the scale-independent part at low $\xi$ in 
Section~\ref{sec:Coefficientfunctions} below.

\pagebreak

\begin{figure}[t!]
\centering
 {
 \includegraphics[width=15.5cm]{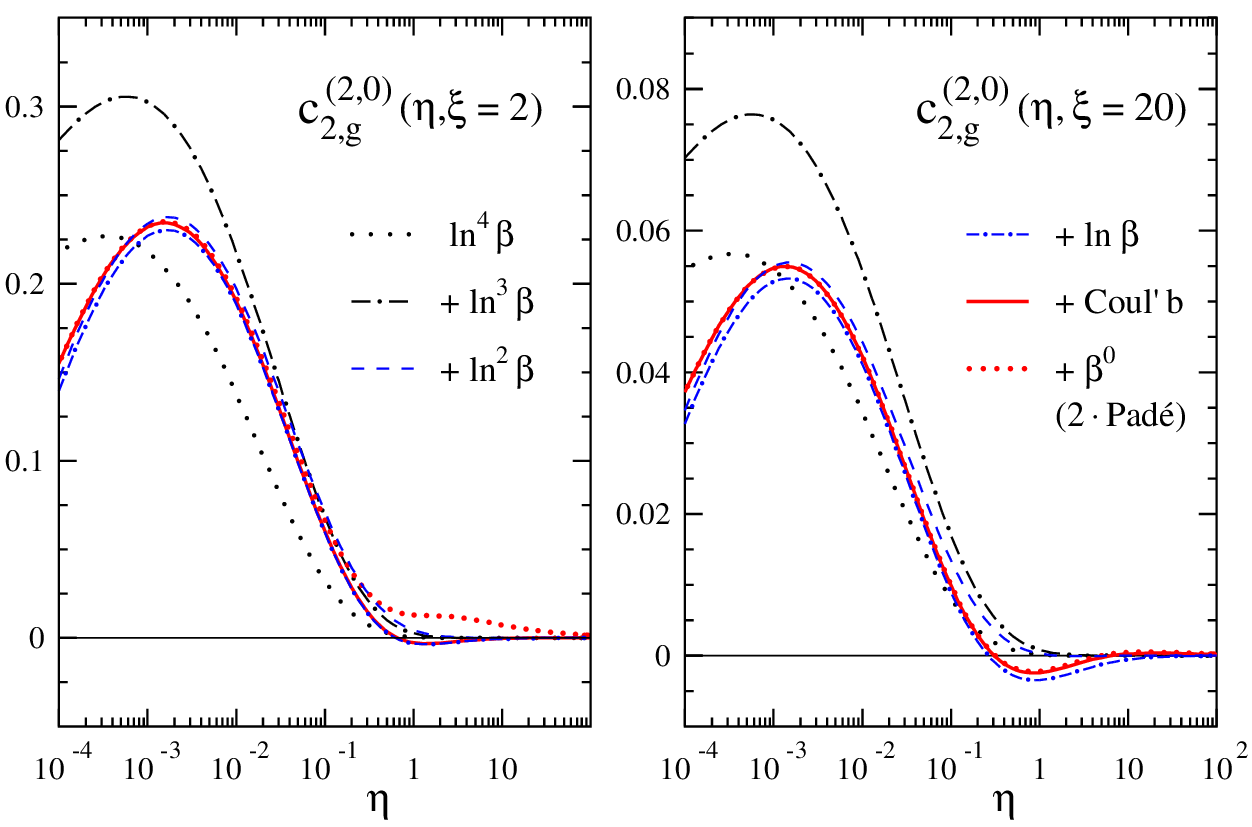}
 }
\vspace*{-1mm}
 \caption{ \small
 \label{fig:cg20thresh}
 The successive logarithmic threshold approximations for the NNLO gluon 
 coefficient function $c_{2,g}^{\,(2)}$ in Eq.~(\ref{eq:c2g2-thresh}) and the 
 $1/\beta$ Coulomb contributions at two typical values of $\xi = \Qs/m^2$ 
 depending on $\eta$ defined in Eq.~(\ref{eq:eta-xi-def}). 
 The two highest logarithms were already known from Ref.~\cite{Laenen:1998kp}.
 }
 \vspace*{-2mm}
\end{figure}
The main new NNLO result of this section, the scale-independent part of 
Eq.~(\ref{eq:c2g2-thresh}), is illustrated in Fig.~\ref{fig:cg20thresh} for 
two values of $\Qs$. 
It is clear that a stable low-$\eta$ result is established only by including
our new results for the third and fourth logarithms. As a first estimate of the
remaining uncertainly at $\eta \gsim 1$ for low values $\xi \gsim 1$
we have included the Pad{\'e} estimate (\ref{eq:c2g2-const}), assigning a
100\% uncertainty (its small value at $\xi = 20$ is accidental and irrelevant
in what follows).

\subsection{High-energy expansion}
\label{sec:small-x}

Let us next discuss the high-energy limit of the coefficient functions 
$c_{2,g}^{}$ and $c_{2,q\,}^{}$. To that end we build on the results of 
Ref.~\cite{Catani:1990eg} from the small-$x$ resummation, which had been 
derived even before NLO corrections were available and thus served as a check 
on Ref.~\cite{Laenen:1992zk}. 
At fixed $\Qs$ the high-energy limit $s = \Qs(1/x-1) \to \infty$ implies 
$x \to 0$, thus $\eta \to \xi/(4 x) + {\cal O}(1)$ due to 
Eq.~(\ref{eq:eta-xi-def}). Therefore we consider the coefficient function 
$c_{2,g}^{}$ as a function of $x$ throughout this section. 
In Mellin $N$-space, cf.~Eq.~(\ref{eq:resummation}) for the definition, the 
dominant contributions in this region behave as $\alpha_{\rm s}^{\,n}/N^n$ 
and can be resummed to all orders.

One can define the partonic cross section $\hat{\sigma}_{2,N}^{}$ in Mellin 
space corresponding to the gluon coefficient functions $c_{2,g}^{}$ 
(see Eq.~(\ref{eq:resummation})) as
\begin{equation}
  \label{eq:partonic-crs}
  m^2 \hat{\sigma}_{2,N}^{} \;=\;
  \alpha\,\alpha_s\, e_q^2\, c_{2,g}^{}(\alpha_s,N)
  \; ,
\end{equation}
which contributes to the structure function $F_2$ via
Eqs.~(\ref{eq:cross-section}) and (\ref{eq:totalF2c}). In complete analogy, if 
multiplied with the Mellin moments of the gluon PDF $f_{g,N}$, 
$\sigma_{2,N}^{}$ also defines the gluonic contribution to the total hadronic 
cross section. 
It is an interesting observation of Ref.~\cite{Catani:1990eg} that the 
$\Qs$-dependence in this quantity can be factorized if one considers 
the photo-production limit, i.e., the process $\gamma \,g \to 
q_h^{} {\bar q}_h^{} X$ with a real photon. This leads to the ansatz
\begin{equation}
  \label{eq:Kxi-factor}
  \sigma_{2,N}^{}(\Qs,m^2) \; = \;
  K_{2,N}(\xi) \: \sigma_{\,\gamma\,g,N}(m^2)
  \; .
\end{equation}
The function $K_{2,N}$ summarizes all $\xi$-dependence originating
from the momentum transfer $\Qs$ of the off-shell photon in DIS
electro-production of heavy quarks, and $\sigma_{\,\gamma\,g,N}$
denotes the Mellin moments of the gluonic contribution to the total
hadronic cross section for photo-production of heavy quarks, see, e.g.,
Ref.~\cite{Ellis:1988sb,Ball:2001pq}.

The resummation of the logarithms at high energy is based on the
framework of un-integrated PDFs in transverse momentum $k_t$ and the
concept of $k_t$-factorization, a procedure which involves two
steps: first computing amplitudes with the initial particles
off-shell in $k_t$, and second performing the convolution with a
gluon PDF which has the small-$x$ corrections included. For
heavy-quark photo-production, this allows to express
$\sigma_{\,\gamma\,g,N}$ in Eq.~(\ref{eq:Kxi-factor}) as the
product~\cite{Catani:1990eg} (cf.~also Ref.~\cite{Ball:2001pq}),
\begin{equation}
  \label{eq:photoproduction-general}
  m^2\, \sigma_{\gamma\,g,N}(m^2) \; = \;
  f_{g,N}(m^2)\: h_N^{}(\gamma_N^{})
  \; ,
\end{equation}
that is, in terms of the gluon PDF and the impact factor
$h_N^{}(\gamma_N)$ depending on the anomalous dimension $\gamma_N^{}$. We
are interested in the perturbative regime with $\gamma_N^{}$ given by
\begin{equation}
\label{eq:gammaN-def}
  \gamma_N^{}\left(\frac{a}{N}\right) \;=\;
  \frac{a}{N}
  \,+\,  2\,\zeta_3 \left( \frac{a}{N}  \right)^4
  \,+\: \dots
\end{equation}
with $a \equiv \ca \alpha_s/\pi$. This well-known result for $\gamma_N^{}$ 
arises as the solution to the following condition on $\chi(\gamma)$,
\begin{equation}
\label{eq:BFKL}
  \frac{a}{N}\, \chi\left(\gamma\left(\frac{a}{N}\right) \right) \,= \, 1
  \, ,
\end{equation}
in a perturbative expansion for $\gamma \ll 1$ and requiring
$\gamma(0) = 0$. Here $\chi(\gamma)$ is the Mellin transform (with
respect to the transverse momentum $k_t^2$) of the lowest order BFKL
kernel expressed through standard $\psi$-functions as
\begin{equation}
  \label{eq:lo-bfkl}
  \chi(\gamma) \;=\;
  2\,\psi(1) - \psi(\gamma) - \psi(1-\gamma)
  \;=\; \frac{1}{\gamma}\,+\,2\,\zeta_3 \, \gamma^{\:2} 
        + {\cal O}(\gamma^{\;4})
  \; .
\end{equation}
The anomalous dimension $\gamma_N^{}$ governs the high-energy behavior
of the gluon PDF in Eq.~(\ref{eq:photoproduction-general}) as
\begin{equation}
\label{eq:small-x-gluon}
  f_{g,N}(m^2) \; = \;
  \left( \frac{m^2}{\mu^2} \right)^{\!\gamma_N^{}} \!  f_{g,N}(\mus)
  \; ,
\end{equation}
which up to terms relevant at NNLO simply gives
\begin{equation}
  \label{eq:fg-taylor}
  \left( \frac{m^2}{\mu^2} \right)^{\!\gamma_N^{}} \:=\:
    1
  + \gamma_N^{}\,\Lmmu
  + \frct{1}{2}\, \gamma_N^{\:2}\, \Lmmus
  + {\cal O}(\gamma_N^{\:3})
  \; .
\end{equation}

In the perturbative regime the impact factor is needed only in the
limit $N \ll \gamma_N^{\:-1}$, and can thus be approximated as
\begin{equation}
  \label{eq:impactfactor-Nllg}
  h_N^{}(\gamma_N^{}) \;=\; h(\gamma_N^{})\, \left(1 + {\cal O}(N) \right)
\end{equation}
because $h_N^{}(\gamma_N)$ is free of singularities and terms of order
$N$ are subleading. The calculation of the impact factor $h$ is
performed by considering off-shell amplitudes in a
$k_t$-factorization scheme. For photo-production of heavy quarks in
$\gamma\, g \to q_h^{}{\bar q_h^{}}$, and with the incoming photon and gluon
being off-shell by an amount $k_t$, this results
in~\cite{Catani:1990eg,Ball:2001pq,Ellis:1990hw}
\begin{eqnarray}
\label{eq:h-gamma}
  h(\gamma_N^{}) &\!=\!&
  \frac{\pi}{3}\,\alpha\, \alpha_s\, e_q^2\:
  \frac{7-5\,\gamma_N^{}}{3-2\,\gamma_N^{}}\:
  B(1-\gamma_N^{}, 1-\gamma_N^{})\,B(1+\gamma_N^{},1-\gamma_N^{})
\end{eqnarray}
in terms of standard Euler Beta-functions. For $\gamma_N^{} \ll 1$ this
expression leads to
\begin{eqnarray}
\label{eq:h-gamma-exp}
  h(\gamma_N) &\!=\!&
  \pi\, \alpha\, \alpha_s\, e_q^2\, \left\{
  \frct{7}{9} + \frct{41}{27}\, \gamma_N^{} + \frct{244}{81}\, \gamma_N^{\:2}
  + {\cal O}(\gamma_N^{\:3})
  \right\}
  \; .
\end{eqnarray}
Note that the precise definition of the $k_t$-factorization scheme
for the impact factor and the conversion to the \MSbar\ scheme
affects the result in the perturbative regime only at order
$\gamma_N^{\:3}$, see Ref.~\cite{Ball:2001pq}. This is beyond the NNLO
accuracy in $\as$ we are aiming at.

Finally the ratio $K_{2,N}$ between the cross sections for DIS
heavy quark electro-production and photo-production in
Eq.~(\ref{eq:Kxi-factor}) has been computed in closed analytical
form in Ref.~\cite{Catani:1990eg},
\begin{eqnarray}
  \label{eq:K2xi-general}
  K_{2, N}(\xi) &\!=\!&
  \bigg(1 + \frct{\xi}{4}\,\bigg)^{-N}\:
  \frac{3}{(7 - 5\:\! \gamma_N^{})(1 + 2\:\! \gamma_N^{})}\:
  \bigg\{ \,\frct{2}{\xi}(1+\gamma_N)
  \nonumber\\ && \!\mbox{\hspn}
  + \bigg( 1 + \frct{\xi}{4}\, \bigg)^{\!\gamma_N^{} - 1}
  \left(
    2 + 3\:\! \gamma_N - 3\:\! \gamma_N^{\:2} - \frct{2}{\xi}\,(1+\gamma_N)
  \right)
  \,
  {}_2F_1^{}\bigg(1-\gamma_N^{},\,\frct{1}{2}\,; \frct{3}{2}\,;\, 
  \frct{\xi}{\xi+4} \bigg)
  \bigg\}
  \; ,\qquad
\end{eqnarray}
where $_2F_1^{}$ denotes the hypergeometric function which can be conveniently 
expanded for small $\gamma_N^{}$ in terms of harmonic polylogarithms 
${\rm H}_{\vec{m}}(x)$ \cite{Remiddi:1999ew} (see e.g., Refs.\ 
\cite{Moch:2001zr,Weinzierl:2004bn,Kalmykov:2006hu,Kalmykov:2008ge}
and Appendix~\ref{sec:appB}). To that end, we have used the {\sc HypExp2}
package~\cite{Huber:2007dx} in {\sc  Mathematica}, which gives us
\begin{eqnarray}
  \label{eq:2F1-exp}
  2\:\!z\; 
  {}_2F_1^{}\bigg(1-\gamma_N^{},\,\frct{1}{2}\,; \frct{3}{2}\,;\,z^2 \bigg)
  &\!\!=\!&
  L(z)
  + \gamma_N^{}
  \Big\{
      H(-,+;z)
    + L(z)\, \ln(1-z^2)
  \Big\}
  - \frct{1}{2}\, \gamma_N^{\:2}\, \Big\{ 
    H(-,+,-;z)
\nonumber\\
& & \mbox{}
  - H(-,+;z)\, \ln(1-z^2)
  - L(z)\, \ln^2(1-z^2)
  \Big\}
  \,+\, {\cal O}(\gamma^{\:3})
  \; ,
  \qquad
\end{eqnarray}
where $L(z) = \ln((1+z)/(1-z))$ and $H(-,+;z)$, $H(-,+,-;z)$ are
short-hand notations (cf.~\cite{Maitre:2007kp}) for combinations of 
harmonic polylogarithms 
\begin{eqnarray}
  \label{eq:Hpm-def}
  H(-,+;z) &\!=\!& 
  H_{1,1}(z) \,+\, H_{1,-1}(z)\,-\, H_{-1,1}(z)\,-\, H_{-1,-1}(z)
  \; , \\[1mm]
  H(-,+,-;z) &\!=\!&
  H_{1,1,1}(z)\,-\,H_{1,1,-1}(z)\,+\,H_{1,-1,1}(z)\,-\,H_{1,-1,-1}(z)
  \nonumber\\ && \mbox{}
  -\, H_{-1,1,1}(z) \,+\, H_{-1,1,-1}(z) \,-\, H_{-1,-1,1}(z) 
  \,+\, H_{-1,-1,-1}(z)
  \; .
\end{eqnarray}
For the expansion coefficients of Eq.~(\ref{eq:2F1-exp}) we define the
functions
\begin{eqnarray}
  \label{eq:Ifunction}
  I(\xi) &=&
  \frct{4}{\xi}\, \sqrt{\xi/(\xi+4)}\, H\bigg(-,+;\sqrt{\xi/(\xi+4)}\:\bigg)
  \; ,
  \\
  \label{eq:Jfunction}
  J(\xi) &=&
  \frct{4}{\xi}\, \sqrt{\xi/(\xi+4)}\, L\bigg(\sqrt{\xi/(\xi+4)}\:\bigg)
  \; ,
  \\
  \label{eq:Kfunction}
  K(\xi) &=&
  \frct{4}{\xi}\, \sqrt{\xi/(\xi+4)}\, H\bigg(-,+,-;\sqrt{\xi/(\xi+4)}\:\bigg)
  \; ,
\end{eqnarray}
of which $I(\xi)$ and $J(\xi)$ have been introduced already in 
Ref.~\cite{Catani:1990eg} where also an expression for $I(\xi)$ in terms of 
normal (di-)logarithms can be found.

At this stage, all that remains is to perform the expansion of
Eqs.~(\ref{eq:photoproduction-general}) and (\ref{eq:K2xi-general})
to second order for small $\gamma_N^{}$ by assembling all formulae
provided so far. Converting from $N$-space back to $x$-space we
obtain for the gluon coefficient function,
\begin{eqnarray}
  \label{eq:c2g1-smallx}
  c_{2,g}^{(1)}(x,\xi) &\!=\!&
  \ca\, \* \frct{1}{12\*\pi}\, \*
  \bigg\{
  \frct{10}{3\*\xi}
  + \left(1 - \frct{1}{\xi} \right) \* I(\xi)
  + \left( \frct{13}{6}
  - \frct{5}{3\*\xi} \right) \* J(\xi)
  + \Lmmu \* \bigg[
  \frct{2}{\xi} + \left(1 - \frct{1}{\xi} \right) \* J(\xi)
  \bigg]
  \bigg\}
  + {\cal O}(x)
  \, , \nn \\ & & 
\\[-1mm]
  \label{eq:c2g2-smallx}
c_{2,g}^{(2)}(x,\xi) \!&=\!&
  \cas\, \* \lnx\, \*  \frct{1}{32\*\pi^3}\, \*
  \bigg\{
  - \frct{184}{27\*\xi}
  - \frct{1}{3} \* \left(1 - \frct{1}{\xi}\right) \* I(\xi) \* 
    \ln\bigg(1 + \frct{\xi}{4}\bigg)
  - \frct{1}{9} \* \left(13 - \frct{10}{\xi}\right) \* I(\xi)
  \nonumber\\[1mm] && \mbox{}
  - \frct{1}{27} \* \left(71 - \frct{92}{\xi}\right) \* J(\xi)
  + \frct{1}{3} \* \left(1 - \frct{1}{\xi}\right) \* K(\xi)
  + \Lmmu \* \bigg[
  - \frct{20}{9\*\xi}
  - \frct{2}{3} \* \left(1 - \frct{1}{\xi}\right) \* I(\xi)
 \nonumber\\ && \mbox{}
  - \frct{1}{9} \* \left(13 - \frct{10}{\xi}\right) \* J(\xi)
  \bigg]
  + \Lmmus \* \bigg[- \frct{2}{3\*\xi}
  - \frct{1}{3} \* \left(1 - \frct{1}{\xi}\right) \* J(\xi) \bigg]
  \bigg\}
  + {\cal O}(x^0)
  \; ,
\end{eqnarray}
where Eq.~(\ref{eq:c2g1-smallx}) agrees with the expression given in 
Ref.~\cite{Riemersma:1994hv} (cf.~Eqs.~(19) -- (22) there),
while the result for the NNLO quantity $c_{2,g}^{(2)}$ is a new result.

For further reference (in Section~\ref{sec:asy}, where we will also provide
graphs of the above results), we have finally derived explicit results for the 
$\xi \gg 1$ limits, i.e., $\Qs \gg m^2$, of Eqs.~(\ref{eq:c2g1-smallx}) and 
(\ref{eq:c2g2-smallx}). We obtain, at leading-logarithmic small-$x$ accuracy,
\begin{eqnarray}
  \label{eq:c2g1-smallx-asy}
  \xi \, c_{2,g}^{(1)}(x,\xi) &\!=\!&
  \ca\, \* \frac{1}{16\,\*\pi}\, \*
  \bigg[\,
  \frct{8}{3} \* \LQms
  + \frct{104}{9} \* \LQm
  + \frct{40}{9} - \frct{16}{3} \* \z2
  + \bigg(\frct{16}{3} \* \LQm + \frct{8}{3} \bigg) \* \Lmmu
  \bigg]
  \,+\, {\cal O}\left(\xi^{-1}\right)
  \; ,\qquad
  \\[2mm]
  \label{eq:c2g2-smallx-asy}
  \xi \, c_{2,g}^{(2)}(x,\xi) &\!=\!&
  - \cas\, \* \lnx\, \* \frct{1}{256\,\*\pi^3}\, \*
  \bigg[\,
  \frct{32}{9} \* \LQmt
  + \frct{208}{9} \* \LQms
  + \left(\frct{2272}{27}- \frct{64}{3} \* \z2 \right) \* \LQm
  \nonumber\\[1mm] && \mbox{}
  + \frct{1472}{27}
  - \frct{416}{9} \* \z2
  + \frct{128}{3} \* \z3
  + \bigg(\frct{32}{3} \* \LQms
  + \frct{416}{9} \* \LQm
  + \frct{160}{9} - \frac{64}{3} \* \z2
  \bigg) \* \Lmmu
  \nonumber\\ && \mbox{}
  + \left(\frct{32}{3} \* \LQm + \frct{16}{3}\right) \* \Lmmus
  \bigg]
  + {\cal O}\left(\xi^{-1}\right)
  \; .
  \qquad
\end{eqnarray}
Here the constant terms at NNLO, i.e., the terms independent of $\LQm$ and 
$\Lmmu$ (recall Eq.~(\ref{eq:LQmLmmdef}) above) in 
Eq.~(\ref{eq:c2g2-smallx-asy}) are new. All other terms in this result
agree with independent calculations. The scale-dependent $\Lmmu$ terms
can, of course, be obtained by the standard renormalization-group
behaviour, cf.~Section~\ref{sec:Coefficientfunctions}.
More importantly, we are also able to compare all $\LQm$
terms with the results obtained in the asymptotic limit $\Qs \gg
m^2$ below, cf.~Section~\ref{sec:asy}. All powers of $\LQm$ and $\Lmmu$ 
match exactly with those results, which constitutes a strong check.
Another test of Eq.~(\ref{eq:c2g2-smallx}) is possible by
comparing with the small-$x$ limit for $c_{2,g}^{}$ derived
in~Ref.~\cite{Thorne:2006qt} for the kinematic region $\Qs \simeq m^2$
(see also Eqs.~(10) and (11) in Ref.~\cite{Alekhin:2008hc}), and we find
agreement at the level of a few parts in a thousand for $\Qs/m^2 = 1$.

In concluding this section, we briefly comment on the
quark coefficient function for heavy-quark DIS
when the photon couples to the heavy quark, that is, $c_{2,q}^{}$ in
Eq.~(\ref{eq:totalF2c}) proportional to $e_h^{\,2}$ and the quark-singlet PDFs. 
Its leading small-$x$ term is related to that of $c_{2,g}^{}$ in
Eqs.~(\ref{eq:c2g1-smallx}) -- (\ref{eq:c2g2-smallx-asy}) by a simple
colour factor replacement, $c_{2,q}^{} = \cf/\ca\: c_{2,g}^{}$, as pointed
out already in Refs.~\cite{Catani:1990eg,Thorne:2006qt}.

\subsection{High-scale limit $\Qs \gg m^2$}
\label{sec:asy}

The results in the limit $\Qs \gg m^2$ build on an exact factorization of the 
heavy-quark coefficient functions into the respective coefficient functions 
with massless quarks and heavy-quark operator matrix elements (OMEs). 
The variable $\eta$ in Eq.~(\ref{eq:eta-xi-def}) factorizes as 
\mbox{$\eta \to \Qs/(4 m^2) (1/x-1) + {\cal O}(1)$}, and, to denote this limit,
we will be considering the coefficient functions as a function of $x$ and the 
ratio $\Qs/m^2$ throughout this section. 
At LO, one can quickly verify for $c_{2,g}^{}$ with the help of 
Eq.~(\ref{eq:Born2}) that this limit reads
\begin{eqnarray}
  \label{eq:cg20-asy}
  c_{2, g}^{(0)}(x,\Qs/m^2) &\!=\!&
  \pi\, x\, \frac{m^2}{\Qs}\,
  \Big\{
  2\*(1-2 \* x+2 \* x^2) \*\LQm
  \nonumber\\[-2mm] && \mbox{\hspp\hspp\hspp\hspp}
  \,+\,
  6 - 2 \* (1 - 2x + 2x^2) \* (4 + \H(0)(x) + \H(1)(x))
  \Big\}
  + {\cal O}\left(\frac{m^4}{Q^4}\right)
  \; ,\qquad
\end{eqnarray}
where the term multiplying $\LQm$ is proportional to the lowest-order splitting
function $P_{qg}^{(0)}$, while the $\LQm$ independent term 
(here written in terms of harmonic polylogarithms ${\rm H}_{\vec{m}}(x)$
\cite{Remiddi:1999ew}, see Appendix~\ref{sec:appB}) is proportional to the 
one-loop gluon coefficient function with massless quarks.

Formally the underlying factorization may be expressed through massive 
partonic OMEs $A_{ij}$ (i.e., with massive quarks) and the
DIS coefficient functions with massless quarks $C^{\,\rm light}_{k,
j}$ together with their exact scale dependence from renormalization
group invariance, see~Ref.~\cite{vanNeerven:2000uj}, i.e., at scales $\mu
\neq Q$ and $\mu \neq m$, in the form~\cite{Buza:1995ie},
\beq
  \label{eq:hqfact}
    c_{k, i}(\eta,\xi,\mu^2) \;\: \to \;\:
    c_{k, i}\left(x,\frac{\Qs}{\mu^2},\frac{m^2}{\mu^2}\right) \:=\:
    \bigg[
    A_{ji}\left(\frac{m^2}{\mu^2}\right)\,
    \otimes\,
    C^{\,\rm light}_{k, j}\left(\frac{\Qs}{\mu^2}\right)
    \bigg](x)
  + {\cal O}\left(\frac{m^2}{\Qs}\right)
  \, .
  \qquad
\eeq
Here $\otimes$ denotes the standard Mellin convolution, 
cf.~Eq.~(\ref{eq:totalF2c}). 
The dependence on the scale $\mu$ has been made explicit together with the
separation of scales between the OMEs $A_{ij}$ and the coefficient functions 
$C^{\,\rm light}_{k,j}$, which is realized through $\ln\Qs/\mu^2 = \LQm+\Lmmu$.
At NLO the formalism in Eq.~(\ref{eq:hqfact}) has first been applied in
Ref.~\cite{Buza:1995ie} to compute the corresponding two-loop OMEs for 
(unpolarized) heavy-quark DIS at asymptotic values $\Qs \gg m^2$, see also the
re-calculation in~Ref.~\cite{Bierenbaum:2007qe}.
In the wider context, Eq.~(\ref{eq:hqfact}) implements the matching conditions 
for QCD with one massive and $\nf$ light quarks to QCD with $\nf+\!1$ light 
quarks (note that $\nf$ denotes the number of massless quarks throughout 
this article). 
The matching of course also affects the strong coupling $\as$, for which the
decoupling formulae are well known~\cite{Larin:1994va,Chetyrkin:1997sg}.
Unless indicated otherwise, we are always working in the scheme with 
$\as(\nf)$, i.e., $\nf = 3$ for charm quarks.
Moreover, on the basis of Eq.~(\ref{eq:hqfact}), also the all-order resummation
of the logarithms in $\Qs/m^2$ has been discussed in the literature and has led
to definitions of a so-called variable flavour number scheme (VFNS)
\cite{Buza:1996wv}, see Refs.~\cite{Alekhin:2009ni,Forte:2010ta} for 
implementations in phenomenological analyses.

Here we will use Eq.~(\ref{eq:hqfact}) to derive the NNLO expression
$c^{(2)}_{2,g}$ and $c^{(2)}_{2,q}$ at large $\Qs \gg m^2$. To that end we rely
on the extensions of the two-loop massive OMEs to higher orders in $\epsilon$ 
(the parameter of dimensional regularization)~\cite{Bierenbaum:2008yu}, on the 
two-loop gluonic OMEs~\cite{Bierenbaum:2009zt} and, most importantly, on the
even-integer Mellin moments of the three-loop heavy-quark 
OMEs~\cite{Bierenbaum:2009mv} along with their complete
$\nf$-dependence~\cite{Ablinger:2010ty}. 
The other important input consists of the results for the anomalous dimensions 
\cite{Moch:2004pa,Vogt:2004mw} and coefficient functions for DIS with massless
quarks~\cite{Zijlstra:1992qd,Moch:1999eb,Vermaseren:2005qc} up to three loops. 
The necessary convolutions, Mellin transforms and inverse Mellin transforms are
all performed with algorithms for harmonic sums~\cite{Vermaseren:1998uu} and 
harmonic polylogarithms~\cite{Remiddi:1999ew} as implemented in the symbolic 
manipulation program {\sc Form}~\cite{Vermaseren:2000nd}.

\begin{figure}[t!]
\vspace*{-1mm}
\begin{center}
  \includegraphics[width=3.5cm]{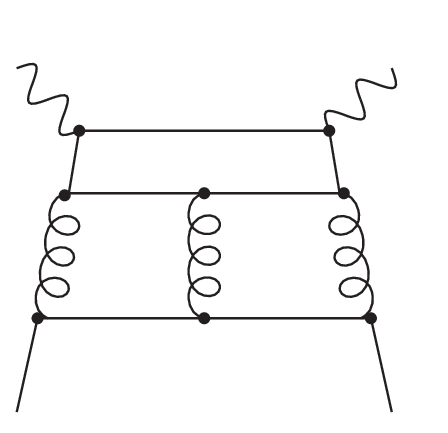}
  \hspace*{2mm}
  \includegraphics[width=3.25cm]{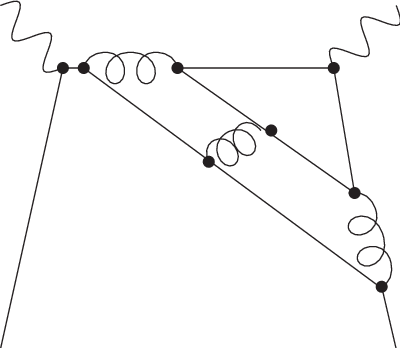}
  \hspace*{9mm}
  \includegraphics[width=3.5cm]{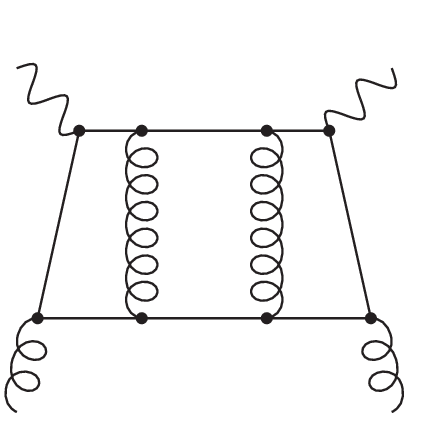}
  \hspace*{2mm}
  \includegraphics[width=3.5cm]{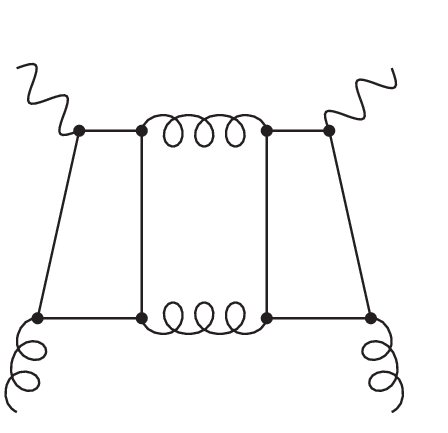}
\end{center}
\vspace*{-4mm}
\caption{ \small
  \label{fig:flavor}
  Representative three-loop diagrams of the flavour topologies 
  $fl_{02}$ and $fl_{11}$ (left) and $fl^{\,g}_{2}$ and $fl^{\,g}_{11}$ (right)
  for photon-quark (left) and photon-gluon scattering (right).}
\vspace*{-2mm}
\end{figure}
The coefficient functions $C^{\,\rm light}_{k, j}$ of the massless theory 
display one interesting feature which requires a comment in the context of the
factorization in Eq.~(\ref{eq:hqfact}).
Starting from three-loop order $C^{\,\rm light}_{k, j}$ contain new quark 
flavour topologies, which are denoted $fl_{11}$ and $fl^{\,g}_{11}$, with both 
photons in a squared matrix element coupling to distinct quark lines, see 
Fig.~\ref{fig:flavor}.
They extend beyond the standard electromagnetic coupling of the heavy-quark, 
proportional to its squared charge $e_h^{\,2}$ in Eq.~(\ref{eq:totalF2c}), as 
implied by the topologies $fl_{02}$ and $fl^{\,g}_{2}$. 
These new flavour classes $fl_{11}$ and $fl^{\,g}_{11}$ 
(being proportional to the colour factor $d^{abc} d_{abc}$) 
affect the factorization for $\Qs \gg m^2$ in Eq.~(\ref{eq:hqfact}) in the 
following way.
The available results for the massless coefficient functions in 
Ref.~\cite{Vermaseren:2005qc} (derived via the optical theorem) contain the 
sum of the contributions from all final state cuts which contribute to the 
inclusive massless structure function, say $F_2^{}$.
However, these are not identical to the contributions for the heavy-quark 
structure function in Eq.~(\ref{eq:totalF2c}), the latter being semi-inclusive 
and requiring a heavy-quark pair in the final state.
This is because some admissible final state cuts go through the quark line
while others do not, e.g., the cut of the two gluon lines in the right diagram 
of Fig.~\ref{fig:flavor}.
Currently available information is, unfortunately, not sufficient to 
disentangle the different terms. 
For that reason, we have decided to omit the (usually small) three-loop 
contributions proportional to $fl_{11}$ and $fl^{\,g}_{11}$ in 
$C^{\,\rm light}_{k, j}$ in the following, cf. also Appendix~\ref{sec:appB}.

The general expansion of the heavy-quark OMEs in Eq.~(\ref{eq:hqfact}) in 
powers of $\as$ reads
\begin{eqnarray}
  \label{eq:OMEexp}
  A_{ij}
  \; = \;
  \delta_{ij} +
  \sum\limits_{\ell=1}^{\infty}\, a_{\rm s}^{\,\ell} \, A^{(\ell)}_{ij}
  \; = \;
  \delta_{ij} +
  \sum\limits_{\ell=1}^{\infty}\, a_{\rm s}^{\,\ell} \,
  \sum\limits_{k=0}^{\ell} L_\mu^{\,k}\, a^{(\ell,k)}_{ij}
  \; ,\qquad
\end{eqnarray}
where, at each order, the terms proportional to powers of $\Lmmu$ are 
determined by lower order OMEs and splitting functions (anomalous dimensions). 
The genuinely new $\ell$-th order information resides in the expressions for 
$a^{(\ell,0)}_{ij}\!$.
The available results for those have been given in the literature for
the heavy-quark mass in the on-shell scheme (pole mass) as well as in the 
\MSbar\ scheme, see~Ref.~\cite{Bierenbaum:2009mv}.
Note, that in the latter case, certain constants such as $\z2$ and $\z2 \ln2$ 
are absent in Mellin space, similar to DIS structure functions with massless 
quarks~\cite{Vermaseren:2005qc}.

For the NNLO gluon coefficient function $c^{\,(2)}_{2,g}$, we specifically need
the heavy-quark OME $A_{{q_h^{}}g}$ ($= A_{Qg}$ in the notation of
Ref.~\cite{Bierenbaum:2009mv}); and for the NNLO light-quark coefficient 
function $c^{(2)}_{2,q}$ the pure-singlet heavy-quark OME 
$A^{\rm ps}_{{q_h^{}}q}$ ($= A^{\rm ps}_{Qq}$ in the notation
of Ref.~\cite{Bierenbaum:2009mv}). 
The three-loop expressions $A^{(3)}_{Qg}$ and $A^{(3),{\rm ps}}_{Qq}$ 
are given in Eq.~(4.39) and Eq.~(4.26) of Ref.~\cite{Bierenbaum:2009mv},
respectively, with the $\mu$-independent terms denoted by $a^{(3)}_{Qg}$
and $a^{(3),{\rm ps}}_{Qq}$.
They can be decomposed in powers of $\nf$ as
\begin{eqnarray}
  \label{eq:aQg30nf-exp}
a^{(3)}_{Qg}\;\;
  &\!=\!&
  a^{(3)\,0}_{Qg\!}
  \;\; +\, \nf\: a^{(3)1}_{Qg}
  \; ,\qquad
\\
  \label{eq:aQqps30nf-exp}
a^{(3),{\rm ps}}_{Qq} \!\!
  &\!=\!&
  a^{(3)\,0}_{Qq,\,\rm ps}
  \,+\, \nf\: a^{(3)1}_{Qq,\,\rm ps}
  \; ,\qquad
\end{eqnarray}
where the $\nf$ terms are now known exactly from Ref.~\cite{Ablinger:2010ty}.
For the $\nf=0$ terms, on the other hand, only a number of integer Mellin 
moments have been computed so far~\cite{Bierenbaum:2009mv}. These are thus the 
only quantities missing for the construction of $c^{\,(2)}_{2,g}$ and 
$c^{\,(2)}_{2,q}$ via Eq.~(\ref{eq:hqfact}),
see also Eqs.~(2.14) and~(2.15) in~\cite{Bierenbaum:2009mv}.

Therefore, leaving the functions $a^{(3)\,0}_{Qg}(x)$ and 
$a^{(3)\,0}_{Qg,\,\rm ps}(x)$ unspecified for the moment, we have constructed
the explicit results for the heavy-quark coefficient functions to NNLO.
The lengthy exact expressions are provided in Appendix~\ref{sec:appB}, see 
Eqs.~(\ref{eq:H2g1}) -- (\ref{eq:H2q3}).
There are a number of checks on these results. First of all, up to NLO
we agree with the previous results~\cite{Buza:1995ie,Bierenbaum:2007qe}.
Next, in the convolution Eq.~(\ref{eq:hqfact}) for $c^{(2)}_{2,g}$ and 
$c^{(2)}_{2,q}$ all cubic powers $L_{\mu}^{\,3}$ from the individual
renormalization of the OMEs $A_{Qg}$ and $A^{{\rm ps}}_{Qq}$ cancel, as they 
have to, against those from the massless coefficient functions 
$C^{\,\rm light}_{2, g}$ and $C^{\,\rm light}_{2, q}$ at three loops.
Moreover, all remaining powers of $\Lmmu$ in $c^{(2)}_{2,g}$ and $c^{(2)}_{2,q}$
agree numerically with the exact results for the $\mu$-dependence
derived by renormalization-group methods in Refs.~\cite
{Laenen:1998kp,Alekhin:2010sv} which are valid for all ratios of $\Qs/m^2$,
if the latter are evaluated in the present limit $\Qs \gg m^2$, see also 
Section~\ref{sec:Coefficientfunctions}.

As a last step, in order to arrive at phenomenologically useful results, it 
thus remains to use the available information on the Mellin moments of
$a^{(3)\,0}_{Qg}(x)$ and $a^{(3)\,0}_{Qq,\,\rm ps}(x)$ for constructing
(hopefully sufficiently accurate) approximate expressions together with 
estimates of their residual uncertainty. This proceeds along the lines of,
e.g., Refs.~\cite{vanNeerven:2000wp,vanNeerven:2001pe,Moch:2001im},
where Mellin moments of three-loop splitting and coefficient functions were
successfully employed to derive useful approximation of these quantities prior 
to their exact computations~\cite
{Moch:2004pa,Vogt:2004mw,Vermaseren:2005qc,Vogt:2005dw,Moch:2008fj}.

A vital constraint on the functions $a^{(3)\,0}_{Qg}(x)$ and 
$a^{(3)\,0}_{Qq,\,\rm ps}(x)$ is provided by the leading-logarithmic small-$x$
behaviour of $c^{\,(2)}_{2,g}$ and $c^{\,(2)}_{2,q}$ determined above, 
see Eq.~(\ref{eq:c2g2-smallx-asy}) and the last paragraph of 
Section~\ref{sec:small-x}. 
Since all other contributions to $c^{\,(2)}_{2,g}$ are known, that result can 
be used to deduce
\beq
  \label{eq:aQg30lnx}
  a_{Qg}^{(3)\,0}(x) \;=\;
  x^{\,-1} \lnx\; \* \cas\, \* \biggl( \frct{41984}{243} 
  + \frct{160}{9}\,\*\z2 - \frct{224}{9}\,\*\z3 \biggr) \:+\: {\cal O}(x^{-1})
  \; .
\eeq
Taking into account also the double-logarithmic large-$x$ terms, the
OME $a^{(3)\,0}_{Qg}$ can be written as
\beq
  \label{eq:aQg30-form}
  a_{Qg}^{\,(3)\,0}(x) \;=\;
  \sum_{m=1}^{5}\,A_m\,\ln^{\,m}(1-x)
  \,+\, f(x)
  \,+\,\frac{B}{x}
  \,+\,
  \biggl( \frct{41984}{27} + 160\,\*\z2 - 224\,\*\z3 \biggr)\, \frac{\ln x}{x}
\eeq
in QCD ($\ca = 3$), 
where the `smooth' function $f(x)$ is a complicated combination of harmonic
polylogarithms which approaches a constant for $x \to 1$ and includes powers of
$\lnx$ at small $x$. 
For approximations based of the five known Mellin moments, $N = 2,\,4,\,
\ldots,\, 10$, of Ref.~\cite{Bierenbaum:2009mv} we select two of the large-$x$ 
constants $A_m$, the subleading small-$x$ parameter $B$, and a two-parameter 
smooth function build from $\ln^{\,1,2}x$ and various linear and quadratic 
functions in $x$. 

The comparison of the Mellin transform of Eq.~(\ref{eq:aQg30-form}) with the 
known moments results in a system of linear equations for the chosen
coefficients $A_m$, $B$ and the parameters of $f(x)$. Varying the selected two
$A_m$ and the two-parameter form of $f(x)$, we arrive at a large number of
approximations which indicate the remaining uncertainty of 
$a_{Qg}^{(3)\,0}(x)$. 
As in Refs.~\cite{vanNeerven:2000wp,vanNeerven:2001pe,Moch:2001im}, we discard 
a small number of functions for which the  system of equations is 
almost singular, resulting in huge numerical coefficients and unrealistically 
large oscillations of the function. 
In this manner we have determined 50 to 100 acceptable approximations and 
finally selected two representatives, which reflect the error band for most of 
the $x$-range (and particularly at small $x$), for the final estimate of 
$a_{Qg}^{(3)\,0}$ and its residual uncertainty. 

\begin{figure}[p!]
\centering
  {
  \includegraphics[width=14.5cm]{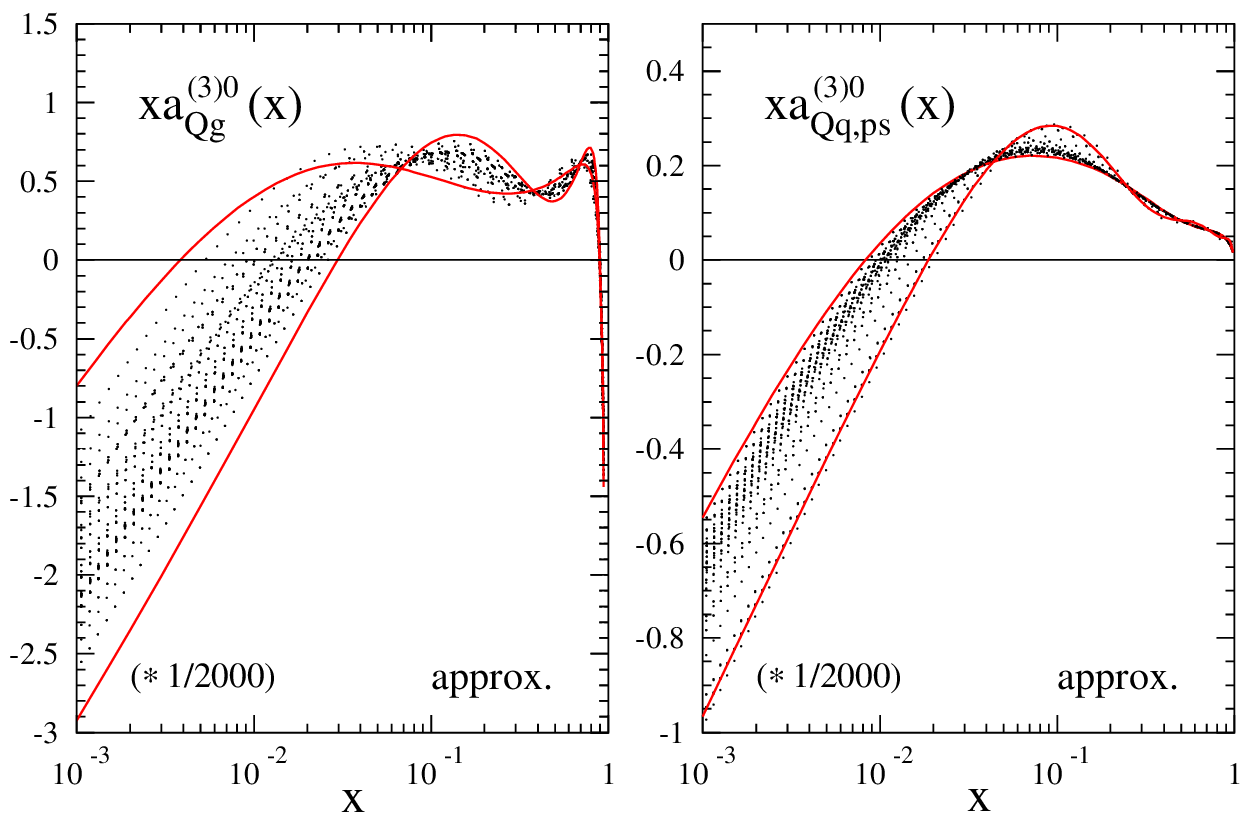}
  }
\vspace*{-2mm}
  \caption{ \small
  \label{fig:aijfits}
   A large number of approximations for the OMEs $a_{Qg}^{\,(3)\,0}$ (left) 
   and $a^{(3)\,0}_{Qq,\,\rm ps}$ (right) compatible with the 
   leading-logarithmic small-$x$ behaviour (\ref{eq:aQg30lnx}) and the known 
   first five even-integer Mellin moments \cite{Bierenbaum:2009mv}. 
   The solid curves show the respective selected representatives 
   (\ref{eq:fitA}), (\ref{eq:fitB}) and (\ref{eq:fitAq}), (\ref{eq:fitBq}).
   }
\end{figure}
\begin{figure}[p!]
\centering
\includegraphics[width=14.5cm]{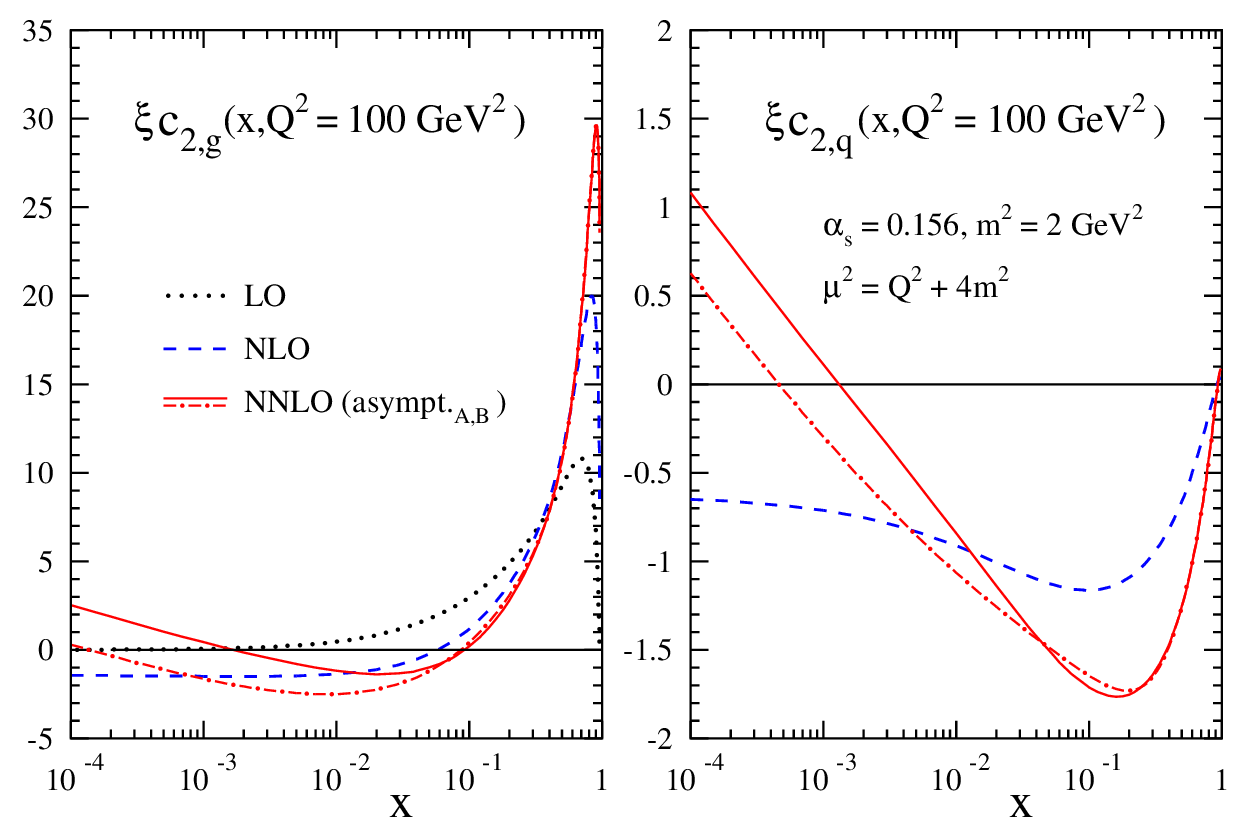}
  \vspace*{-2mm} 
  \caption{\small
  \label{fig:c2as100}
The perturbative expansion of the coefficient functions $c_{2,g}^{}$ (left) 
and $c_{2,q}^{}$ (right) for heavy-quark DIS at a large scale $\Qs \gg m^2$,
with the respective uncertainty bands due to the OMEs in Eqs.~(\ref{eq:fitA}), 
(\ref{eq:fitB}) and (\ref{eq:fitAq}), (\ref{eq:fitBq}). 
The results are shown for three light flavours and a typical value of $\as$ at 
$\mus = \Qs + 4\,m^2$. 
}
\end{figure}

The results of this process are shown in the left part of 
Fig.~\ref{fig:aijfits}. The chosen two approximations (solid lines) are given
by
\begin{eqnarray}
  \label{eq:fitA}
   a_{Qg,A}^{(3)\,0}(x) &\!=\!&  
        354.1002 \, \ln^3(1-x)
     \,+\, 479.3838 \, \ln^2(1-x)  
     \,-\, 7856.784 \,(2-x)    
\nn\\&& \mbox{}
     \,-\, 6233.530 \,\ln^2 x
     \,+\, 9416.621 \,x^{-1}
     \,+\, 1548.891 \,\,x^{-1}\,\ln x 
     \; ,
\\[1mm]
  \label{eq:fitB}
   a_{Qg,B}^{(3)\,0}(x) &\!=\!&
       -2658.323\,\ln^2(1-x)
     \,-\,7449.948\, \ln(1-x)  
     \,-\,7460.002\,(2-x)      
\nn\\&&
     \,+\,  3178.819\,\ln^2 x
     \,+\, 4710.725 \,x^{-1}
     \,+\, 1548.891 \,\,x^{-1}\,\ln x 
     \; ,
\end{eqnarray}
where the coefficient of $\,x^{-1}\ln x\,$ is a truncation of the exact result 
(\ref{eq:aQg30lnx}). The average of the two extremes can be used as the central
result of our approximation procedure.

The same procedure has been applied to the function 
$a_{Qq,\,\rm ps}^{(3)\,0}(x)$ in Eq.~(\ref{eq:aQqps30nf-exp}). 
Here the lowest six even-integer moments have been computed in 
Ref.~\cite{Bierenbaum:2009mv}. Taking into account that this pure-singlet 
quantity vanishes for $x \to 1$, our approximations are built from a subset of 
the functions
\begin{equation}
  \label{eq:aQq30-trialfun}
  (1-x)\ln^{\,\{3,2,1\}} (1-x) \, ,\quad
  (1-x)\,x^{\,\{-1,0,1,2\}} \, ,\quad
  x\,\ln^{\,\{2,1\}} x  \, ,\quad
  \ln^{\,\{2,1\}}x 
\end{equation}
together with the small-$x$ limit given by $\cf/\ca\: a_{Qg}^{(3)\,0}$ of  
Eq.~(\ref{eq:aQg30lnx}). The results are shown in the right part of 
Fig.~\ref{fig:aijfits}, and the extremal representatives have been chosen as
\begin{eqnarray}
  \label{eq:fitAq}
   a_{Qq,\,{\rm ps},\,A}^{\,(3)\,0}(x) &\!=\!&  
     \,(1-x)\,
     \big\{ 232.9555\,\ln^3(1-x) 
     \,+\,1309.528\,\ln^2(1-x) 
     \,-\, 31729.716\,x^2 
\\&& \mbox{\hspn}   
     \,+\, 66638.193\,x 
     \,+\, 2825.641\,x^{-1}\big\}
     \,+\, 41850.518\,x\,\ln x              
     \,+\, 688.396\,\,x^{-1}\,\ln x 
\; , \quad\nn
\\[1mm]
  \label{eq:fitBq}
   a_{Qq,\,{\rm ps},\,B}^{\,(3)\,0}(x) &\!=\!&
     \,(1-x)\,
     \big\{ 126.3546\,\ln^2(1-x)
     \,+\, 353.8539\,\ln(1-x)  
     \,+\, 6787.608\,x                 
\\&& \mbox{\hspn} 
     \,+\, 3780.192\,x^{-1}\big\}
     \,+\, 8571.165\,x\,\ln x  
     \,-\, 2346.893\,\ln^2 x    
     \,+\, 688.396\,\,x^{-1}\,\ln x 
\; , \nn
\end{eqnarray}
where, again, the coefficient of $\,x^{-1}\ln x\,$ arises from the truncated 
exact result, and the average of Eqs.~(\ref{eq:fitAq}) and (\ref{eq:fitBq})
provides the central result.

Using these approximations, the high-scale coefficient functions $c_{2,g}^{}$
and $c_{2,q}^{}$ can now be assembled up to NNLO using Eqs.~(\ref{eq:H2g1}) --
(\ref{eq:H2q3}). The successive LO, NLO and NNLO results are shown in
Fig.~\ref{fig:c2as100} at $\Qs = 100 \mbox{ GeV}^2$, a scale that is low
enough to be relevant to the HERA measurements and sufficiently high for the
safe applicability of the $\Qs \gg m^2$ approximation (\ref{eq:hqfact}).
We observe a satisfactory convergence, especially of the dominant quantity
$c_{2,g\,}^{}$, over a wide range of $x$, together with the well-understood
large-$x$ region of soft-gluon enhancement for $c_{2,g\,}^{}$,
cf.~Ref.~\cite{Moch:2005ba}.

For both coefficient functions the
uncertainty band due to the heavy-quark OMEs $A_{Qg}$ and $A^{\rm ps}_{Qq}$ as 
given by Eqs.~(\ref{eq:fitA}), (\ref{eq:fitB}) and (\ref{eq:fitAq}), 
(\ref{eq:fitBq}), respectively, is phenomenologically relevant only at 
small $x$, up to $x \approx 10^{\,-2}$. 
The uncertainty in this region is due to that of the (large) coefficient of 
the $1/x$ terms in Eq.~(\ref{eq:aQg30-form}) and its pure-singlet counterpart.
The computation of more Mellin moments along lines of Ref.~\cite{Bierenbaum:2009mv} would certainly help to improve the 
approximation also here; however the issue would be settled for all practical 
purposes by extending the results of Ref.~\cite{Catani:1990eg} to the 
next-to-leading small-$x$ terms at order $\alst$.

\begin{figure}[b!]
\vspace*{-1mm}
\centering
\includegraphics[width=16.2cm]{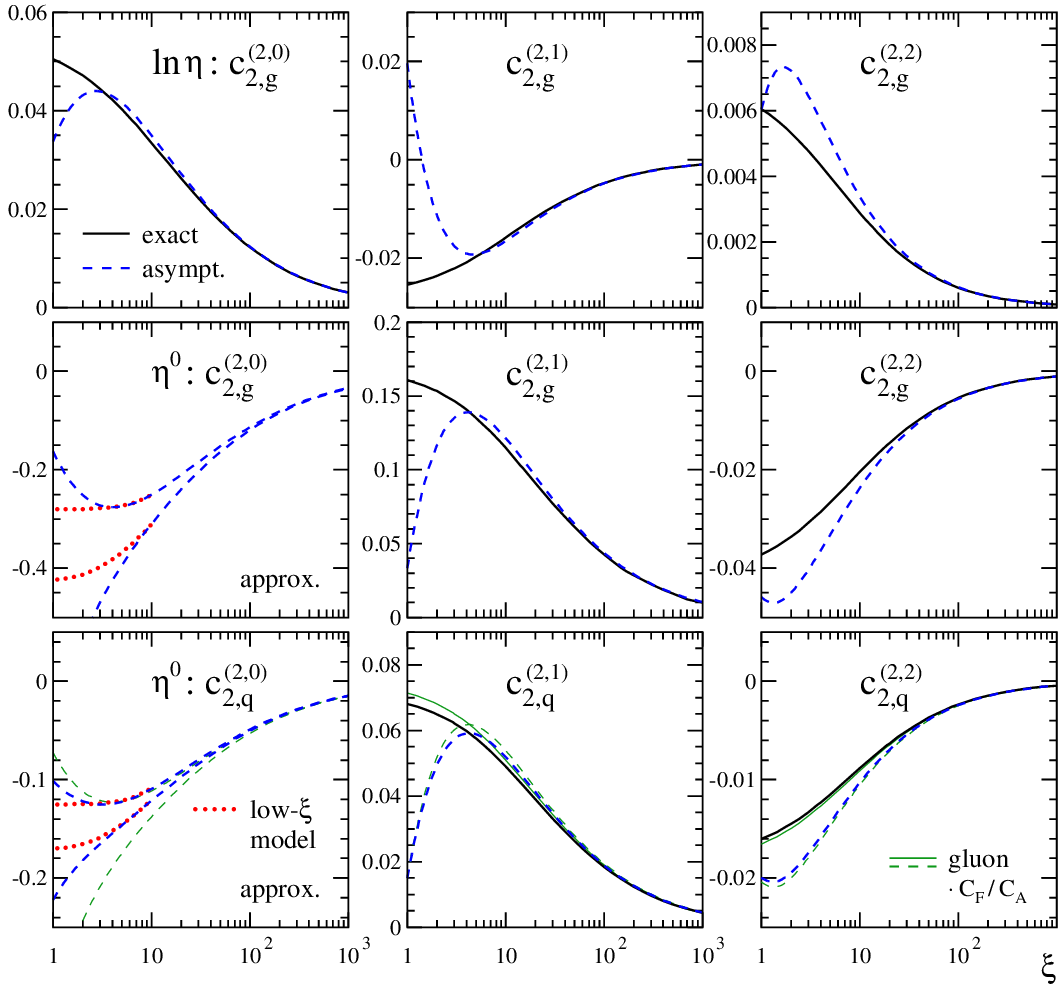}
\vspace*{-3mm}
\caption{\small
  \label{fig:c2i-nllx}
  Top panels: the coefficients of the leading small-$x\,/\,$high-$\eta$
  logarithm for the contributions $c_{2,g}^{(2,\ell)}$, $\ell = 0, 1, 2$,
  to the NNLO gluon coefficient function defined in Eq.~(\ref{eq:coeff-exp}).
  Middle and bottom panels: the respective next-to-leading $\eta^{\,0}$ 
  coefficients for $c_{2,g}^{\,(2)}$ and $c_{2,q}^{\,(2)}$.
  The solid (black) lines are the exact all-$\Qs$ results, the dashed (blue) 
  ones the high-scale asymptotic results; the dotted (red) low low-scale 
  extrapolations are discussed in Section~\ref{sec:Coefficientfunctions} below.
  Also illustrated, by the thin (green) lines in the bottom panels, is the 
  small next-to-leading high-$\eta$ deviation of $c_{2,q}^{\,(2)}$ from the 
  `Casimir-scaled' results for $c_{2,g}^{\,(2)}$. 
}
\end{figure}
A crucial question is, obviously, down to which values of $\xi = \Qs/m^2$
these high-scale asymptotic results are applicable. For the important
high-energy (small-$x$) contributions, which one may expect to be least
affected by $m^2/\Qs$ corrections, this issue is addressed in Fig.~\ref
{fig:c2i-nllx} for the quantities $c_{2,i}^{\,(2,\ell)}$, $i = q, g$, in
Eq.~(\ref{eq:coeff-exp}).  
Here the exact leading, cf.~Eqs.~(\ref{eq:c2g2-smallx}) and 
(\ref{eq:c2g2-smallx-asy}), and next-to-leading high-$\eta$ results are 
compared to the respective high-scale expressions. Except for the $\Lmmus$
quantities $c_{2,i}^{\,(2,2)}$ in Eq.~(\ref{eq:coeff-exp}) fixed by LO 
information, the asymptotic results provide good approximations of the 
$\ln \eta$ parts of $c_{2,i}^{\,(2,0)}$ and the $\ln \eta$ and $\eta^{\,0}$ 
parts of $c_{2,i}^{\,(2,1)}$ down to $\xi \simeq 4$, but then deteriorate
dramatically towards $\xi = 1$.
This behaviour can be used to extrapolate the (unfortunately still rather 
uncertain) asymptotic coefficients of $\eta^{\,0}$ for $c_{2,i}^{\,(2,0)}$ to 
small $\xi$. This extrapolation will be discussed in the next section, but the 
corresponding curves have already been included in the figure.

\renewcommand{\theequation}{\thesection.\arabic{equation}}
\setcounter{equation}{0}
\section{Approximate coefficient functions at NNLO}
\label{sec:Coefficientfunctions}

We are now in a position to construct improved NNLO approximations for the
genuinely new parts of the NNLO gluon and pure-singlet ($e_h^{\,2}$) coefficient
functions, $c_{2,g}^{(2,0)}$ and $c_{2,q}^{(2,0)}$, for heavy-quark DIS. 
Specifically, we will discuss how to combine the above (approximate) expressions
in the three kinematic limits, i.e., threshold, high-energy and large $\Qs$, 
in order to arrive at functional forms which smoothly interpolate over 
the full relevant range in $\eta$ and $\xi$ (recall, again, 
Eq.~(\ref{eq:eta-xi-def})).

All logarithmic terms proportional to powers of $\Lmmu$ in 
Eq.~(\ref{eq:coeff-exp}) are known exactly from standard renormalization-group 
methods.  For completeness, we give their explicit form in the \MSbar\ 
scheme. Suppressing all arguments, these coefficients read up to 
NNLO~\cite{Laenen:1998kp,vanNeerven:2000uj,Alekhin:2010sv}
\begin{eqnarray}
\label{eq:cq11}
  c_{2,q}^{\,(1,1)}&\!\!=\!&
  (4 \pi)^{-2}\, c_{2,g}^{(0,0)} \otimes P_{gq}^{\,(0)}
  \; ,
\\[1mm]
\label{eq:cq21}
  c_{2,q}^{\,(2,1)}&\!\!=\!& 
  (4 \pi)^{-4}\, c_{2,g}^{\,(0,0)} \otimes P_{gq}^{\,(1)} 
  \,+\,
  (4 \pi)^{-2}\, c_{2,g}^{\,(1,0)} \otimes P_{gq}^{\,(0)}  
  \,+\,
  (4 \pi)^{-2}\, c_{2,q}^{\,(1,0)} \otimes \left[ 
    P_{qq}^{\,(0)}\,-\, 2\,\beta_0 
  \right]
  \; ,
\\[1mm]
\label{eq:cq22}
  c_{2,q}^{\,(2,2)}&\!\!=\!&
  (4 \pi)^{-4}\, c_{\,2,g}^{(0,0)} \otimes \left[
   \frct{1}{2} \left( P_{gg}^{(0)} + P_{qq}^{\,(0)} \right) 
     \otimes P_{gq}^{\,(0)}
   - \frct{3}{2}\, \beta_0 \,  P_{gq}^{\,(0)} 
  \right]
  \; ,
\\[3mm]
\label{eq:cg11}
  c_{2,g}^{\,(1,1)}&\!\!=\!&
  (4 \pi)^{-2}\, c_{2,g}^{\,(0,0)} \otimes 
     \left[ P_{gg}^{\,(0)} - \beta_0 \right]
  \; ,
\\[1mm]
\label{eq:cg21}
  c_{2,g}^{\,(2,1)}&\!\!=\!& 
  (4 \pi)^{-4}\, c_{2,g}^{\,(0,0)} \otimes \left[ 
    P_{gg}^{(1)} - \beta_1\right]
  \,+\,
  (4 \pi)^{-2}\, c_{2,q}^{\,(1,0)} \otimes P_{qg}^{\,(0)} 
  \,+\,
  (4 \pi)^{-2}\, c_{2,g}^{\,(1,0)}  \otimes \left[ 
    P_{gg}^{\,(0)}\,-\, 2\,\beta_0 \right]
  , \qquad
\\[1mm]
\label{eq:cg22}
  c_{2,g}^{\,(2,2)}&\!\!=\!&
  (4 \pi)^{-4}\, c_{2,g}^{\,(0,0)}  \otimes \left[
    \frct{1}{2}\, P_{gg}^{\,(0)}\otimes P_{gg}^{\,(0)}
    + \frct{1}{2}\, P_{gq}^{\,(0)}\otimes P_{qg}^{\,(0)}
    - \frct{3}{2}\,  \beta_0 \* P_{gg}^{\,(0)} 
    + \beta_0^2 
  \right] 
  \; ,
\end{eqnarray}
where, as in Eq.~(\ref{eq:hqfact}) above, $\otimes$ denotes the standard 
convolution.
The splitting functions $P_{ij}^{\,(l)}(x)$ can be taken 
from~\cite{Moch:2004pa,Vogt:2004mw} and $\beta_0$ and $\beta_1$ are the 
standard coefficients of the QCD beta-function, normalized such that 
$\beta_0 = 11/3\: C_A - 2/3\: \nf$. 
The powers of $4\:\!\pi$ account for the normalization of Refs.~\cite
{Laenen:1992zk,Riemersma:1994hv} adopted in this article, 
cf.~Eq.~(\ref{eq:coeff-exp}).
Lacking more exact third-order information, in particular $c_{2,g}^{(2,1)}$ in 
Eq.~(\ref{eq:cg21}) will provide important guidance below for the assembly of 
the approximate NNLO result for the most important quantity $c_{2,g}^{(2,0)}$.

Two issues need to be considered for the combination of the results from 
the various kinematic regions. 
Firstly, one has to merge the available information from threshold at low 
$\beta$ (and $\eta\,$) with the high-energy limit at large $\eta$. 
Since we use the complete Born coefficient function (\ref{eq:Born2}) which 
vanishes for $\eta \to \infty$ as the prefactor in the threshold expansion
(\ref{eq:c2g2-thresh}), this can be done by multiplying the high-energy terms 
by sufficiently (but not excessively) high powers of $\beta$.
Secondly, the asymptotic expressions of the large-$\xi$ limit need to be 
joined with the low-$\xi$ region.  To that end we employ the function
\begin{equation}
  \label{eq:eqxicut}
  f\,(\xi) \;=\;\frac{1}{1 + e^{\,2(\xi - 4)}}
  \:\: ,
\end{equation}
which provides a smooth transition between these two regimes.
The parameters 2 and 4 in the exponent are chosen in such a way that the
transition is sufficiently rapid at values close to $\xi = 5$. This choice
is motivated by the above discussion of Fig.~\ref{fig:c2i-nllx} and the
finding of Ref.~\cite{Buza:1995ie} that, e.g.~for $c_{2,g}^{(1,0)}$, the
asymptotic limit (\ref{eq:H2g2asnl}) represents the exact result with a
high accuracy already at $\xi \gsim 10$.

These considerations lead the following ansatz for $c_{2,g}^{(1)}$ 
(cf.~also the NLO parametrizations in Ref.~\cite{Riemersma:1994hv}) 
which we test below,
\begin{eqnarray}
  \label{eq:assembly1}
  c_{2,g}^{\,(1)} &\!\simeq\!& 
  c_{2,g}^{\,(1)\,\rm thr}
  \:+\: \left(1 - f(\xi) \right)\, \beta^k\, c_{2,g}^{\,(1)\,\rm asm}
  \:+\: f(\xi)\, \beta^3\,
  \left( c_{2,g}^{\,(1){\,\rm LLx}}\,
  \frct{\eta^{\gamma}}{C + \eta^{\gamma}} \right)
  \; .
\end{eqnarray}
Here $c_{2,g}^{\,(1)\,\rm thr}$ and $c_{2,g}^{\,(1)\,\rm asm}$ denote the
threshold expansion in Eq.~(\ref{eq:c2g1-thresh}) and the asymptotic result in 
Eq.~(\ref{eq:H2g2asnl}), respectively.
The leading term in the small-$x$ limit in Eq.~(\ref{eq:c2g1-smallx}) defines 
$c_{2,g}^{(1){\rm LLx}}$ and provides the high-$\eta$ tail in the low-$\xi$ 
region which is smoothly matched with the factor 
$\eta^{\gamma}/(C + \eta^{\gamma})$.
The values of $\gamma,\: C$ and $k$ in Eq.~(\ref{eq:assembly1}) will be 
specified below.

Correspondingly, for $c_{2,g}^{(2)}$ we use
\begin{eqnarray}
  \label{eq:assembly2}
  c_{2,g}^{\,(2)} &\!\simeq\!& 
  c_{2,g}^{\,(2){\,\rm thr}}
  \,+\, \left(1 - f(\xi) \right) \beta^k\, c_{2,g}^{(2)\,{\rm asm}}
  \,+\, f(\xi)\, \beta^3\,
  \left( - c_{2,g}^{(2){\rm LLx}}\: \frct{\ln \eta}{\ln x}
  \,+\, c_{2,g}^{\,(2){\,\rm NLL}}\,\frct{\eta^{\gamma}}{C + \eta^{\gamma}} 
  \right) \quad
\end{eqnarray}
with $c_{2,g}^{\,(2)\,{\rm thr}}$ and $c_{2,g}^{\,(2)\,{\rm asm}}$ of 
Eqs.~(\ref{eq:c2g2-thresh}) and~(\ref{eq:H2g3asnl}).
The description of the low-$\xi$ region in Eq.~(\ref{eq:assembly2}) deserves 
particular attention.
Here $c_{2,g}^{\,(2){\rm LLx}}$ is the leading contribution 
(\ref{eq:c2g2-smallx}) in the small-$x$ limit, and we have divided out the 
factor $\ln x$ in order to be able to substitute $\ln x \to -\ln \eta$, which 
is valid at high energies and determines the slope in $\eta$ at all values of 
$\xi$.
The next-to-leading large-$\eta$ term, denoted by $c_{2,g}^{(2){\rm NLL}}$ 
in Eq.~(\ref{eq:assembly2}), is related (but, due to $\eta = \xi/(4 x) 
+ {\cal O}(1)$ not identical) to the ${\cal O}(x^{\,0})$ contribution in 
Eq.~(\ref{eq:c2g2-smallx}). 
It is currently unknown in the low-$\xi$ region. We will derive constraints 
on $c_{2,g}^{(2){\rm NLL}}$ below, although, even at large $\xi$, we still 
have to cope with the uncertainty of the heavy-quark OME $A_{Qg}$ estimated
by Eqs.~(\ref{eq:fitA}) and (\ref{eq:fitB}).
Recall also that the unknown next term in the low-$\eta$ threshold
expansion (\ref{eq:c2g2-thresh}) consists of terms proportional to the Born 
result as discussed in Section~\ref{sec:Thresresum}.

Therefore we adopt the following strategy for applying 
Eqs.~(\ref{eq:assembly1}) and (\ref{eq:assembly2}). 
When merging the contributions from the various regions, we account for the
residual uncertainties and construct two enveloping curves designed to span
the uncertainty band in the entire kinematic plane.  Thus at NLO we define

\pagebreak
\vspace*{-12mm}
\begin{eqnarray}
  \label{eq:assembly10A}
  c_{2,g}^{\,(1,0),\,A} &\!\!=\!\!& 
  c_{2,g}^{\,(1,0)\,{\rm thr}} 
  - c_{2,g}^{(1,0)\,{\rm const}}
  + \left(1 - f(\xi) \right) \beta\,\,\,\, c_{2,g}^{\,(1,0){\,\rm asm}}
  + f(\xi)\, \beta^3
  \left( c_{2,g}^{(1,0)\,{\rm LLx}}\,\frct{\eta^{\gamma}}{C + \eta^{\gamma}} 
  \right)
  , \qquad
  \\[1mm]
  \label{eq:assembly10B}
  c_{2,g}^{\,(1,0),B} &\!\!=\!\!& 
  c_{2,g}^{\,(1,0)\,{\rm thr}} 
  \phantom{
    - c_{2,g}^{\,(1,0){\rm const}}~
  }
  + \left(1 - f(\xi) \right) \beta^3\, c_{2,g}^{\,(1,0){\,\rm asm}}
  + f(\xi)\, \beta^3
  \bigg( c_{2,g}^{\,(1,0)\,{\rm LLx}}\,\frct{\eta^{\delta}}{D + \eta^{\delta}} 
  \bigg)
  .  \qquad
\end{eqnarray}
Here $c_{2,g}^{\,(1,0)\,{\rm thr}}$ and $c_{2,g}^{\,(1,0)\,{\rm asm}}$
are given by the $\Lmmu$-independent parts of Eqs.~(\ref{eq:c2g1-thresh}) and 
(\ref{eq:H2g2}), while the term $c_{2,g}^{\,(1,0){\rm const}}$ subtracted 
in Eq.~(\ref{eq:assembly10A}) summarizes the NLO constants in $\beta$ in 
Eq.~(\ref{eq:c2g1-const}).
The suppression parameters take the values
\begin{equation}
  \label{eq:suppr-param10}
  \gamma = 1.0\: ,\quad C = 42.5\: ,
  \qquad {\rm and}\qquad
  \delta = 0.8\: ,\quad D = 19.4
  \: .
\end{equation}
They are needed in order to suppress the coefficients of $\eta^0$ at low $\eta$ 
and have been determined by fitting the high-$\eta$ tail of $c_{2,g}^{(1,0)}$ 
using the {\sc Minuit} package of the CERN {\sc Fortran} library. It is not
possible to achieve agreement with the exact result without such
an additional suppression factor.

\begin{figure}[p!]
\centering
   {
   \includegraphics[width=16cm]{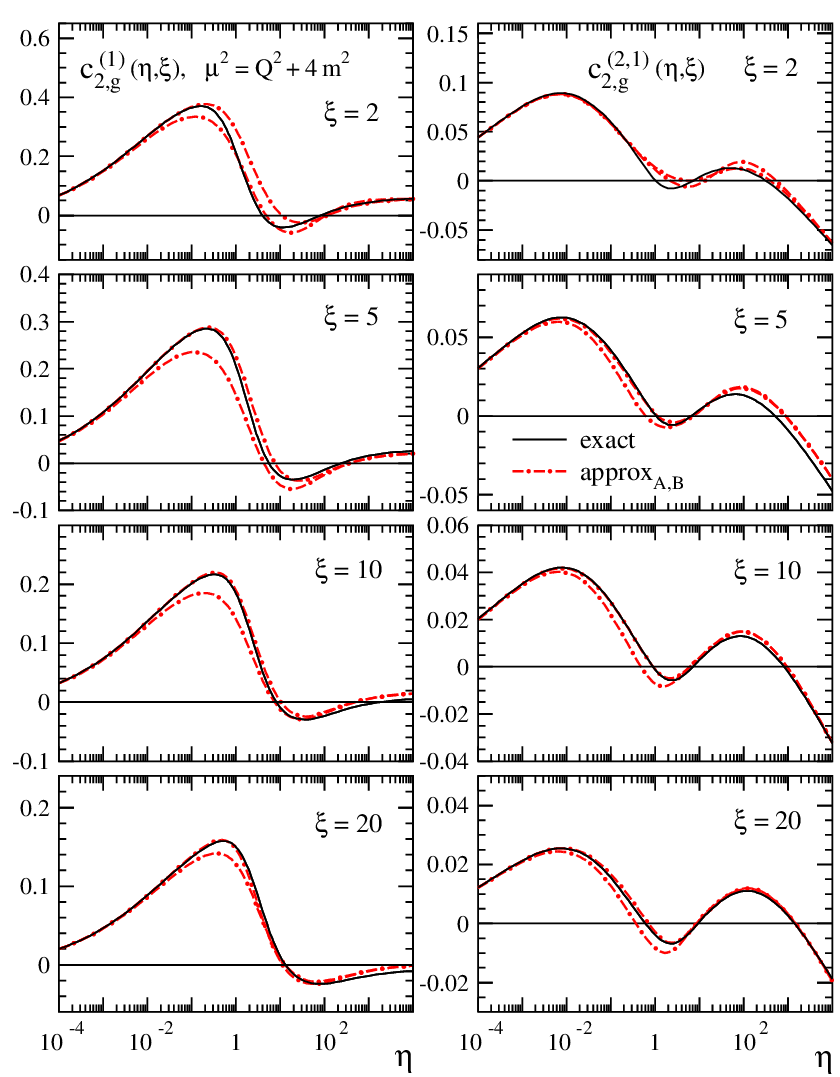}
   }
\vspace*{-1mm}
   \caption{ \small
   \label{fig:cg10+21}
Comparison of the exact result for $c_{2,g}^{\,(1)}$ and $c_{2,g}^{\,(2,1)}$,
at four representative values of $\xi = \Qs/m^2$,
with the approximations based on the threshold, high-$\eta$ and high-$\Qs$
limits as respectively specified in Eqs.~(\ref{eq:assembly10A}) and 
(\ref{eq:assembly10B}) and Eqs.~(\ref{eq:assembly21A}) and 
(\ref{eq:assembly21B}).  The first-order results are shown at a physically
relevant scale $\mu$ after adding the exact scale-dependent contribution to all
three curves.
    }
\end{figure}

The resulting two approximations are compared, after adding the exact 
scale-dependent contribution, in the left parts of Fig.~\ref{fig:cg10+21} to 
the parametrized \cite{Riemersma:1994hv} exact result of Ref.~\cite
{Laenen:1992zk} at four typical values of $\xi$ for the standard choice 
$\mus = \Qs + 4\:\!m^2$ of the (identical) renormalization and factorization 
scale. 
It is worthwhile to note not only that the approximations (\ref{eq:assembly10A})
and (\ref{eq:assembly10B}) indeed span the exact results, but also that there
is, at this scale, a considerable cancellation between the high-energy
constants in $c_{2,g}^{\,(1,0)}$ and $c_{2,g}^{\,(1,1)}$.

At NNLO we proceed in a similar manner. As indicated below Eq.~(\ref{eq:cg22}),
we first consider the known quantity $c_{2,g}^{\,(2,1)}$ and define
\begin{eqnarray}
  \label{eq:assembly21A}
  c_{2,g}^{\,(2,1),\,A} &\!\!=\!\!& 
  c_{2,g}^{\,(2,1){\,\rm thr}}
  \,+\, \left(1 - f(\xi) \right) \beta\,\,\,\, c_{2,g}^{(2,1)\,{\rm asm}}
  \,+\, f(\xi)\, \beta^3
  \bigg( - c_{2,g}^{\,(2,1)\,{\rm LLx}}\: \frct{\ln \eta}{\ln x}
  \,+\, c_{2,g}^{\,(2,1){\,\rm NLL}}\,\frct{\eta^{\gamma}}{C + \eta^{\gamma}} 
  \bigg) \, , \quad
  \nn \\[1mm] & & \\ 
  \label{eq:assembly21B}
  c_{2,g}^{\,(2,1),\,B} &\!\!=\!\!& 
  c_{2,g}^{\,(2,1){\,\rm thr}}
  \,+\, \left(1 - f(\xi) \right) \beta^3\, c_{2,g}^{(2,1)\,{\rm asm}}
  \,+\, f(\xi)\, \beta^3
  \bigg( - c_{2,g}^{\,(2,1)\,{\rm LLx}}\: \frct{\ln \eta}{\ln x}
  \,+\, c_{2,g}^{\,(2,1){\,\rm NLL}}\,\frct{\eta^{\delta}}{D + \eta^{\delta}} 
  \bigg) \quad
  \nn \\[1mm] & & 
\end{eqnarray}
with $c_{2,g}^{\,(2,1)\,{\rm thr}}$, $c_{2,g}^{\,(2,1)\,{\rm asm}}$ and 
$c_{2,g}^{\,(2,1)\,{\rm LLx}}$ respectively given by terms linear in $\Lmmu$ in 
Eqs.~(\ref{eq:c2g2-thresh}), (\ref{eq:H2g3asnl}) and (\ref{eq:c2g2-smallx}).
For the next-to-leading term at small-$x$ denoted by 
$c_{2,g}^{\,(2,1)\,{\rm NLL}}$ we can derive from Eq.~(\ref{eq:cg21}) the 
parametrizations
\begin{equation}
  \label{eq:c2g21-nllx}
  c_{2,g}^{\,(2,1)\,{\rm NLL}}(\xi) \;\simeq\; 
  0.16086 \,-\, 0.00711\,\ln\xi \,-\, 0.00549\,\ln^2\xi
  \; .
\end{equation}
The damping of this NLL high-energy tail can be approximated by (again using 
{\sc Minuit})
\begin{equation}
  \label{eq:suppr-param21}
  \gamma = 1.0\: ,\quad C = 20.0\: ,
  \qquad {\rm and}\qquad
  \delta = 0.8\: ,\quad D = 10.7
  \: .
\end{equation}
These approximations for $c_{2,g}^{\,(2,1)}$ are compared in the right parts
of Fig.~\ref{fig:cg10+21} with the exact result (\ref{eq:cg21}), for which 
these graphs supersede previous studies in 
Refs.~\cite{Laenen:1998kp,Alekhin:2008hc}.
Also these approximations are completely satisfactory, taking into account that 
a term corresponding to $c_{2,g}^{(1,0)\,{\rm const}}$ in 
Eq.~(\ref{eq:assembly10A}) has not been included in Eqs.~(\ref{eq:assembly21A})
and (\ref{eq:assembly21B}). 

Encouraged by these results, we now apply the same procedure to the main new
NNLO coefficient function $c_{2,g}^{\,(2,0)}$ and write
\begin{eqnarray}
  \label{eq:assembly20A}
  c_{2,g}^{\,(2,0),\,A} &\!\!=\!\!& 
  c_{2,g}^{\,(2,0)\,{\rm thr}} 
  \,+\, \left(1 - f(\xi) \right) \beta\,\,\,\, c_{2,g}^{\,(2,0){\,\rm asm},\,A}
\nn \\ & & \mbox{}
  +\, f(\xi)\, \beta^3
  \bigg( - c_{2,g}^{\,(2,0)\,{\rm LLx}}\; \frct{\ln \eta}{\ln x}
  \:+\: c_{2,g}^{\,(2,0){\,\rm NLL},\,A}
  \,\frct{\eta^{\gamma}}{C + \eta^{\gamma}} 
  \bigg)
\end{eqnarray}
and
\begin{eqnarray}
  \label{eq:assembly20B}
  c_{2,g}^{\,(2,0),\,B} &\!\!=\!\!& 
  c_{2,g}^{\,(2,0)\,{\rm thr}} 
  \,+\, f(\xi)\: 2\:\!c_{2,g}^{(2,0)\,{\rm const}}
  \,+\, \left(1 - f(\xi) \right) \beta^3\, c_{2,g}^{\,(2,0){\,\rm asm},\,B}
\nn \\ & & \mbox{}
  +\, f(\xi)\, \beta^3
  \bigg( - c_{2,g}^{\,(2,0)\,{\rm LLx}}\; \frct{\ln \eta}{\ln x}
  \:+\: c_{2,g}^{\,(2,0){\,\rm NLL},\,B}
  \,\frct{\eta^{\delta}}{D + \eta^{\delta}} 
  \bigg) \;\; 
\end{eqnarray}
with $c_{2,g}^{\,(2,0)\,{\rm thr}}$ and $c_{2,g}^{\,(2,0)\,{\rm asm}}$ 
given by the $\Lmmu$-independent terms of Eqs.~(\ref{eq:c2g2-thresh}) and 
(\ref{eq:H2g3asnl}).
It is understood that in the latter case the results for $a_{Qg,A}^{(3)\,0}$
and $a_{Qg,B}^{(3)\,0}$ in Eqs.~(\ref{eq:fitA}) and (\ref{eq:fitB}) are used to
account for the uncertainty band due to the heavy-quark OME $A_{Qg}$. 
The term $c_{2,g}^{\,(2,0)\,{\rm const}}$ denotes the scale-independent part
of the  Pad{\'e} estimate Eq.~(\ref{eq:c2g2-const})  
of the NNLO constant in $\beta$.  The factor of two in front originates from 
assigning a 100\% uncertainty to this estimate, i.e., we add zero in 
Eq.~(\ref{eq:assembly20A}) and $2\:\!c_{2,g}^{\,(2,0)\,{\rm const}}$ in 
Eq.~(\ref{eq:assembly20B}).
The values for the suppression parameters $\gamma,\: \delta,\: C$ and $D$ are 
taken from Eq.~(\ref{eq:suppr-param21}), and the functions 
$c_{2,g}^{\,(2,0)\,{\rm NLL}}$ in the low-$\xi$ region at high-$\eta$ are 
estimated via the error band
\bea
  \label{eq:c2g20-nllxA}
  c_{2,g}^{\,(2,0)\,{\rm NLL},\:A}(\xi)
  &\!=\!& 
  0.007\,\left(\frac{\ln\xi}{\ln5}\right)^{\!4} \,-\;0.28 \; ,
\\[1mm]
  \label{eq:c2g20-nllxB}
  c_{2,g}^{\,(2,0)\,{\rm NLL},\:B}(\xi)
  &\!=\!& 
  0.055\,\left(\frac{\ln\xi}{\ln5}\right)^{\!2} \,-\,0.423 \; .
\eea
These are the extrapolations already shown in Fig.~\ref{fig:c2i-nllx}. They
smoothly continue the large-scale results at $5 \lsim \xi \lsim 10$ to the
lower value of $\xi$ considered here, $\xi = 1$, in a manner suggested by the
corresponding low-$\xi$ behaviour of $c_{2,g}^{\,(1,0)}$ and 
$c_{2,g}^{\,(2,1)}$. The fine details of Eqs.~(\ref{eq:c2g20-nllxA}) and 
(\ref{eq:c2g20-nllxB}) do not matter, of course, due to the large uncertainty
of the underlying high-$\xi$ results (\ref{eq:fitA}) and (\ref{eq:fitB}).

\begin{figure}[p!]
\centering
  {
  \includegraphics[width=16cm]{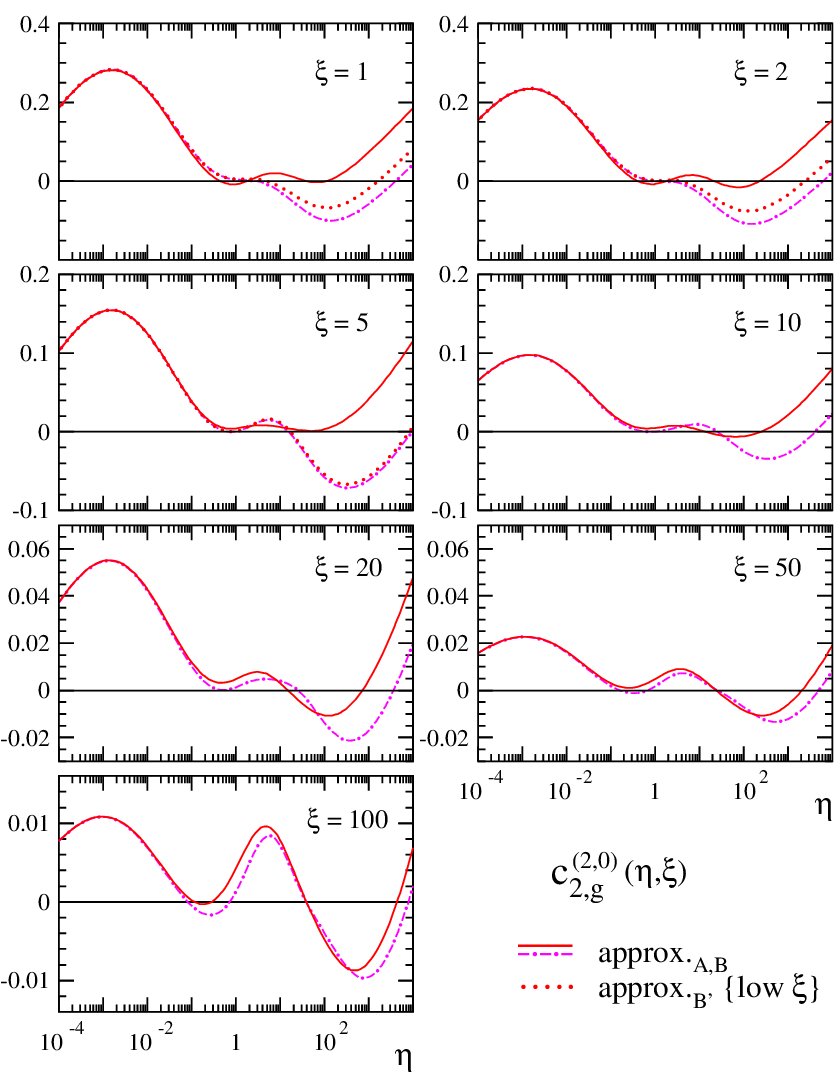}
  }
\vspace*{-1mm}
  \caption{ \small
   \label{fig:cg20}
  The two representative approximations (\ref{eq:assembly20A}) and
  (\ref{eq:assembly20B}) for the `irreducible' scale-independent contribution 
  $c_{2,g}^{\,(2,0)}$ to the NNLO gluon coefficient function for the structure
  function $F_2^{}$ in heavy-quark production in DIS in 
  Eq.~(\ref{eq:cross-section}),
  shown for the experimentally most relevant range of $\xi = \Qs/m^2$, where
  $m$ denotes the heavy-quark mass, for $\nf = 3$ light flavours.
  Also shown, for $\xi \leq 5$, is the low-$\xi$ improvement of the 
  approximation `B' as discussed at the end of this section.
  }
\end{figure}

\begin{figure}[p!]
\centering
  {
  \includegraphics[width=16cm]{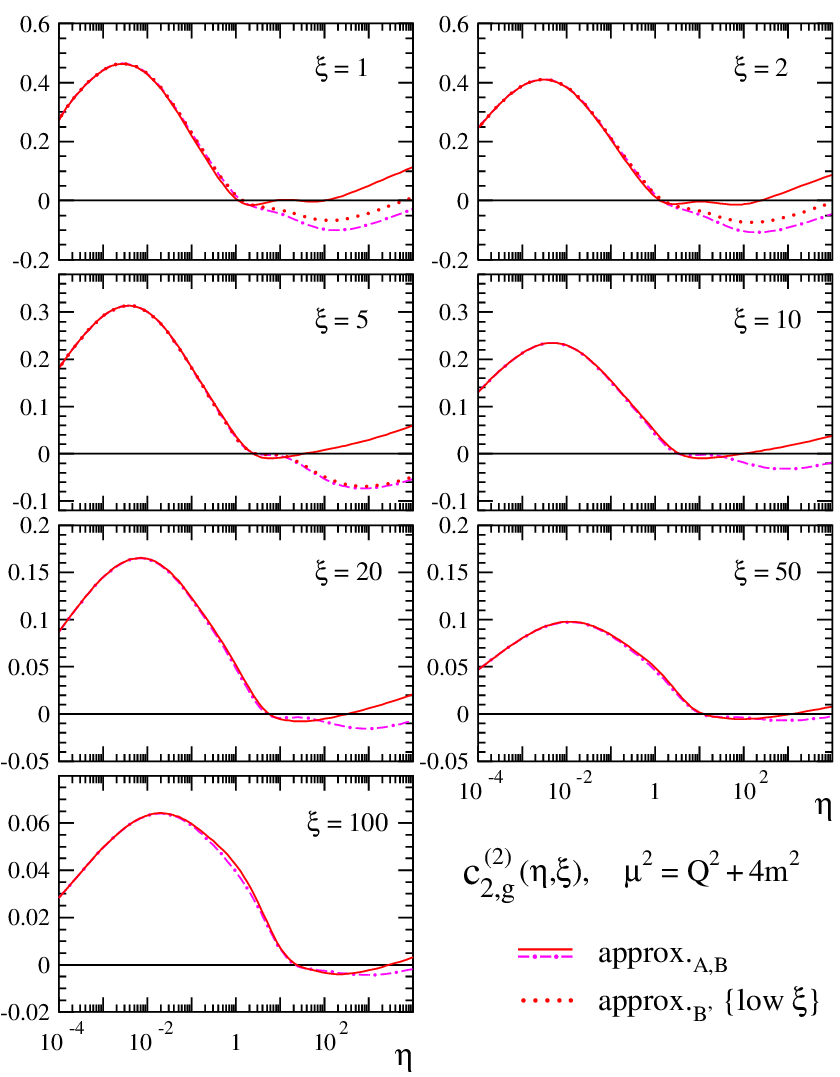}
  }
\vspace*{-1mm}
  \caption{ \small
    \label{fig:cg2}
  As Fig.~\ref{fig:cg20}, but for the complete coefficient function
  $c_{2,g}^{\,(2)}$, obtained by adding the known scale-dependent parts,
  for a standard choice of the renormalization and mass-factorization scale 
  $\mus \equiv\, \mufs = \murs$. 
  }
\end{figure}

In Fig.~\ref{fig:cg20} we display these approximate NNLO results for 
$c_{2,g}^{\,(2,0)}$ for seven values of $\xi$ in the range $1\leq \xi\leq 100$.
The known scale-dependent terms with $c_{2,g}^{\,(2,1)}$ and 
$c_{2,g}^{\,(2,2)}$ are then added in Fig.~\ref{fig:cg2} to these expressions 
for the standard scale $\mus = \Qs + 4\:\!m^2$ which we will also use in 
Section~\ref{sec:Phenomenology} below. 
The lack of a stronger constraint on the next-to-leading high-$\eta$
coefficient leads to a large uncertainty in particular at low $\xi$, where it
extends down to $\eta \approx 10$ in the construction detailed above. We 
have investigated other approaches to this important mid-$\xi$ region, but
have been unable to derive a reliable stronger constraint with the available
information on $c_{2,g}^{\,(2,0)}$. 
The present NNLO results in Fig.~\ref{fig:cg2} are consistent, albeit if with 
a large uncertainty, with the `high-energy censorship' already seen for 
$c_{2,g}^{\,(1)}$ in Fig.~\ref{fig:cg10+21}, i.e., a small contribution of the 
high-energy tail to the total coefficient function despite large contributions 
to the individual components $c_{2,g}^{\,(2,\ell)}$, $\ell = 0,\,1,\,2$.


\begin{figure}[p!]
\centering
  {
  \includegraphics[width=16cm]{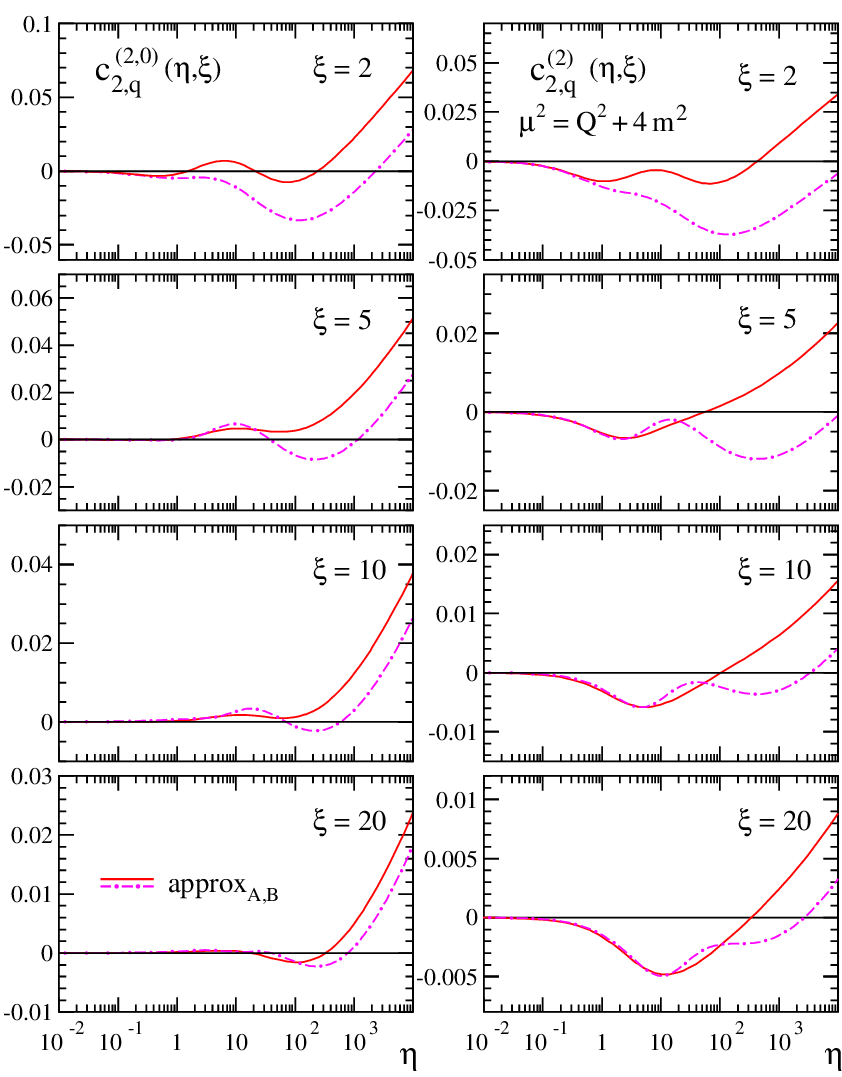}
  }
\vspace*{-1mm}
  \caption{ \small
    \label{fig:cq20}
  As Figs.~\ref{fig:cg20} and \ref{fig:cg2}, but for the pure-singlet 
  (light-)$\,$quark coefficient function $c_{2,q}^{\,(2)}$ in terms of the 
  approximations (\ref{eq:assembly20qA}) and (\ref{eq:assembly20qB}) at some 
  low and intermediate values of $\xi$.
  }
\end{figure}

For completeness, we apply the same procedure to the pure-singlet coefficient 
function $c_{2,q}^{}$ which exhibits a closely related high-energy behaviour, 
recall the end of Section~\ref{sec:small-x} and the bottom parts of 
Fig.~\ref{fig:c2i-nllx}, but no low-$\eta$ threshold enhancement. 
Suppressing the quantities corresponding to Fig.~\ref{fig:cg10+21} for 
brevity, we directly turn to the new NNLO quantity $c_{2,q}^{\,(2,0)}$ which 
thus we approximate by 
\begin{eqnarray}
  \label{eq:assembly20qA}
  c_{2,q}^{\,(2,0),\,A} &\!=\!& 
  \left(1 - f(\xi)\right)\beta\,\,\,\, c_{2,q}^{\,(2,0)\,{\rm asm},A}
  \,+\, f(\xi)\,\beta^3\,
  \bigg( -\,\frct{\cf}{\ca}\: c_{2,g}^{\,(2,0)\,{\rm LLx}}\;
  \frct{\ln \eta}{\ln x}
  \:+\: c_{2,q}^{\,(2,0){\,\rm NLL},\,A}
  \,\frct{\eta^{\gamma}}{C + \eta^{\gamma}} 
  \bigg) \,, \quad
\nn \\ & & \\[-2mm] 
  \label{eq:assembly20qB}
  c_{2,q}^{\,(2,0),\,B} &\!=\!& 
  \left(1 - f(\xi)\right) \beta^3\, c_{2,q}^{\,(2,0)\,{\rm asm}\,B}
  \,+\, f(\xi)\, \beta^3
  \bigg( -\,\frct{\cf}{\ca}\: c_{2,g}^{\,(2,0)\,{\rm LLx}}\; 
  \frct{\ln \eta}{\ln x}
  \:+\: c_{2,q}^{\,(2,0){\,\rm NLL},\,B}
  \,\frct{\eta^{\delta}}{D + \eta^{\delta}} 
  \bigg) \,. \quad 
\nn \\[-1mm] & & 
\end{eqnarray}
The high-$\xi$ asymptotic result $c_{2,q}^{\,(2,0)\,\rm asm}$ is given by the
$\mu$-independent part of Eq.~(\ref{eq:H2q3}), which presently includes the
two approximations (\ref{eq:fitAq}) and (\ref{eq:fitBq}) for the pure-singlet
operator
matrix element $a_{Qq,\,\rm ps}^{(3)\,0}$. The values $\gamma$, $\delta$, $C$
and $D$ are again taken from Eq.~(\ref{eq:suppr-param21}). Analogous to 
Eqs.~(\ref{eq:c2g20-nllxA}) and (\ref{eq:c2g20-nllxB}), the low-$\xi$
extrapolations of the high-$\eta$ constant, constrained by the 
$1/x$-coefficients in Eqs.~(\ref{eq:fitAq}) and (\ref{eq:fitBq}), are 
chosen as (see again also Fig.~\ref{fig:c2i-nllx})
\bea
  \label{eq:c2q20-nllxA}
  c_{2,q}^{\,(2,0){\rm NLL},A}(\xi)
  &\!=\!& \;
  0.004\,\left(\frac{\ln\xi}{\ln5}\right)^4
  \,-\,0.125
  \; ,
\\[1mm] 
  \label{eq:c2q20-nllxB}
  c_{2,q}^{\,(2,0){\rm NLL},B}(\xi)
  &\!=\!& 
  0.0245\,\left(\frac{\ln\xi}{\ln5}\right)^2
  \,-\,0.17
  \; .
\eea
The resulting approximations for $c_{2,q}^{\,(2,0)}$ and the (small -- compare 
the scales of the right parts to those of Fig.~\ref{fig:cg2}) complete
coefficient function $c_{2,q}^{\,(2)}$ at $\mus = \Qs + 4\:\!m^2$ are 
illustrated in Fig.~\ref{fig:cq20}.

So far we have separately considered approximations to the NNLO gluon and
pure-singlet quark coefficient functions. However, as mention above, the 
small-$x/\,$large-$\eta$ limits of these two functions are closely related; and
while their $\cf/\ca$ relation does not hold exactly beyond the leading 
logarithms, it still approximately holds for the NNLO $\eta^{0}$-terms as
shown already in the bottom panels of Fig.~\ref{fig:c2i-nllx} in Section
\ref{sec:asy}. In particular, the numerical breaking of this relation
for these terms is much smaller than the current uncertainties of
$c_{2,g}^{\,(2,0)}$ and $c_{2,q}^{\,(2,0)}$ in the low-$\xi$ region.

This opens up the possibility of one final improvement, i.e., using the
$\cf/\ca$ relation and the better constrained coefficient function to reduce
the uncertainty of its more uncertain counterpart. For both the OME and the
coefficient function the high-energy uncertainty is smaller in the quark
case, as can be seen from Figs.~\ref{fig:aijfits} and \ref{fig:c2i-nllx},
which is unsurprising given that one more high-$\xi$ moment is known for the
OME in this case \cite{Bierenbaum:2009mv}, which is moreover an `easier' 
function with $a_{Qq,\,{\rm ps}}^{\,(3)\,0}(x) \to 0$ for $x \to 1$. 
Since the `$A$' approximations are very close to the $\cf/\ca$ relation,
cf.~Fig.~\ref{fig:c2i-nllx}, this leads to replacing Eq.~(\ref{eq:c2g20-nllxB})
by
\beq
  \label{eq:c2g20-nllxBs}
  c_{2,g}^{\,(2,0)\,{\rm NLL},B'}
  \;=\;
  \frct{\ca}{\cf}\: c_{2,q}^{\,(2,0)\,{\rm NLL},B}
\eeq
with the right-hand side given by Eq.~(\ref{eq:c2q20-nllxB}). The resulting
improved low-$\xi$ estimate has been included in the first three bins in 
Figs.~\ref{fig:cg20} and \ref{fig:cg2}, and it is this reduced uncertainty
estimate that we will use when we study the approximate NNLO results for
$F_2^{\,\rm charm}$ and $F_2^{\,\rm bottom}$ in the next section.
  
\renewcommand{\theequation}{\thesection.\arabic{equation}}
\setcounter{equation}{0}
\section{Charm and bottom structure functions at NNLO}
\label{sec:Phenomenology}

We finally illustrate the impact of the above approximate NNLO coefficient
functions on the heavy-quark contribution to the structure function $F_2^{}$
in Eqs.~(\ref{eq:cross-section}) and (\ref{eq:totalF2c}). Since present-day
parton distributions are rather well constrained for most of the kinematic
range relevant to the charm and bottom structure functions, we confine
ourselves to one set, that of Ref.~\cite{Alekhin:2012du}. 
See Refs.~\cite{Alekhin:2009ni,JimenezD:2008hf,JimenezD:2009tv,Martin:2009iq,%
Martin:2010db,Ball:2011uy} for other recent PDF fits including NNLO 
corrections. 
For the heavy-quark pole masses we use
\beq
\label{eq:masses}
  m_{\:\!\rm c}^{\:\!2} \;=\; 2 \mbox{ GeV}^2
  \quad \mbox{ and } \quad
  m_{\:\!\rm b}^{} \;=\; 4.5 \mbox{ GeV} \;\; ,
\eeq
and our default choice for the renormalization and mass-factorization scales
is 
\beq
\label{eq:mu-std}
  \mus \:\equiv\: 
  \mufs \,=\, \murs 
  \:=\: \Qs + 4\, m_h^{\:\!2} \;\; , \quad h \,=\, \rm c,\: b \;\; .
\eeq

\begin{figure}[p!]
\centering
  {
  \includegraphics[width=16cm]{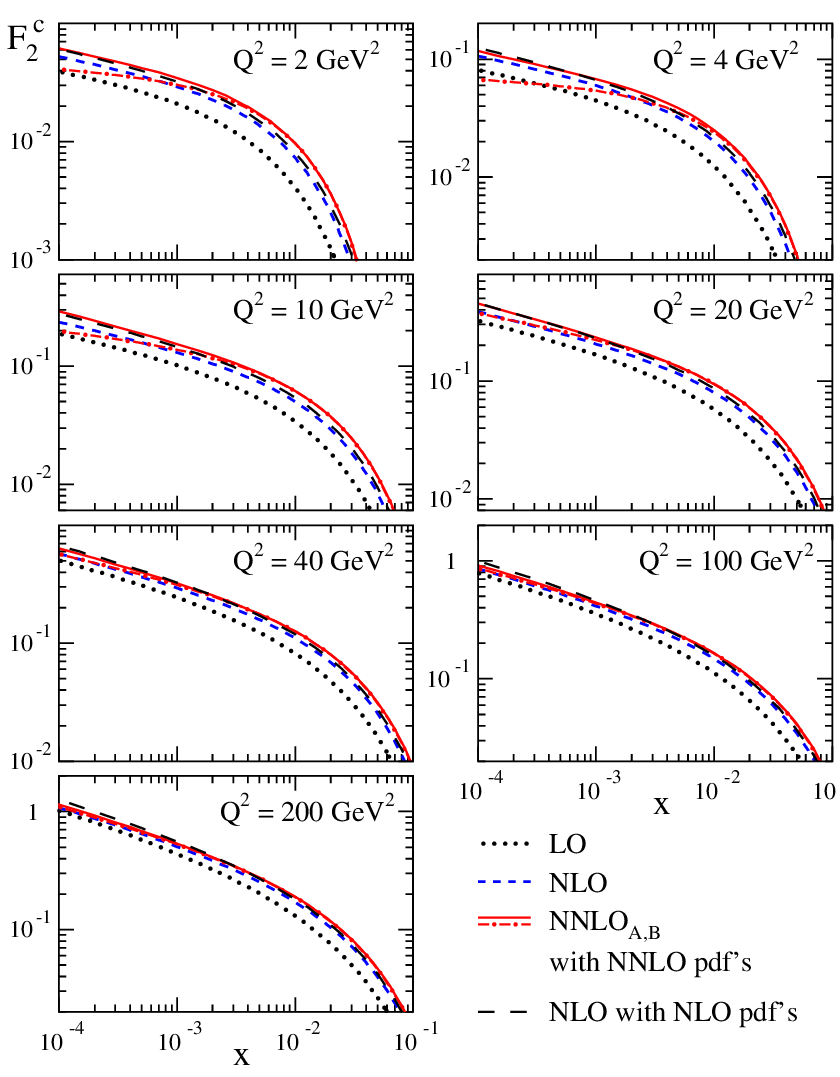}
  }
\vspace*{-2mm}
  \caption{ \small
    \label{fig:f2c-xQ2}
  The charm-production structure function $F_2^{\,c}(x,\Qs)$ in the 
  experimentally important region $2 \mbox{ GeV}^2 \leq \Qs \leq 200 
  \mbox{ GeV}^2$. The effects of the LO, NLO and approximate NNLO coefficient 
  functions are illustrated by using in Eq.~(\ref{eq:totalF2c}) the NNLO values
  of $\as$ and the parton distributions of the recent ABM$\,$11 fit
  \cite{Alekhin:2012du}. 
  Also shown are the NLO results obtained for the corresponding NLO values of
  $\as$ and the PDFs.
  All curves are shown for a pole mass of $m^{\:\!2} = 2 \mbox{ GeV}^2$ and the
  standard scale $\mus = \Qs + 4\:\!m^{\:\!2}$.
  }
\end{figure}

\vspace{-2mm}
In Fig.~\ref{fig:f2c-xQ2} the resulting charm structure function 
$F_2^{\,\rm c}$ is shown for seven scales $\Qs$ which correspond to the 
$\xi$-values chosen in Figs.\ 7$\,$--$\,$9 above. 
The dotted and short-dashed curves are the contributions of the LO and NLO
coefficient functions to the NNLO results, i.e., these quantities have been
calculated using the NNLO values of the strong coupling and the parton 
distributions as determined in Ref.~\cite{Alekhin:2012du}. The solid and
dash-dotted curves show the NNLO corrections for the approximations `A' and
`B' constructed in the previous section, including the improvement 
(\ref{eq:c2g20-nllxBs}) of the latter. 
Where the approximations are sufficiently accurate, roughly speaking at
$x \gsim 10^{\:\!-3}$, the overall effect of the NNLO coefficient functions is 
positive but considerably smaller than that of the NLO contribution, indicating
a good convergence of the perturbation series. 
At $x\lsim 10^{\:\!-3}$ the present results are inconclusive in this respect in
particular, unfortunately, in the important low- and medium-$\Qs$ region.
Also shown, by the long-dashed curves, are the physical NLO results involving
the NLO coupling constants and parton distributions according to 
Ref.~\cite{Alekhin:2012du}.

\begin{figure}[p!]
\centering
  {
  \includegraphics[width=16cm]{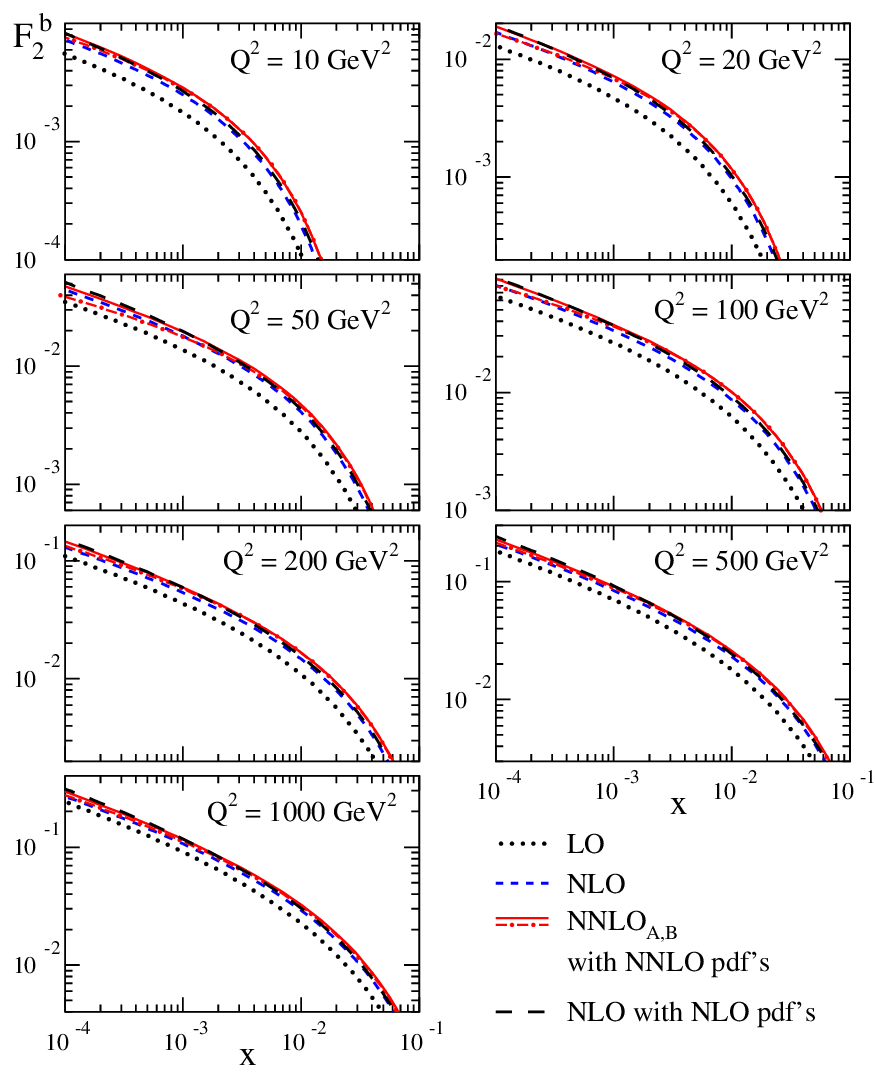}
  }
\vspace*{-1mm}
  \caption{ \small
    \label{fig:f2b-xQ2}
  As Fig.~\ref{fig:f2c-xQ2}, but for the bottom-production part of $F_2$ using 
  $m = 4.5$ GeV. \hspace*{3.8cm}
  }
\end{figure}

The corresponding results for the bottom-production contribution to $F_2^{}$
are illustrated in Fig.~\ref{fig:f2b-xQ2} for the slightly shifted region
$0.5 \lsim \xi \lsim 50$. 
The pattern of the corrections is qualitatively similar to that for charm
production in Fig.~\ref{fig:f2c-xQ2}. Again the NNLO approximation 
uncertainty is largest at $x < 10^{\:\!-3}$ and $1 \lsim \xi \lsim 5$, here 
corresponding to $20 \mbox{ GeV}^2 \lsim \Qs \lsim 100 \mbox{ GeV}^2$, where a 
pure threshold approximation can be insufficient and the mid-$\eta$ constraints
of the high-scale expressions are not applicable yet. Note that this figure
includes a bin at $\Qs\! < m_h^{\:\!2}$, where the impact of the medium- and 
large-$\eta$ parts of the coefficient functions is smaller than at higher 
scales.
In general, the size and uncertainty of the NNLO corrections are smaller for 
$F_2^{\,\rm b}$ than for $F_2^{\,\rm c}$ at the same values of $\xi$ due to 
the smaller values of $\as$ and the steeper small-$x$ PDFs at the 
corresponding higher scales.

\begin{figure}[p!]
\centering
  {
  \includegraphics[width=14.5cm]{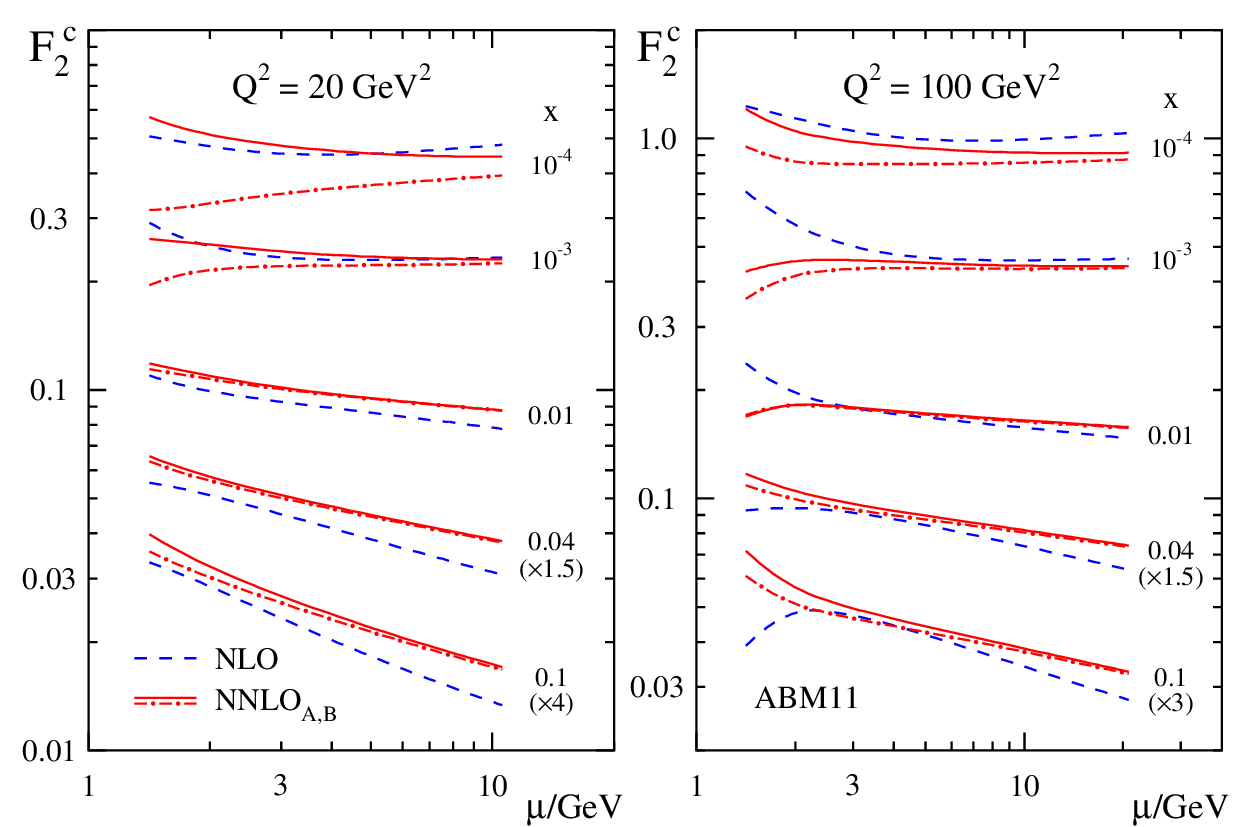}
  }
\vspace*{-1mm}
  \caption{ \small
   \label{fig:f2c-muvar}
  The dependence of the NLO and approximate NNLO results for $F_2^{\,c}$, 
  shown for $\mu = \sqrt{\Qs \!+\! 4\:\!m^{\:\!2}}$ in Fig.\ \ref{fig:f2c-xQ2},
  on the scale $\mu \equiv \muf = \mur$ for the wide range $m \leq \mu \leq 
  2 \sqrt{ \Qs \!+\! 4\:\!m^{\:\!2} }$ at selected values of $x$ and~$\Qs$.
  }
\end{figure}
\begin{figure}[p!]
\centering
  {
  \includegraphics[width=14.5cm]{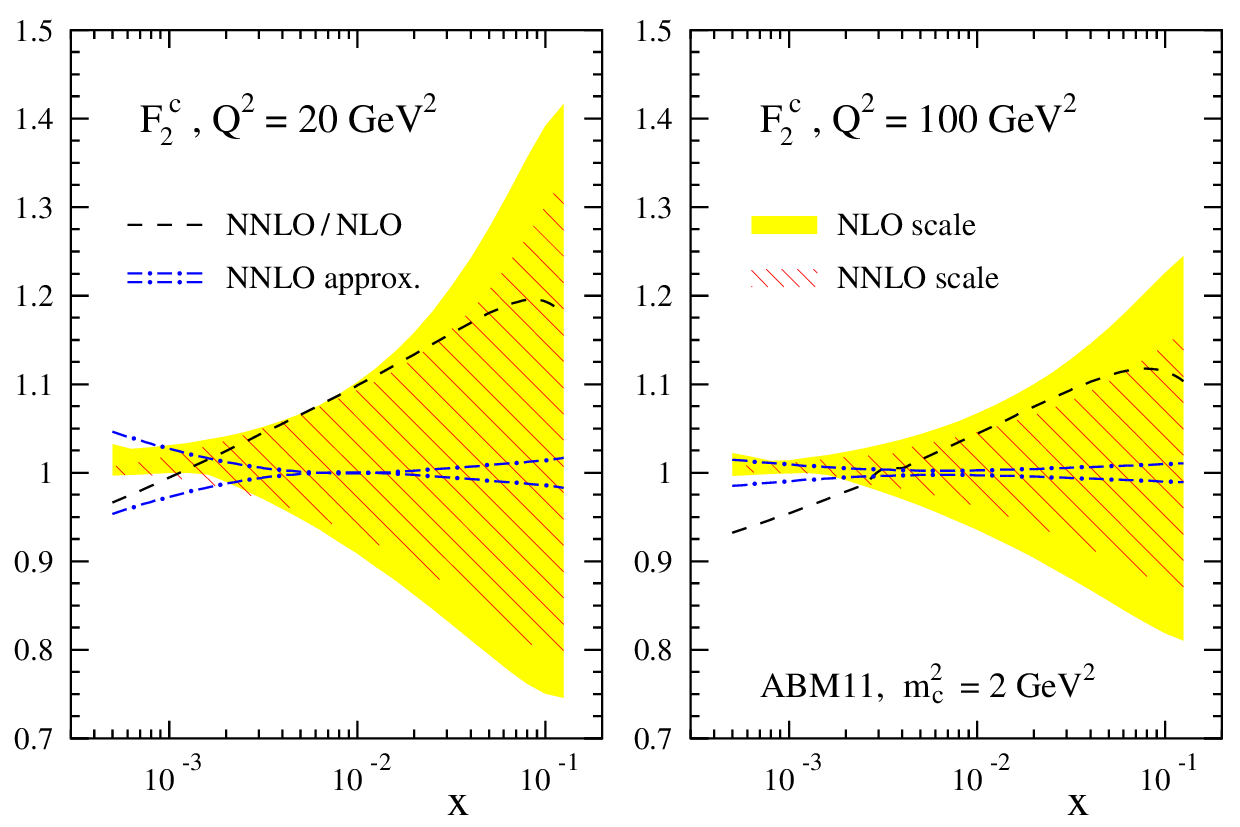}
  }
\vspace*{-1mm}
  \caption{ \small
   \label{fig:f2c-ratios}
  The relative NLO and NNLO scale uncertainties, for the standard
  range $\frac{1}{2} \leq \mu/\sqrt{\Qs \!+\! 4\:\!m^{\:\!2} } \leq 2$,
  and the relative NNLO approximation uncertainties of the results in 
  Fig.\ \ref{fig:f2c-xQ2} at two representative values of~$\Qs$.
  Also shown is the relative size of the average NNLO corrections at 
  the default scale $\mu = \sqrt{\Qs \!+\! 4\:\!m^{\:\!2}}$.
  }
\end{figure}

Returning to the case of charm production which has a far larger impact on the
determination of especially the gluon distribution from DIS data,
Figs.~\ref{fig:f2c-muvar} and \ref{fig:f2c-ratios} provide a more detailed
look at the NNLO results and their remaining uncertainties for two
typical values of $\Qs$.
The former figure shows the dependence on the scale $\mu$ for the range
$m_{\:\!\rm c} \leq \mu \leq 2\, \mu_{\,\rm std}$ with $\mu_{\,\rm std} 
= \mu$ of Eq.~(\ref{eq:mu-std}) which includes both hard scales entering the
problem, $2\:\!m_{\:\!\rm c}$ and $Q$.
Except at very low scales, $\mu < 2\:\!m_{\:\!\rm c}$, (and possibly  at very
small $x$) the NNLO scale variation of $F_2^{\,\rm c}$ is smaller than its
NLO counterpart given at $\mu = \mu_{\,\rm std}$ by the long-dashed curves in
Fig.~\ref{fig:f2c-xQ2}.
This is also obvious from Fig.~\ref{fig:f2c-ratios}, where the NLO and average
($\frac{1}{2}$(`A'+`B')) NNLO
scale uncertainties, estimated via the maximal and minimal values in the
conventional interval $[\frac{1}{2}\:\mu_{\,\rm std}, 2\,\mu_{\,\rm std\,}]$, 
are compared with the present approximation uncertainty which becomes important
at $x \simeq 10^{\:\!-3}$. The relatively small NNLO improvement at large $x$
indicates the need for yet higher orders. Here a first step would be to
include the (almost complete) NNLL threshold resummation discussed in
Section~\ref{sec:Thresresum}.

\vspace*{-2mm}
\section{Summary and outlook}
\label{sec:Conclusions}

\vspace*{-1mm}
The production of heavy quarks, in particular charm, contributes a sizeable
fraction of the proton structure function $F_2^{}$ measured at HERA, thus
affecting fits of the parton distributions and hence the predictions for hard 
processes at other colliders such as the LHC.
Unlike the case of massless quarks, the corresponding partonic cross sections
(coefficient functions) are not fully known yet at the next-to-next-to-leading
order (NNLO) of perturbative QCD.
In this article, we have collected and extended the partial NNLO results for 
three kinematic limits, i.e., the threshold region with $s \gsim 4\:\!m^2$,
the limit of high partonic CM energies, $s \gg m^2$, and the high-scale
region $Q^2\gg m^2$, and then combined these results into approximate 
expressions covering most of the kinematic plane.  

For the NNLO gluon coefficient function $c_{2,g}^{\,(2)}$ we have determined
all threshold contributions which are logarithmically enhanced in the 
heavy-quark velocity $\beta$, the so-called Sudakov logarithms, as well as the 
Coulomb corrections leading to powers of $1/\beta$ (Eq.~(\ref{eq:c2g2-thresh}) 
in Section.~\ref{sec:Thresresum}). 
In~the high-energy limit, we have employed results of the corresponding
`small-$x$' resummation \cite{Catani:1990eg} to derive explicit expression for 
the (closely related) leading-logarithmic $\ln s$ contribution to the 
gluon and quark coefficient functions at order $\alst$ 
(Eqs.~(\ref{eq:c2g2-smallx}) and (\ref{eq:c2g2-smallx-asy}) in 
Section~\ref{sec:small-x}). \linebreak
Finally, in the high-scale limit where $m^2/Q^2$ power corrections 
can be disregarded, we have utilized the mass-factorization 
formula to derive exact expressions for the NNLO gluon and (pure-singlet) 
quark coefficient functions $c_{2,g}^{\,(2)}$ and $c_{2,q}^{\,(2)}$ 
(Section~\ref{sec:asy} and Eqs.~(\ref{eq:H2g3}) and (\ref{eq:H2q3}) in 
Appendix~B).

The latter results are complete for the contributions proportional to $\nf$,
the number of light flavours, and rely on the three-loop operator-matrix 
elements (OMEs) of Ref.~\cite{Ablinger:2010ty} for the $\Qs$ independent 
contribution.
Their counterparts for the $\nf$-independent contributions are not fully
known yet, though, so we had to rely on the low even-$N$ Mellin moments
computed in Ref.~\cite{Bierenbaum:2009mv} and our above small-$x$ result for 
deriving the approximate expressions given in Eqs.~(\ref{eq:fitA})
-- (\ref{eq:fitBq}) in Section \ref{sec:asy}.
The largest uncertainty in these results is due to the next-to-leading
small-$x$ (at NNLO behaving as $s^{\,0}$ at large $s$) contributions which are 
not tightly constrained at this point.

We have then combined the results of the various regions to provide the best
possible approximations for the coefficient functions for heavy-quark
production in DIS at all values of $s$ and $\Qs$, carefully modeling and
estimating the impact of missing information such as the NNLO threshold 
constant and of the uncertainty of the high-$s$ constants and their dampening 
towards medium values of $s$. The corresponding results are given in 
Eq.~(\ref{eq:assembly20A}) -- (\ref{eq:c2g20-nllxBs}) in Section
\ref{sec:Coefficientfunctions}. The implications of these uncertainties for
the charm and bottom structure functions $F_2^{\,c}(x,\Qs)$ and 
$F_2^{\,b}(x,\Qs)$ have been illustrated in Section \ref{sec:Phenomenology}, 
where we have also addressed the dependence on the renormalization and 
factorization scale $\mu$ which can signal the importance of yet higher orders.
The uncertainty due to the missing information on the NNLO coefficient 
functions is large at $x \lsim 10^{-3}$ especially at low and medium scales 
$\Qs$, but irrelevant at much larger values of~$x$. The dominant uncertainty 
in that region (where the absolute values of $F_2^{\,c}$ and $F_2^{\,b}$ are 
much smaller) is due to higher orders, as shown by the rather modest NNLO 
improvement of the $\mu$-dependence. 

At high scales, $\Qs \gsim 10\: m^2$, the remaining small-$x$ problem can be
solved by extending the $k_t$-factorization results of 
Ref.~\cite{Catani:1990eg} to the next-to-leading high-energy contributions at 
order $\alst$ or by extending the calculations of \cite{Ablinger:2010ty} to the
$\nf$-independent OMEs. 
Already the extension of the fixed Mellin-$N$ computations of 
Ref.~\cite{Bierenbaum:2009mv} by a couple of moments, however, would facilitate
a considerable reduction of the uncertainties. These calculations could also 
be employed to considerably reduce, but presumably not remove, the large 
small-$x$ uncertainties at low and medium scales, $\Qs < 10\: m^2$. 
It is also worthwhile to note that there is a close relation between the 
next-to-leading high-energy contributions in the present case and in 
heavy-quark hadro-production, see Refs.~\cite{Catani:1990eg,Ball:2001pq}. 
This has been exploited in Ref.~\cite{Moch:2012mk} to estimate the high-energy 
tail of the coefficient functions for top-pair production at the LHC, which 
complements numerically determined results on NNLO heavy-quark production 
in hadron-hadron collisions~\cite{Baernreuther:2012ws} valid in the region 
where the partonic CM energy $s$ is not too large, $s/4\:\!m^2 \lsim 100$.

The uncertainties at low $\Qs$ may be removed to a fully satisfactory extent 
only by an exact third-order calculation of heavy-quark production in DIS. This
is a formidable task, including three-loop Feynman diagrams with two scales, 
i.e., an internal mass and an off-shell leg. 
A useful, but conceptually easier intermediate step would be a calculation in 
the threshold limit determining the missing threshold `constant' (i.e., the 
contribution proportional to the Born cross section) at order $\alst$. 
This result would also provide the only piece missing for extending our results
in Section \ref{sec:Thresresum} to a NNLO+NNLL (next-to-next-to-leading 
logarithmic) threshold resummation. This resummation, even without the NNLO
constant, provides the natural framework for a first step beyond the third
order, e.g., for addressing the still rather large scale uncertainty at large $x$.
 
At this point, though, our approximate third-order results represent the best 
and most complete predictions of heavy-quark DIS in the important context of 
NNLO determinations of parton distributions from inclusive and heavy-quark 
structure functions measured at HERA. 
Given the wider implications of those analyses on LHC physics, however, further
improvements in particular at small $x$ would be extremely welcome.
For use until then,
{\sc Form} files and {\sc Fortran} subroutines with our main results can be
obtained from the preprint server {\tt http://arXiv.org} by downloading the
source of this article. Furthermore they are available from the authors upon
request.
%
 
\subsection*{Acknowledgments}
We thank S.~Alekhin, T.~Huber and E.~Laenen for useful discussions. 
We are very grateful to E.~Laenen for providing us with the {\sc Fortran} code 
of Ref.~\cite{Laenen:1992zk} as a basis for the calculation of the NLO matching
coefficient in Eq.~(\ref{eq:nloconst}), and we have used the latest version of 
{\sc FORM}~\cite{Vermaseren:2000nd} for the analytic calculations.
This work has partially been  partially supported by the UK Science \& 
Technology Facilities Council (STFC) under grant numbers PP/E007414/1 and
ST/G00062X/1, by the Deutsche Forschungsgemeinschaft (DFG) in 
Sonderforschungs\-be\-reich/Transregio~9 and by the European Commission through
the contract PITN-GA-2010-264564 ({\it LHCPhenoNet}). 
H.K.~acknowledges the Grant-in-Aid for Scientific Research on Primary Areas
'Elucidation of New Hadrons with a Variety of Flavors (E01: 21105006)'.

\newpage
\appendix
\renewcommand{\theequation}{\ref{sec:appA}.\arabic{equation}}
\setcounter{equation}{0}
\section{Gluon coefficient function at NLO for $s \to 4 m^2$}
\label{sec:appA}

In this first appendix we derive the functions $c_0^{}(\xi)$ and 
$\bar{c}_0^{}(\xi)$ in Eqs.~(\ref{eq:nloconst}) and (\ref{eq:c0bar}) for the 
gluon coefficient function $c_{2,g}^{}$ at NLO. 
This derivation starts from the coefficient function in the differential 
kinematics of one-particle inclusive DIS~\cite{Laenen:1992zk},
\begin{eqnarray}
  \label{eq:integral}
  c_{2,g}^{\,(n)}(\xi,\beta,\mus) &\!=\!&
  \int_{\frac{\sspr}{2}(1-\beta)}^{\,\frac{\sspr}{2}(1+\beta)}\, 
  d(-t_1)\, \int_0^{\,s_4^{\,\rm max}} \!\! ds_4\;
  \frac{d^{\,2} c_{2,g}^{\,(n)}(\sspr,t_1,u_1,\mu^2)}{dt_1 \, ds_4}
  \:\; ,
\end{eqnarray}
where the kinematic variables for the subprocess
$g(zp)+\gamma^{\,\ast}(q) \,\rightarrow\, q_h(p_1)+X[\bar{q}_h]$ are defined 
as $t_1=(zp-p_1)^2-m^2$, $u_1=(q-p_1)^2-m^2$ and $\sspr=s+Q^2$.
The variable $s_4=\sspr+t_1+u_1=M_X^{\,2}-m^2\geq 0$ measures the inelasticity 
of the subprocess, with $s_4\rightarrow 0$ corresponding to the limit of soft 
gluon emission. The upper limit of $s_4$ is given by
\begin{eqnarray}
  \label{eq:s4max}
  s_4^{\,\rm max} \; =\;
  \frac{s}{\sspr t_1}
  \left(t_1+\frac{\sspr(1-\beta)}{2}\right)
  \left(t_1+\frac{\sspr(1+\beta)}{2}\right)
  \: .
\end{eqnarray}
The LO differential coefficient function is given by 
\cite{Laenen:1992zk,Witten:1975bh}
\begin{eqnarray}
  \label{eq:c2diff0}
  \frac{d^{\,2} c_{2,g}^{(0)}(\sspr,t_1,u_1)}{dt_1\, ds_4} \; = \;
  \delta(s_4) \: \sigma^{\,\rm Born}(\sspr,t_1,u_1) 
\end{eqnarray}
with
\begin{eqnarray}
\label{eq:born}
  \sigma^{\,\rm Born}(\sspr,t_1,u_1)
  &=&
  \frac{\pi}{2}\:\tf\:\frac{4m^2}{{s^{\prime}}^2}
  \left[\frac{t_1}{u_1}+\frac{u_1}{t_1}
  +4\:\frac{m^2\sspr}{t_1u_1}\left(1-\frac{m^2\sspr}{t_1u_1}\right)\right.
  -2\:\frac{\sspr \Qs}{t_1u_1}+2\frac{Q^{\,4}}{t_1u_1}
\nn\\[0.5mm] &&
\left. \qquad\qquad \mbox{}
  -2\:\frac{m^2\Qs}{t_1u_1}\left(2-\frac{s^{\prime 2}}{t_1u_1}\right)
  -12\:\frac{\Qs}{\sspr}\left(\frac{m^2\sspr}{t_1u_1}-\frac{s}{\sspr}\right)
\right]
\; . \qquad
\end{eqnarray}
The NLO differential coefficient function was calculated in 
Ref.~\cite{Laenen:1992zk}.
In the threshold region $s_4\simeq 0$, it can be simplified 
as~\cite{Laenen:1992zk,Laenen:1998kp,Alekhin:2008hc}
\begin{eqnarray}
  \label{eq:c2diff1}
  \frac{d^{\,2} c_{2,g}^{\,(1)}(\sspr,t_1,u_1,\mu^2)}{dt_1\, ds_4}
  &\!\simeq\!&
  4\, K^{(1)}(\sspr,t_1,u_1,\mu^2) \, \sigma^{\,\rm Born}(\sspr,t_1,u_1)
  \, .
\end{eqnarray}
Here the factor $K^{(1)}$ (in the MOM scheme for the strong coupling 
$\alpha_s$) is given by~\cite{Laenen:1992zk}
\begin{eqnarray}
  \label{eq:K1}
  K^{\,(1)}(\sspr,t_1,u_1,\mu^2) &\!=\!&
  2\:\! \ca \left[\frac{\ln(s_4/m^2)}{s_4}\right]_+
  \nn \\[0.5mm] &&
  + \left[\frac{1}{s_4}\right]_+
  \left\{
    \ca \left[
      \ln\left(\frac{t_1}{u_1}\right) + {\rm Re}\,L_\beta + \Lmmu
    \right]
    - 2 \cf ({\rm Re}\,L_\beta+1)
  \right\}
\nn\\[0.5mm] &&
\left. \mbox{}
  + \delta(s_4)
  \left\{
    R(\sspr,t_1,u_1)
    - \ca \ln\left(\frac{-u_1}{m^2}\right) \Lmmu
  \right\}\right]
\; ,
\end{eqnarray}
with
\begin{eqnarray}
  L_\beta &=&
  \frac{1-2\:\!m^2/s}{\beta}\,
  \left\{\ln\left(\frac{1-\beta}{1+\beta}\right)+i\pi\right\}
\end{eqnarray}
and the $+$-distribution in $s_4$ defined as
\begin{eqnarray}
\label{eq:+-dist}
 \left[\frac{\ln^n(s_4/m^2)}{s_4}\right]_+
 &\!=& 
 \lim_{\Delta\rightarrow 0}\left[\frac{\ln^{\,n}(s_4/m^2)}{s_4} 
 \theta(s_4-\Delta)
  +\frac{1}{n+1}\ln^{\,n+1}\left(\frac{\Delta}{m^2}\right) \delta(s_4)\right]
\; . \quad
\end{eqnarray}
The conversion of $K^{(1)}$ to the standard \MSbar\ scheme for $\alpha_s$ 
proceeds via the well-known relation
\begin{eqnarray}
  \label{eq:mom2ms}
  \alpha_{\rm s}^{\:\!\rm MOM}(\mus) &\!=\!&
  \as(\mus)
  \left[
    1
    + \frac{4}{3}\, \tf\, \frac{\as(\mus)}{4\pi} \Lmmu
    + {\cal O}(\alss)
  \right] \; ,
\end{eqnarray}
and affects the function $\bar{c}_0(\xi)$ for the scale-dependent matching 
constant, see Eq.~(\ref{eq:c0barApp}) below. 
The scale-independent function $R(\sspr, t_1, u_1)$ in Eq.~(\ref{eq:K1})
originates from the soft- and virtual- gluon contributions, for which the 
analytic expression is quite lengthy and available only via a {\sc Fortran} 
code by the authors of Ref.~\cite{Laenen:1992zk}.

In order to identify the leading contribution for $s \to 4\:\! m^2$, i.e., in 
the limit $\beta \to 0$, we first consider Eq.~(\ref{eq:integral}) for the LO 
coefficient function with the integrand given by Eq.~(\ref{eq:c2diff0}).
After the trivial $s_4$-integral, we have $u_1=-\sspr-t_1$, and the leading 
contribution in the $t_1$-integral can be obtained by the simple replacement
\begin{eqnarray}
  \label{eq:replace}
  t_1^{} \;\rightarrow\; -\frac{\sspr}{2}
\, ,\qquad\qquad
  u_1^{} \;\rightarrow\; -\frac{\sspr}{2}
\; ,
\end{eqnarray}
and multiplying with an overall factor $\sspr\beta$.
This can be understood by observing
\begin{eqnarray}
  \label{eq:t1-int}
  \int_{\frac{\sspr}{2}(1-\beta)}^{\frac{\sspr}{2}(1+\beta)}\,d(-t_1)\, f(t_1)
  &=&
  \sspr\beta\: f\left(-\frac{\sspr}{2}\right)
  + \frac{1}{24} {\sspr}^3\beta^3 f^{\prime\prime}\left(-\frac{\sspr}{2}\right)
  + \cdots \:\; ,
\end{eqnarray}
where the first term on the right-hand side gives the leading contribution in 
the $\beta$ unless the derivative of $f(t_1)$ at $t_1=-\sspr/2$ contains 
negative powers of $\beta$.
Furthermore, we perform expansion in $\beta$ such that 
$s = 4\:\!m^2(1+{\cal O}(\beta^2))$ and we recover from Eq.~(\ref{eq:born})
the threshold approximation of the LO coefficient function in 
Eq.~(\ref{eq:c2g0-thresh}).

At NLO we first note that constants and positive powers in
$s_4$, which are ignored in factorization ansatz of
Eq.~(\ref{eq:c2diff1}), yield corrections of higher order in $\beta$
after integration due to $s_4^{\rm max}\simeq{\cal O}(\beta^2)$.
Then the replacement of (\ref{eq:replace}) can effectively be
applied to $\sigma^{\,\rm Born}$ in (\ref{eq:c2diff1}), so that the
$s_4$-integrals become trivial with the $+$-distribution from
Eq.~(\ref{eq:+-dist}). Care needs to be taken in the subsequent
$t_1$-integration, though, if we cannot rely on the replacement rule
(\ref{eq:replace}). This exception only arises if the terms
$\ln^n(s_4^{\,\rm max}/m^2)~(n=1,2)$ appear via Eq.~(\ref{eq:+-dist}),
because the coefficients of their Taylor expansion at
$t_1=-\sspr/2$ contain negative power of $\beta$, thus
invalidating the assumptions underlying Eq.~(\ref{eq:t1-int}). These
terms give rise to the threshold logarithms, and we obtain by
explicit calculation
\begin{eqnarray}
  \int_{\frac{\sspr}{2}(1-\beta)}^{\frac{\sspr}{2}(1+\beta)}d(-t_1)
  \ln\left(\frac{s_4^{\,\rm max}}{m^2}\right)\,
  &\!=\!&
  \sspr\,\beta\left[\ln(8\beta^2)-2\right]+{\cal O}(\beta^3)
  \, ,
  \\[1mm]
  \int_{\frac{\sspr}{2}(1-\beta)}^{\frac{\sspr}{2}(1+\beta)}d(-t_1)
  \ln^2\left(\frac{s_4^{\,\rm max}}{m^2}\right)
  &\!=\!&
  \sspr\,\beta\left[\ln^2(8\beta^2)-4\ln(8\beta^2)+8-\frac{\pi^2}{3}\right]
  + {\cal O}(\beta^3)
  \, .
\end{eqnarray}
Using these formulae together with Eq.~(\ref{eq:replace}) and the expansions 
in $\beta$, we recover the threshold approximation of the gluon coefficient 
function $c_{2,g}^{(1)}$ in Eq.~(\ref{eq:c2g1-thresh}), now with explicit 
results for the the constant terms $c_0^{}(\xi)$ and $\bar{c}_0^{}(\xi)$ at NLO.

The scale-dependent term $\bar{c}_0(\xi)$ which is also fixed by 
renormalization-group constraints can be easily read off from 
Eq.~(\ref{eq:K1}),
\begin{eqnarray}
\label{eq:c0barApp}
  \bar{c}_0^{}(\xi) &\!=\!& 4\:\! \ca
  \left[2+\ln\left(1+\frac{\xi}{4}\right)\right]-\frac{4}{3} \tf
  \; ,
\end{eqnarray}
where the last term is due to the transformation of the strong
coupling from the MOM scheme in~\cite{Laenen:1992zk} to $\alpha_s$
in the standard $\overline{\rm MS}$ 
scheme~\cite{Buza:1996wv,Bierenbaum:2009zt}, cf.~Eq.~(\ref{eq:mom2ms}).
The calculation of the scale-independent term $c_0(\xi)$ on the other hand 
requires some automated manipulations of $R(\sspr,t_1,u_1)$, based on the 
{\sc Fortran} code of~\cite{Laenen:1992zk}, which finally lead to the new
result (\ref{eq:nloconst}) in Section \ref{sec:Thresresum}.

\def\ca{{\textcolor{blue}{C^{}_A}}}
\def\cas{{\textcolor{blue}{C^{\,2}_A}}}
\def\cf{{\textcolor{blue}{C^{}_F}}}
\def\cfs{{\textcolor{blue}{C^{\,2}_F}}}
\def\nf{{\textcolor{blue}{n^{}_{\! f}}}}
\def\tf{\textcolor{blue}{T_F}}
\renewcommand{\theequation}{\ref{sec:appB}.\arabic{equation}}
\setcounter{equation}{0}
\section{Exact results at asymptotic values $Q^2 \gg m^2$}
\label{sec:appB}

We close by presenting the exact expressions for the heavy-quark coefficient 
function $H_{2,g}$ and $H^{\,\rm ps}_{2,q}$ at high scales, $\Qs \gg m^2$.
In order to shorten the results, we have always separated the contribution
of the coefficient functions with massless quarks in photon-exchange DIS, 
denoted as $c_{2,k}^{\,(\ell)}$ below, as computed in 
Ref.~\cite{Vermaseren:2005qc} at $\ell \leq 3$. 
See also Refs.~\cite{Zijlstra:1992qd,Moch:1999eb} for earlier results up to
the second order.

For easier comparison with the literature, we will use the following 
established normalization of these asymptotic coefficient function, 
cf.~Ref.~\cite{Buza:1995ie,Bierenbaum:2007qe},
\begin{eqnarray}
  \label{eq:Hcoeff-exp}
  H_{2,g}(x,\Qs,m^2) &\!=\!&
    \ar\, H_{2,g}^{(1)}(x,\Qs,m^2)
  \,+\, \ars\, H_{2,g}^{(2)}(x,\Qs,m^2)
  \,+\, \art\, H_{2,g}^{(3)}(x,\Qs,m^2)
  \,+\, {\cal O}(a_{\rm s}^4)
  \qquad
\end{eqnarray}
and correspondingly for $H^{\,\rm ps}_{2,q}$. As above we expand in terms of
$\ar = \as/(4 \pi)$ and, again following 
Refs.~\cite{Buza:1995ie,Bierenbaum:2007qe}, this expansion is performed in the 
\MSbar\ scheme with $\alpha_s(\nf+1)$. All terms originating from subsequent 
matching $\alpha_s(\nf+1) \to \alpha_s(\nf)$ with the help of the decoupling 
formulae~\cite{Larin:1994va,Chetyrkin:1997sg} will be presented separately as 
well.
The relation of, e.g., the quantity (\ref{eq:Hcoeff-exp}) to the gluon 
coefficient function $c_{2,g\,}^{}$, defined in Eq.~(\ref{eq:totalF2c}) in the
normalization of Refs.~\cite{Laenen:1992zk,Riemersma:1994hv}, is given by
\begin{eqnarray}
  \label{eq:Hcoeff-norm}
  c_{2, g}^{}(\eta,\xi) &\!=\!&
  \frac{\pi\, x}{\xi}\,
  \bigg\{
  H_{2,g}^{(1)}(x,\Qs,m^2)
  \,+\, \ar\, H_{2,g}^{(2)}(x,\Qs,m^2)
  \,+\, \ars\, H_{2,g}^{(3)}(x,\Qs,m^2)
  \,+\, {\cal O}(\art)
  \bigg\}
  \qquad
\end{eqnarray}
where we have suppressed the additional dependence on $\mu = \muf = \mur$.

The explicit results are presented in terms of harmonic polylogarithms 
${\rm H}_{m_1,\ldots,m_w}(x)$ with $m_j = 0,\pm 1$, where our notation follows 
Ref.\ \cite{Remiddi:1999ew} to which the reader is referred for a detailed 
discussion. For chains of indices zero we again employ the abbreviated 
notation
\begin{equation}
  \label{eq:habbr}
  {\rm H}_{{\footnotesize \underbrace{0,\ldots ,0}_{\scriptstyle m} },\,
    \pm 1,\, {\footnotesize \underbrace{0,\ldots ,0}_{\scriptstyle n} },
    \, \pm 1,\, \ldots}(x) \: = \: {\rm H}_{\pm (m+1),\,\pm (n+1),\, \ldots}
\end{equation}
in which also the argument of ${\rm H}_{\vec{m}}$ has been suppressed for 
brevity. The numerical evaluation of the the harmonic polylogarithms relies on 
the {\sc Fortran} package of Ref.~\cite{Gehrmann:2001pz} and its weight-five 
extension provided by the authors.

The leading-order coefficient $H_{2,g}^{(1)}$ in the expansion 
(\ref{eq:Hcoeff-exp}) reads
\begin{eqnarray}
\label{eq:H2g1}
  & H_{2,g}^{(1)}(x,Q^2,m^2) & \!= \;
  2\*(1-2 \* x+2 \* x^2) \*\LQm
  \,+\, c_{2,g}^{\,(1)}(x, \nf +1)/(\nf +1)
\, ,
\end{eqnarray}
where $c_{2,g}^{\,(1)} (x, \nf +1)$ denotes the one-loop gluon coefficient 
function with massless quarks at $\mu = Q$ as given in Eq.~(B.4) 
of~Ref.~\cite{Vermaseren:2005qc}, and $\LQm$ had been defined in Eq.\
(\ref{eq:LQmLmmdef}) above.
At NLO we find, in agreement with Refs.~\cite{Buza:1995ie,Bierenbaum:2007qe},
\begin{eqnarray}
\label{eq:H2g2}
{\lefteqn{
H_{2,g}^{(2)}(x,Q^2,m^2) \, = \,}}
\nonumber\\&&
  \cf   \* \Big[
  2 \* (1 - 12 \* x + 10 \* x^2) \* \H(1,0)
  - 4 \* (1 - 2 \* x) \* \H(2,0)
  - 4 \* (1 - 2 \* x - 2 \* x^2) \*  \z3
\nonumber\\&& \mbox{}
  + 4 \* (1 + 2 \* x - 3 \* x^2) \* \H(1,1)
  - 4 \* (1 + 6 \* x - 3 \* x^2) \*  \z2
  + 4 \* (1 + 6 \* x - 3 \* x^2) \* \H(2)
\nonumber\\&& \mbox{}
  - (1 + 12 \* x - 20 \* x^2) \* \H(0,0)
  - (8 + 9 \* x + 24 \* x^2) \* \H(0)
  + (13 - 41 \* x + 40 \* x^2)
  + 2 \* (13 \* x - 12 \* x^2) \* \H(1)
\nonumber\\&& \mbox{}
  + 2 \* (1-2 \* x+4 \* x^2)  \*\H(0,0,0)
  +  4 \* (1-2 \* x+2 \* x^2)  \*  (
  \H(1,0,0)
  - \H(1,1,1)
  - \H(2,1)
  )\Big]
\nonumber\\&& \mbox{}
  +  \ca   \* \Big[
  - 4 \* (1 + 2 \* x) \* \H(0,0,0)
  - 4 \* (1 + 2 \* x + 2 \* x^2) \* \H(-1) \*  \z2
  - 8 \* (1 + 2 \* x + 2 \* x^2) \* \H(-1,-1,0)
\nonumber\\&& \mbox{}
  + 4 \* (1 + 2 \* x + 2 \* x^2) \* \H(-1,0,0)
  + 8 \* (1 + 4 \* x) \* \H(2,0)
  - 2 \* (1 + 4 \* x - 5 \* x^2) \* \H(1,1)
  - 2 \* (1 + 4 \* x - 4 \* x^2) \* \H(1)
\nonumber\\&& \mbox{}
  + 8 \* (2 + 7 \* x) \*  \z3
  + 2/3 \* (3 + 12 \* x + 23 \* x^2) \* \H(0,0)
  + 2/3 \* (9 + 8 /x + 48 \* x - 65 \* x^2) \* \H(1,0)
\nonumber\\&& \mbox{}
  - 2/27 \* (9 + 112 /x + 1413 \* x - 1588 \* x^2)
  - 2/9 \* (42 + 129 \* x + 400 \* x^2) \* \H(0)
\nonumber\\&& \mbox{}
  + 8 \* (x + x^2) \* \H(-1,0)
  + 2 \* (4 \* x - x^2) \*  \z2
  + 2 \* x^2 \* \H(2)
  +  4 \* (1-2 \* x+2 \* x^2)  \*  (
  - \H(1,1,0)
  + \H(1,1,1)
  - \H(1,2)
  )\Big]
\nonumber\\ & & \mbox{\hspn} +
  \LQm  \* \Big\{ \cf  \* \Big[
  4 \* (1 - 6 \* x + 10 \* x^2) \* \H(0)
  + 2 \* (7 - 24 \* x + 20 \* x^2) \* \H(1)
  + 2 \* (9 - 17 \* x + 4 \* x^2)
\nonumber\\ \mbox{}&&
  +  4 \*(3-6 \* x+8 \* x^2)  \*  (\H(2) - \z2 )
  +  8 \* (1-2 \* x+4 \* x^2)  \* \H(0,0)
  +  16 \* (1-2 \* x+2 \* x^2)  \*  (\H(1,0) + \H(1,1))
  \Big]
\nonumber\\&& \mbox{}
  +  \ca  \* \Big[ (
  - 8 \* (1 + 2 \* x + 2 \* x^2) \* \H(-1,0)
  + 4/3 \* (3 - 4 /x - 60 \* x + 67 \* x^2) \* \H(1)
\nonumber\\&& \mbox{}
  - 2/9 \* (165 - 52 /x + 276 \* x - 407 \* x^2)
  - 16 \* (3 \* x - x^2) \* \H(2)
  - 4 \* (24 \* x - 25 \* x^2) \* \H(0)
  )
\nonumber\\&& \mbox{}
  +  16 \*(2 \* x-x^2)  \* \z2
  - 16 \*  (1+3 \* x)  \* \H(0,0)
  +  8 \* (1-2 \* x+2 \* x^2)  \*  ( \H(1,0) + \H(1,1) )
  \Big]\Big\}
\nonumber\\ & & \mbox{\hspn} +
  \LQms  \*\Big\{\cf  \* \Big[
  -   1 + 4 \* x
  -   2 \*  (1-2 \* x+4 \* x^2)  \* \H(0)
  -   4 \*   (1-2 \* x+2 \* x^2)  \*  \H(1)
  \Big]
\nonumber\\&& \mbox{}
  +  \ca  \* \Big[
  4 \* (1 + 4 \* x) \* \H(0)
  + 2/3 \* (3 + 4 /x + 24 \* x - 31 \* x^2)
  - 4 \*(1-2 \* x+2 \* x^2)  \*  \H(1)
  \Big]\Big\}
\nonumber\\ & & \mbox{\hspn} +
  \Lmmu  \* \Big\{  \ca  \* \Big[
  - 8 \* (1 + 4 \* x) \* \H(0,0)
  + 4/3 \* (3 - 4 /x - 72 \* x + 79 \* x^2) \* \H(1)
  - 4/3 \* (3 + 96 \* x - 31 \* x^2) \* \H(0)
\nonumber\\&& \mbox{}
  +  (1-2 \* x+2 \* x^2)  \*  (
  8 \* \H(1,0) + 16 \* \H(1,1) )
  - 2/3 \* (43 - 4 /x + 242 \* x - 281 \* x^2)
  + 16 \* (3 \* x - x^2) \*  \z2
\nonumber\\&& \mbox{}
  - 16 \* (3 \* x - x^2) \* \H(2)
  \Big]
  - 4/3 \* (1 - 8 \* x + 8 \* x^2)
  - 4/3 \* (1-2 \* x+2 \* x^2)  \*  (\H(0) + \H(1))
\nonumber\\ & & \mbox{\hspn} +
  \LQm \*\Big\{
  \ca  \* \Big[
  8 \* (1 + 4 \* x) \* \H(0)
  + 4/3 \* (3 + 4 /x + 24 \* x - 31 \* x^2)
  - 8 \* (1-2 \* x+2 \* x^2)  \* \H(1)
  \Big]
\nonumber\\&& \mbox{}
  +4/3 \* (1-2 \* x+2 \* x^2)
  \Big\} \Big\}
\,+\, c_{2,g}^{\,(2)}(x, \nf +1)/( \nf +1)
\; .
\end{eqnarray}
Here $c_{2,g}^{\,(2)}(x, \nf +1)$ is the two-loop gluon coefficient function 
with massless quarks at $\mu = Q$ given in Eq.~(B.6) of 
Ref.~\cite{Vermaseren:2005qc}.
Expanding in powers of $\alpha_s(\nf)$, instead, leads to the additional term
\begin{eqnarray}
\label{eq:H2g2asnl}
H_{2,g}^{(2)}(x,Q^2,m^2) 
&\stackrel{\alpha_s(\nf+1) \;\to\; \alpha_s(\nf)}\longrightarrow&
H_{2,g}^{(2)}(x,Q^2,m^2) - \frct{2}{3} \Lmmu \, H_{2,g}^{(1)}(x,Q^2,m^2)
\; .
\end{eqnarray}

The NNLO contribution $H_{2,g}^{(3)}$ is new.
It is exact as far as all $\nf$-dependence as well as all terms proportional 
to $\LQm$ and $\Lmmu$ are concerned, and is given by
\small

\normalsize
Again, $c_{2,g}^{\,(3)}(x, \nf +1)$ denotes the gluon coefficient function with
massless quarks at $\mu = Q$, now at three loops, as determined in Eq.~(B.9) 
of Ref.~\cite{Vermaseren:2005qc}, where all contributions proportional to 
$fl^{\,g}_{11}$ are omitted as discussed in Section~\ref{sec:asy}.
The $\nf$-independent part of the massive OME has been approximated in 
Eqs.~(\ref{eq:fitA}) and (\ref{eq:fitB}) and is denoted by $a_{Qg}^{(3)\,0}$.
Decoupling of the heavy quark in $\alpha_s$ in the \MSbar\ scheme leads to the
additional matching terms~\cite{Larin:1994va,Chetyrkin:1997sg} 
\begin{eqnarray}
\label{eq:H2g3asnl}
H_{2,g}^{(3)}(x,Q^2,m^2) 
&\stackrel{\alpha_s(\nf+1) \;\to\; \alpha_s(\nf)}\longrightarrow&
H_{2,g}^{(3)}(x,Q^2,m^2) 
- \frct{4}{3} \Lmmu \, H_{2,g}^{(2)}(x,Q^2,m^2)
- \biggl\{
  \biggl(\frct{16}{9}\,\ca - \frct{15}{2}\,\cf\biggr) 
\nonumber \\ && 
  +\biggl(\frct{10}{3}\,\ca + 2\cf\biggr) \Lmmu 
  - \frct{4}{9} \Lmmus 
\biggr\} \:
H_{2,g}^{(1)}(x,Q^2,m^2)
\; .
\end{eqnarray}

The corresponding expansion coefficients $H_{2,q}^{\,(\ell),{\rm ps}}$ at LO
and NLO are given by $H_{2,q}^{\,(1),{\rm ps}} = 0$ and
\begin{eqnarray}
\label{eq:H2q2}
{\lefteqn{
H_{2,q}^{\,(2),{\rm ps}}(x,Q^2,m^2) \, = \,}}
\nonumber\\&&
       \cf \* \Big[
       4\* (1 + x) \* (4 \*   \z3  - \H(0,0,0) + 2  \* \H(2,0) )
      + 4/3 \* (3 + 4/x - 3 \* x - 4 \* x^2) \* \H(1,0) 
\nonumber\\&& \mbox{}
      + 2/3 \* (3 + 15 \* x + 8 \* x^2) \* \H(0,0) 
         - 2/27 \* (9 + 112/x + 279 \* x - 400 \* x^2)
         - 4/9 \* (21 + 33 \* x + 56 \* x^2) \* \H(0) 
         \Big]
\nonumber\\ & & \mbox{\hspn} +
        \LQm \* \cf \* \Big[ 8 \* (1 + x) \* ( \z2  - 2 \* \H(0,0) - \H(2) )
            - 4/3 \* (3 + 4/x - 3 \* x - 4 \* x^2) \* \H(1) 
\nonumber\\&& \mbox{}
       - 8/9 \* (39 - 13/x - 30 \* x + 4 \* x^2) 
       + 16 \* x^2 \* \H(0) 
       \Big]
\nonumber\\ & & \mbox{\hspn} +
        \LQms  \* \cf \* \Big[ 
       4 \* (1 + x) \* \H(0) 
       + 2/3 \* (3 + 4/x - 3 \* x - 4 \* x^2) \Big]
\nonumber\\ & & \mbox{\hspn} +
       \Lmmu \* \Big\{ \cf \* \Big[
       8 \* (1 + x) \* ( \z2  - \H(0,0) -  \H(2) )
        - 4/3 \* (3 + 4/x - 3 \* x - 4 \* x^2) \* \H(1)
\nonumber\\&& \mbox{}
        - 4/3 \* (3  + 15 \* x - 4 \* x^2) \* \H(0) 
        - 8/3 \* (10 - 1/x - x - 8 \* x^2) 
        \Big]
\nonumber\\ & & \mbox{\hspn} +
       \LQm \* \cf \* \Big[ 
          8 \* (1 + x) \* \H(0) 
          + 4/3 \* (3 + 4/x - 3 \* x - 4\* x^2) 
          \Big]
          \Big\}
  \,+\, c_{2,q}^{\,(2),{\rm ps}}(x, \nf +1)/(\nf +1)
\, ,
\end{eqnarray}
where $c_{2,q}^{\,(2),{\rm ps}}(x, \nf +1)$ denotes the two-loop pure-singlet 
quark coefficient function at $\mu = Q$ as written down in Eq.~(B.7) of
Ref.~\cite{Vermaseren:2005qc}.
 
Finally the new third-order (NNLO) pure-singlet coefficient function for
heavy-quark production at $\Qs \gg m^2$ reads
\small

\normalsize
Here $c_{2,q}^{\,(3),{\rm ps}}(x, \nf +1)$ is the three-loop pure-singlet quark
coefficient function at $\mu = Q$ as given in Eq.~(B.10) of 
Ref.~\cite{Vermaseren:2005qc}, while $a_{Qq,\,{\rm ps}}^{\,(3)\,0}$ denotes 
the $\nf$-independent part of the massive OME approximated in 
Eqs.~(\ref{eq:fitAq}) and (\ref{eq:fitBq}). 
Again, the terms proportional to $fl_{11}$ in 
$c_{2,q}^{\,(3),{\rm ps}}(x, \nf +1)$ -- cf.~Eq.~(4.12) in 
Ref.~\cite{Vermaseren:2005qc} -- are omitted following the discussion in 
Section~\ref{sec:asy}.
Expanding in powers of $\alpha_s(\nf)$ instead of $\alpha_s(\nf+1)$
leads to the additional term
\begin{eqnarray}
\label{eq:H2q3asnl}
H_{2,q}^{(3),{\rm ps}}(x,Q^2,m^2) 
&\stackrel{\alpha_s(\nf+1) \;\to\; \alpha_s(\nf)}\longrightarrow&
H_{2,q}^{(3),{\rm ps}}(x,Q^2,m^2) 
- \frct{4}{3} \Lmmu \, H_{2,q}^{(2),{\rm ps}}(x,Q^2,m^2) 
\; .
\end{eqnarray}


\renewcommand{\theequation}{E.\arabic{equation}}
\setcounter{equation}{0}
\begin{titlepage}

\noindent
\hfill August 2016\
\vspace{1.7cm}
\begin{center}
\Large{\bf
Erratum: \\On the next-to-next-to-leading order QCD corrections \\ 
to heavy-quark production in deep-inelastic scattering}\\
\vspace{1.5cm}
\large
H. Kawamura$^{\, a}$, N.A. Lo Presti$^{\, b}$, S. Moch$^{\, c}$ and 
A. Vogt$^{\, d}$\\
\vspace{1.5cm}
\normalsize
{\it $^a$KEK Theory Center \\
Tsukuba, Ibaraki 305-0801, Japan}\\[4mm]

{\it $^b$ Physik-Institut, Universit\"at Z\"urich \\
  Winterthurerstrasse 190, CH--8057 Z\"urich, Switzerland}\\[4mm]

{\it $^c$II. Institut f\"ur Theoretische Physik, Universit\"at Hamburg \\
   Luruper Chaussee 149, D--22761 Hamburg, Germany}\\[4mm]

{\it $^d$Department of Mathematical Sciences, University of Liverpool \\
Liverpool L69 3BX, United Kingdom}\\[2.5cm]

\end{center}

The result for the heavy-quark coefficient function $H_{2,g}^{(3)}$ 
at asymptotic values $Q^2 \gg m^2$ in Eq.~(B.7) of Ref.~\cite{Kawamura:2012cr} 
has to be corrected by adding the following term
\begin{eqnarray}
H_{2,g}^{(3)}(x,Q^2,m^2) \, = \, 
H_{2,g}^{(3)}(x,Q^2,m^2)\biggr|_{\small \mbox{Eq.~(B.7)}}
          + \cf \* \nf \* (1 - 2 \* x + 2 \* x^2)\*(69 - 28 \* \z2)
          \, .
\end{eqnarray}
All numerical results in Ref.~\cite{Kawamura:2012cr} are unchanged. 
The relative effect of the additional term is at most of the order $10^{-3}$.

\noindent
{\bf{Acknowledgments}}: We thank I.~Bierenbaum for useful discussions.


\end{titlepage}


{\footnotesize

\begin{thebibliography}{10}

\bibitem{Chekanov:2009kj}
  S.~Chekanov {\it et al.}  [ZEUS Collaboration],
  Eur.\ Phys.\ J.\ C65 (2010) 65, arXiv:0904.3487 [hep-ex]

\bibitem{Aaron:2009ut}
  F.D.~Aaron {\it et al.}  [H1 Collaboration],
  Eur.\ Phys.\ J.\ C65 (2010) 89, arXiv:0907.2643 [hep-ex]

\bibitem{Aaron:2011gp}
  F.D.~Aaron {\it et al.}  [H1 Collaboration],
  Eur.\ Phys.\ J.\ C71 (2011) 1769, arXiv:1106.1028 [hep-ex]

\bibitem{Laenen:1992zk}
  E.~Laenen, S.~Riemersma, J.~Smith and W.L.~van Neerven,
  Nucl.\ Phys.\ B392 (1993) 162

\bibitem{Riemersma:1994hv}
  S.~Riemersma, J.~Smith and W.L.~van Neerven,
  Phys.\ Lett.\ B347 (1995) 143, hep-ph/9411431

\bibitem{Harris:1995tu}
  B.W.~Harris and J.~Smith,
  Nucl.\ Phys.\ B452 (1995) 109, hep-ph/9503484

\bibitem{Laenen:1998kp}
  E.~Laenen and S.~Moch,
  Phys.\ Rev.\ D59 (1999) 034027, hep-ph/9809550

\bibitem{Catani:1990eg}
  S.~Catani, M.~Ciafaloni and F.~Hautmann,
  Nucl.\ Phys.\ B366 (1991) 135

\bibitem{Buza:1995ie}
  M.~Buza, Y.~Matiounine, J.~Smith, R.~Migneron, W.~van Neerven,
  Nucl.\ Phys.\ B472 (1996) 611, hep-ph/9601302

\bibitem{Bierenbaum:2007qe}
  I.~Bierenbaum, J.~Bl\"umlein and S.~Klein,
  Nucl.\ Phys.\ B780 (2007) 40, hep-ph/0703285

\bibitem{Bierenbaum:2008yu}
  I.~Bierenbaum, J.~Bl\"umlein, S.~Klein and C.~Schneider, arXiv:0803.0273
   Nucl.\ Phys.\ B803 (2008) 1

\bibitem{Bierenbaum:2009zt}
  I.~Bierenbaum, J.~Bl\"umlein and S.~Klein,
  Phys.\ Lett.\ B672 (2009) 401, arXiv:0901.0669

\bibitem{Bierenbaum:2009mv}
  I.~Bierenbaum, J.~Bl\"umlein and S.~Klein,
  Nucl.\ Phys.\ B820 (2009) 417, arXiv:0904.3563

\bibitem{Ablinger:2010ty}
  J.~Ablinger, J.~Bl\"umlein, S.~Klein, C.~Schneider and F.~Wissbrock,
  Nucl.\ Phys.\ B844 (2011) 26, arXiv:1008.3347

\bibitem{Alekhin:2010sv}
  S.~Alekhin and S.~Moch,
  Phys.\ Lett.\ B699 (2011) 345, 1011.5790 

\bibitem{Presti:2010pd}
  N.A.~Lo Presti, H.~Kawamura, S.~Moch and A.~Vogt,
  PoS DIS$\:\!$2010 (2010) 163, arXiv:1008.0951 

\bibitem{Vermaseren:2005qc}
  J.A.M.~Vermaseren, A.~Vogt and S.~Moch,
  Nucl.\ Phys.\ B724 (2005) 3, hep-ph/0504242

\bibitem{Witten:1975bh}
  E.~Witten,
  Nucl.\ Phys.\ B104 (1976) 445

\bibitem{Gluck:1979aw}
  M.~Gl\"uck and E.~Reya,
  Phys.\ Lett.\ B83 (1979) 98

\bibitem{Alekhin:2003ev}
  S.I.~Alekhin and J.~Bl\"umlein,
  Phys.\ Lett.\ B594 (2004) 299, hep-ph/0404034

\bibitem{vanNeerven:2000uj}
  W.L.~van Neerven and A.~Vogt,
  Nucl.\ Phys.\ B588 (2000) 345, hep-ph/0006154

\bibitem{Sterman:1986aj}
  G.~F.~Sterman,
  Nucl.\ Phys.\ B281 (1987) 310

\bibitem{Catani:1989ne}
  S.~Catani and L.~Trentadue,
  Nucl.\ Phys.\ B327 (1989) 323

\bibitem{Contopanagos:1996nh}
  H.~Contopanagos, E.~Laenen and G.F.~Sterman,
  Nucl.\ Phys.\ B484 (1997) 303, hep-ph/9604313

\bibitem{Kidonakis:1997gm}
  N.~Kidonakis and G.F.~Sterman,
  Nucl.\ Phys.\ B505 (1997) 321, hep-ph/9705234

\bibitem{Bonciani:1998vc}
  R.~Bonciani, S.~Catani, M.L.~Mangano and P.~Nason,
  Nucl.\ Phys.\ B529 (1998) 424, hep-ph/9801375

\bibitem{Hoang:2000yr}
  A.H.~Hoang {\it et al.},
  Eur.\ Phys.\ J.\ direct C2 (2000) 1, hep-ph/0001286

\bibitem{Gluck:1993dpa}
  M.~Gl\"uck, E.~Reya and M.~Stratmann,
  Nucl.\ Phys.\ B422 (1994) 37

\bibitem{Vogt:1996wr}
A.~Vogt, proceedings of DIS$\:\!$96, hep-ph/9601352

\bibitem{Alekhin:2008hc}
  S.~Alekhin and S.~Moch,
  Phys.\ Lett.\ B672 (2009) 166, arXiv:0811.1412

\bibitem{Moch:2008qy}
  S.~Moch and P.~Uwer,
  Phys.\ Rev.\ D78 (2008) 034003, arXiv:0804.1476

\bibitem{Moch:2004pa}
  S.~Moch, J.A.M.~Vermaseren and A.~Vogt,
  Nucl.\ Phys.\ B688 (2004) 101, hep-ph/0403192

\bibitem{Vogt:2004mw}
  A.~Vogt, S.~Moch and J.A.M.~Vermaseren,
  Nucl.\ Phys.\ B691 (2004) 129, hep-ph/0404111

\bibitem{Moch:2005ky}
  S.~Moch and A.~Vogt,
  Phys.\ Lett.\ B631 (2005) 48, hep-ph/0508265

\bibitem{Idilbi:2005ni}
  A.~Idilbi, X.~d.~Ji, J.-P.~Ma and F.~Yuan,
  Phys.\ Rev.\ D73 (2006) 077501, hep-ph/0509294

\bibitem{Beneke:2009rj}
  M.~Beneke, P.~Falgari and C.~Schwinn,
  Nucl.\ Phys.\ B828 (2010) 69, arXiv:0907.1443  

\bibitem{Czakon:2009zw}
  M.~Czakon, A.~Mitov and G.F.~Sterman,
  Phys.\ Rev.\ D80 (2009) 074017, arXiv:0907.1790  

\bibitem{Beneke:2009ye}
  M.~Beneke, M.~Czakon, P.~Falgari, A.~Mitov and C.~Schwinn,
  Phys.\ Lett.\ B690 (2010) 483, arXiv:0911.5166  

\bibitem{Becher:2009kw}
  T.~Becher and M.~Neubert,
  Phys.\ Rev.\ D79 (2009) 125004 [E: ibid.\ D80 (2009) 109901],  
  arXiv:0904.1021

\bibitem{Ferroglia:2009ep}
  A.~Ferroglia, M.~Neubert, B.~D.~Pecjak and L.~L.~Yang,
  Phys.\ Rev.\ Lett.\ 103 (2009) 201601, arXiv:0907.4791  

\bibitem{Vogt:2000ci}
  A.~Vogt,
  Phys.\ Lett.\ B497 (2001) 228, hep-ph/0010146  

\bibitem{Moch:2005ba}
  S.~Moch, J.A.M.~Vermaseren and A.~Vogt,
  Nucl.\ Phys.\ B726 (2005) 317, hep-ph/0506288  

\bibitem{Hagiwara:2008df}
  K.~Hagiwara, Y.~Sumino and H.~Yokoya,
  Phys.\ Lett.\ B666 (2008) 71, arXiv:0804.1014  

\bibitem{Kiyo:2008bv}
  Y.~Kiyo, J.H.~K\"uhn, S.~Moch, M.~Steinhauser and P.~Uwer,
  Eur.\ Phys.\ J.\ C60 (2009) 375, arXiv:0812.0919  

\bibitem{Beneke:2011mq}
  M.~Beneke, P.~Falgari, S.~Klein and C.~Schwinn,
  Nucl.\ Phys.\ B855 (2012) 695, arXiv:1109.1536  

\bibitem{Czarnecki:2001gi}
  A.~Czarnecki and K.~Melnikov,
  Phys.\ Rev.\ D65 (2002) 051501, hep-ph/0108233  

\bibitem{Pineda:2006ri}
  A.~Pineda and A.~Signer,
  Nucl.\ Phys.\ B762 (2007) 67, hep-ph/0607239  

\bibitem{Beneke:1999qg}
  M.~Beneke, A.~Signer and V.A.~Smirnov,
  Phys.\ Lett.\ B454 (1999) 137, hep-ph/9903260  

\bibitem{Ellis:1988sb}
  R.K.~Ellis and P.~Nason,
  Nucl.\ Phys.\ B312 (1989) 551

\bibitem{Ball:2001pq}
  R.D.~Ball and R.K.~Ellis,
  JHEP 05 (2001) 053, hep-ph/0101199  

\bibitem{Ellis:1990hw}
  R.K.~Ellis and D.~A.~Ross,
  Nucl.\ Phys.\ B345 (1990) 79  

\bibitem{Remiddi:1999ew}
  E.~Remiddi and J.~A.~M.~Vermaseren,
  Int.\ J.\ Mod.\ Phys.\ A15 (2000) 725, hep-ph/9905237  

\bibitem{Moch:2001zr}
  S.~Moch, P.~Uwer and S.~Weinzierl,
  J.\ Math.\ Phys.\ 43 (2002) 3363, hep-ph/0110083  

\bibitem{Weinzierl:2004bn}
  S.~Weinzierl,
  J.\ Math.\ Phys.\ 45 (2004) 2656, hep-ph/0402131

\bibitem{Kalmykov:2006hu}
  M.Y.~.Kalmykov, B.F.L.~Ward and S.~Yost,
  JHEP 02 (2007) 040, hep-th/0612240  

\bibitem{Kalmykov:2008ge}
  M.Y.~.Kalmykov and B.A.~Kniehl,
  Nucl.\ Phys.\ B809 (2009) 365, arXiv:0807.0567  

\bibitem{Huber:2007dx}
  T.~Huber and D.~Maitre,
  Comput.\ Phys.\ Commun.\ 178 (2008) 755, arXiv:0708.2443  

\bibitem{Maitre:2007kp}
  D.~Maitre,
  Comput.\ Phys.\ Commun.\ 183 (2012) 846, hep-ph/0703052  

\bibitem{Thorne:2006qt}
  R.S.~Thorne,
  Phys.\ Rev.\ D73 (2006) 054019, hep-ph/0601245  

\bibitem{Larin:1994va}
  S.A.~Larin, T.~van Ritbergen and J.A.M.~Vermaseren,
  Nucl.\ Phys.\ B 438 (1995) 278, hep-ph/9411260  

\bibitem{Chetyrkin:1997sg}
  K.~G.~Chetyrkin, B.A.~Kniehl and M.~Steinhauser,
  Phys.\ Rev.\ Lett.\ 79 (1997) 2184, hep-ph/9706430  

\bibitem{Buza:1996wv}
  M.~Buza, Y.~Matiounine, J.~Smith and W.L.~van Neerven,
  Eur.\ Phys.\ J.\ C1 (1998) 301, hep-ph/9612398  

\bibitem{Alekhin:2009ni}
  S.~Alekhin, J.~Bl\"umlein, S.~Klein and S.~Moch,
  Phys.\ Rev.\ D8 (2010) 014032, arXiv:0908.2766  

\bibitem{Forte:2010ta}
  S.~Forte, E.~Laenen, P.~Nason and J.~Rojo,
  Nucl.\ Phys.\ B834 (2010) 116, arXiv:1001.2312  

\bibitem{Zijlstra:1992qd}
  E.B.~Zijlstra and W.L.~van Neerven,
  Nucl.\ Phys.\ B383 (1992) 525  

\bibitem{Moch:1999eb}
  S.~Moch and J.A.M.~Vermaseren,
  Nucl.\ Phys.\ B573 (2000) 853, hep-ph/9912355  

\bibitem{Vermaseren:1998uu}
  J.A.M.~Vermaseren,
  Int.\ J.\ Mod.\ Phys.\ A14 (1999) 2037, hep-ph/9806280  

\bibitem{Vermaseren:2000nd}
  J.A.M.~Vermaseren,
  {\it New features of FORM}, math-ph/0010025  

\bibitem{vanNeerven:2000wp}
  W.L.~van Neerven and A.~Vogt,
  Phys.\ Lett.\ B490 (2000) 111, hep-ph/0007362

\bibitem{vanNeerven:2001pe}
  W.L.~van Neerven and A.~Vogt,
  Nucl.\ Phys.\ B603 (2001) 42, hep-ph/0103123
  
\bibitem{Moch:2001im}
  S.~Moch, J.A.M.~Vermaseren and A.~Vogt,
  Nucl.\ Phys.\ B621 (2002) 413, hep-ph/0110331

\bibitem{Vogt:2005dw}
  A.~Vogt, S.~Moch and J.A.M~Vermaseren,
  Acta Phys.\ Polon.\ B37 (2006) 683, hep-ph/0511112

\bibitem{Moch:2008fj}
  S.~Moch, J.A.M.~Vermaseren and A.~Vogt,
  Nucl.\ Phys.\ B813 (2009) 220, arXiv:0812.4168 

\bibitem{Alekhin:2012du}
  S.~Alekhin, J.~Bl\"umlein, and S.~Moch, arXiv:1202.4642

\bibitem{JimenezD:2008hf} 
  P.~Jimenez-Delgado and E.~Reya,
  Phys.\ Rev.\ D79 (2009) 074023, arXiv:0810.4274

\bibitem{JimenezD:2009tv}
  P.~Jimenez-Delgado and E.~Reya,
  Phys.\ Rev.\ D80 (2009) 114011, arXiv:0909.1711

\bibitem{Martin:2009iq}
  A.D.~Martin, W.J.~Stirling, R.S.~Thorne and G.~Watt,
  Eur.\ Phys.\ J.\ C63 (2009) 189, arXiv:0901.0002

\bibitem{Martin:2010db}
A.D.~Martin, W.J.~Stirling, R.S.~Thorne and G.~Watt,
  Eur.\ Phys.\ J.\ C70 (2010) 51, arXiv:1007.2624

\bibitem{Ball:2011uy}
  R.D.~Ball {\it et al.} [NNPDF Collaboration],
  Nucl.\ Phys.\ B855 (2012) 153, arXiv:1107.2652

\bibitem{Moch:2012mk}
  S.~Moch, P.~Uwer and A.~Vogt,
  arXiv:1203.6282 (Physics Letters B, to appear)

\bibitem{Baernreuther:2012ws}
  P.~Baernreuther, M.~Czakon and A.~Mitov,
  arXiv:1204.5201 

\bibitem{Gehrmann:2001pz}
  T.~Gehrmann and E.~Remiddi,
  Comput.\ Phys.\ Commun.\ 141 (2001) 296, hep-ph/0107173

\end{thebibliography}

}


{\footnotesize

\begin{thebibliography}{10}

\bibitem{Kawamura:2012cr}
H.~Kawamura, N.~Lo~Presti, S.~Moch, and A.~Vogt,
\newblock Nucl.Phys. {\bf B864}, 399 (2012), arXiv:1205.5727v1.

\end{thebibliography}

}
\end{document}